\documentclass[a4paper,english,cleveref,autoref,thm-restate,nolineno]{socg-lipics-v2021}
\usepackage[utf8]{inputenc}
\usepackage{mathtools}
\usepackage{thmtools}

\usepackage{xspace}

\usepackage{url}

\usepackage{color}
\usepackage[usenames,dvipsnames]{xcolor}

\usepackage{microtype}

\newcommand{\Su}{\ensuremath{\mathcal{S}}}%
\newcommand{\TG}{\ensuremath{\mathcal{T}(G)}}%

\newcommand{\cell}{cell\xspace}
\newcommand{\cells}{cells\xspace}

\newcommand{\ERS}{ERS\xspace}
\newcommand{\ERSs}{ERSs\xspace}
\newcommand{\parity}{parity\xspace} 
\newcommand{\parities}{parities\xspace}
\newcommand{\crossing}{\ensuremath\chi}

\newcommand{\curveC}{$\Gamma$\xspace}
\newcommand{\caratheodoryC}{$\Gamma_{ab\cup P}$\xspace}
\newcommand{\caratC}{$\Gamma_{\Delta}$\xspace}

\usepackage{mathtools}

\newtheorem{quest}{Question}
\usepackage{dsfont,etoolbox}

\newtheoremstyle{stepstyle}%
{}{}%
{}{}%
{}{\textcolor{darkgray}{.}}%
{ }%
{\textcolor{darkgray}{\sffamily\bfseries\thmname{#1}\thmnumber{ #2}}}
\theoremstyle{stepstyle}

\newcommand{\myhooktoc}[1]{}
\newtheoremstyle{casestyle}%
{}{}%
{}{}%
{}{\textcolor{darkgray}{.}}%
{ }%
{\textcolor{darkgray}{\sffamily\bfseries\thmname{#1}\thmnumber{ #2}:}\thmnote{ #3}\myhooktoc{\thmname{#1}\thmnumber{ #2}: \thmnote{ #3}}}
\theoremstyle{casestyle}
\newtheorem{case}{Case}

\newtheorem{subcase}{Case}

\newtheorem{subsubcase}{Case}

\AtBeginEnvironment{case}{\renewcommand\qedsymbol{\textcolor{darkgray}{\ensuremath{\triangleleft}}}}
\AtEndEnvironment{case}{\qed}%
\AtBeginEnvironment{subcase}{\renewcommand\qedsymbol{\textcolor{darkgray}{\ensuremath{\triangleleft}}}}
\AtEndEnvironment{subcase}{\qed}%
\AtBeginEnvironment{subsubcase}{\renewcommand\qedsymbol{\textcolor{darkgray}{\ensuremath{\triangleleft}}}}
\AtEndEnvironment{subsubcase}{\qed}%
\AtBeginEnvironment{subsubsubcase}{\renewcommand\qedsymbol{\textcolor{darkgray}{\ensuremath{\triangleleft}}}}
\AtEndEnvironment{subsubsubcase}{\qed}%
\AtBeginEnvironment{subsubsubsubcase}{\renewcommand\qedsymbol{\textcolor{darkgray}{\ensuremath{\triangleleft}}}}
\AtEndEnvironment{subsubsubsubcase}{\qed}%
\makeatletter
\@addtoreset{step}{theorem}
\@addtoreset{case}{theorem}
\@addtoreset{subcase}{case}
\@addtoreset{subsubcase}{subcase}
\@addtoreset{subsubsubcase}{subsubcase}
\@addtoreset{subsubsubsubcase}{subsubsubcase}
\makeatother

\newcommand{\emptyqed}{\renewcommand\qedsymbol{\textcolor{darkgray}{\ensuremath{\triangleleft}}}}
\newcommand{\noqed}{\renewcommand{\qed}{}}

\newcommand{\N}{\ensuremath{\mathds{N}}\xspace}
\newcommand{\R}{\ensuremath{\mathds{R}}\xspace}
\newcommand{\ett}{\ensuremath{\widetilde{e_2}}}
\newcommand{\ered}{\ensuremath{\widetilde{e_2}}}
\newcommand{\eblue}{\ensuremath{e_1}}
\newcommand{\Ha}{H_1}
\newcommand{\Hb}{H_2}
\newcommand{\Hc}{H}

\newcommand{\kmult}{\ensuremath{K_{n_1,\ldots,n_k}}\xspace}

\title{Drawings of Complete Multipartite Graphs Up to Triangle Flips}

\author{Oswin Aichholzer}{Graz University of Technology, Graz, Austria}{oaich@ist.tugraz.at}{https://orcid.org/0000-0002-2364-0583}{Partially supported by the Austrian Science Fund (FWF): W1230.}
\author{Man-Kwun Chiu}{Wenzhou-Kean University, Wenzhou, China}{mchiu@kean.edu}{https://orcid.org/0000-0001-7435-1020}{partially supported by ERC StG 757609. Part of the work was done while Chiu was at FU Berlin.}
\author{Hung P. Hoang}{ETH Z\"urich, Z\"urich, Switzerland}{hung.hoang@inf.ethz.ch}{https://orcid.org/0000-0001-7883-4134}{}
\author{Michael~Hoffmann}{ETH Z\"urich, Z\"urich, Switzerland}{hoffmann@inf.ethz.ch}{https://orcid.org/0000-0001-5307-7106}{supported by the Swiss National Science Foundation within the collaborative D-A-CH project \emph{Arrangements and Drawings} as SNSF project 200021E-171681.}
\author{Jan Kyn\v{c}l}{Charles University, Prague, Czech Republic}{kyncl@kam.mff.cuni.cz}{https://orcid.org/0000-0003-4908-4703}{supported by the grant no. 22-19073S of the Czech Science Foundation (GA\v{C}R).}
\author{Yannic Maus}{Graz University of Technology, Graz, Austria}{yannic.maus@ist.tugraz.at}{https://orcid.org/0000-0003-4062-6991}{}
\author{Birgit Vogtenhuber}{Graz University of Technology, Graz, Austria}{bvogt@ist.tugraz.at}{https://orcid.org/0000-0002-7166-4467}{Partially supported by Austrian Science Fund within the collaborative D-A-CH project \emph{Arrangements and Drawings} as FWF project \mbox{I 3340-N35}.}
\author{Alexandra Weinberger}{Graz University of Technology, Graz, Austria}{weinberger@ist.tugraz.at}{https://orcid.org/0000-0001-8553-6661}{Supported by the Austrian Science Fund (FWF): W1230.}

\authorrunning{Aichholzer, Chiu, Hoang, Hoffmann, Kyn\v{c}l, Maus, Vogtenhuber, Weinberger}

\Copyright{{Oswin Aichholzer and Man-Kwun Chiu and Hung P. Hoang and Michael Hoffmann and Jan Kyn\v{c}l and Yannic Maus and Birgit Vogtenhuber and Alexandra Weinberger}}

\ccsdesc[500]{Mathematics of computing~Combinatorics}
\ccsdesc[500]{Mathematics of computing~Graph theory}
\ccsdesc[500]{Human-centered computing~Graph drawings}

\keywords{Simple drawings, simple topological graphs, complete graphs, multipartite graphs, k-partite graphs, bipartite graphs, Gioan's Theorem, triangle flips, Reidemeister moves}

\hideLIPIcs

\newcommand\blfootnote[1]{%
	\begingroup
	\renewcommand\thefootnote{}\footnote{#1}%
	\addtocounter{footnote}{-1}%
	\endgroup
}

\begin{document}
\maketitle

\enlargethispage{6ex}
\begin{abstract}
For a drawing of a labeled graph, the rotation of a vertex or crossing is the cyclic order of its incident edges, represented by the labels of their other endpoints.
	The extended rotation system (ERS) of the drawing is the collection of the rotations of all vertices and crossings. A drawing is simple if each pair of edges has at most one common point. Gioan's Theorem states that for any two simple drawings of the complete graph~$K_n$ with the same crossing edge pairs, one drawing can be transformed into the other by a sequence of triangle flips (a.k.a.~Reidemeister moves of Type~3).
This operation refers to the act of moving one edge of a triangular \cell formed by three pairwise crossing edges over the opposite crossing of the \cell, via a local transformation.

We investigate to what extent Gioan-type theorems can be obtained for wider classes of graphs.
A necessary (but in general not sufficient) condition for two drawings of a graph to be transformable into each other by a sequence of triangle flips is that they have the same ERS. 
As our main result, we show that for the large class of complete multipartite graphs, this necessary condition is in fact also sufficient. We present two different proofs of this result, one of which is shorter, while the other one yields a polynomial time algorithm 
for which the number of needed triangle flips for graphs on $n$ vertices is bounded by $O(n^{16})$. 
The latter proof uses a Carath{\'e}odory-type theorem for simple drawings of complete multipartite graphs, which we believe to be of independent interest.

Moreover, we show that our Gioan-type theorem for complete multipartite graphs is essentially tight 
in the following sense:
For the complete bipartite graph $K_{m,n}$ minus two edges and $K_{m,n}$ plus one edge for any $m,n \ge 4$, as well as $K_n$ minus a 4-cycle for any $n \ge 5$,
there exist two simple drawings with the same ERS that cannot be transformed into each other using triangle flips.
So having the same ERS does not remain sufficient when removing or adding very few edges.
	\blfootnote{\hspace{-0.3cm}\begin{minipage}[l]{0.99\textwidth} \it 
		This work (without appendix) is available at the 39th International Symposium on Computational Geometry (SoCG 2023).
	\end{minipage}}
\end{abstract}

\section{Introduction}\label{sec:intro}
\begin{figure}[htb] 
	\centering
	\includegraphics[page=1]{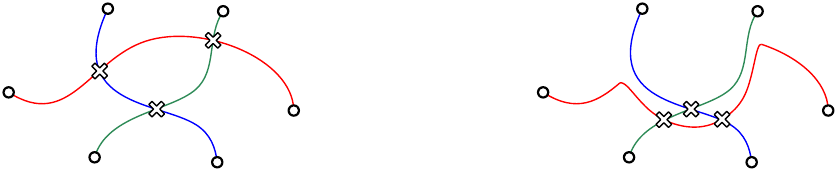}
	\caption{A sketch of a triangle flip.
		\label{fig:flip}}
\end{figure}

Gioan's Theorem states that any two simple drawings of the complete graph~$K_n$ in which the same pairs of edges cross can be transformed into each other (up to strong isomorphism) via a sequence of triangle flips.
Informally, a triangle flip is the act of moving one edge of a triangular \cell formed by three pairwise crossing edges over the opposite crossing of the \cell; see Figure~\ref{fig:flip} for an illustration of this operation and Section~\ref{sec:def} for the formal definition.

\enlargethispage{3ex}

Gioan's Theorem can be seen as a generalization of results on pseudolines by Ringel~\cite{ringel55} from~1955 and Roudneff~\cite{roudneff88} from~1988 to simple drawings of~$K_n$.
Gioan's conference paper~\cite{gioan} from 2005 contained a proof sketch only.
A full proof was first published in 2017 by Arroyo, McQuillan, Richter, and Salazar~\cite{gioan_proof}, who also 
coined the name ``Gioan's Theorem''. 
In 2021, Schaefer~\cite{schaefer21} generalized Gioan's~Theorem
to slightly sparser graphs, namely, simple drawings of~$K_n$ minus any non-perfect matching. 
A full version of Gioan's proof~\cite{gioan_final} finally appeared in 2022.

\begin{figure}[htb]
	\centering
	\includegraphics[page=16]{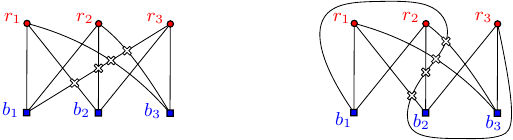}
	\caption {Two simple drawings of $K_{3,3}$ with the same crossing edge pairs and same rotations at all vertices but different rotations at all crossings involving the edge $b_1r_3$ and hence different \ERSs.}
	\label{fig:not_only_rs}
\end{figure}

A priori it is not clear how to generalize Gioan's Theorem beyond Schaefer's result.
For transforming drawings of general graphs via triangle flips, it is not sufficient to only have the same crossing edge pairs.
We should also consider the rotation of a vertex or edge crossing, which is defined as the cyclic order of emanating edges.
For example, Figure~\ref{fig:not_only_rs} shows two simple drawings of
the complete bipartite graph $K_{3,3}$ with the same crossing edge pairs and the same rotations of vertices, but different rotations of the crossings involving~$b_1r_3$. 
Observe that triangle flips do not change the rotations of crossings or vertices. A take-away from this observation is that for a Gioan-type theorem to hold, the rotations of all crossings and vertices must be the same in both drawings. A concept capturing exactly this necessity is the extended rotation system. The extended rotation system (ERS) of a drawing of a graph is the collection of the rotations of all vertices and crossings. In this light, one of the contributions of Gioan's Theorem is that for drawings of the complete graph, having the same crossing edge pairs is equivalent to having the same \ERS (up to global inversion)~\cite{gioan, gioan_final}.
This fact has been first stated by Gioan~\cite{gioan}; the first published proofs are by Kyn\v{c}l~\cite{kyncl2009,crossings_to_extended_rot}. 
An analogous statement for $K_n$ minus any non-perfect matching has been shown by Schaefer~\cite{schaefer21}. 
For complete multipartite graphs, this equivalence does not hold; see again \Cref{fig:not_only_rs}. 

As our main result, we show that having the same \ERS is sufficient to transform simple drawings of complete multipartite graphs into each other via triangle flips. We thus obtain a Gioan-type theorem for a large class of graphs that includes the before studied graphs, namely  complete graphs~\cite{gioan_proof,gioan,gioan_final,schaefer21} and complete graphs minus a non-perfect matching~\cite{schaefer21}.  

\begin{restatable}{theorem}{thmmain}\label{thm:main}
  Let $D_1$ and $D_2$ be two simple drawings of a complete multipartite graph 
  on the sphere~$\Su^2$ 
  with the same \ERS. Then there is a sequence of triangle flips that transforms $D_1$ into~$D_2$.
\end{restatable}

We also show that \cref{thm:main} is essentially tight in the sense that having the same ERS does not remain sufficient when removing or adding very few edges.

\begin{restatable}{theorem}{thmTight}\label{thm:tight}
For any~$m,n\geq 3$ and $K_{m,n}$ minus any two edges, there exist two simple drawings with the same \ERS that cannot be transformed into each other using triangle flips.
	The same holds for any~$n\geq 5$ and $K_n$ minus any four-cycle $C_4$, 
	as well as for any~$m\geq 4, n\geq 1$ and $K_{m,n}$ plus one edge between vertices in the bipartition class of size $m$.
\end{restatable}

The first part of \cref{thm:tight} implies that an analogue to Schaefer's generalization of Gioan's Theorem for $K_n$ minus a non-perfect matching cannot be achieved for complete bipartite graphs, not even for~$K_{m,n}$ minus a matching of size two. 
Note that $K_{m,n}$ with $m\geq 4$ and $n\geq 1$ is a subgraph of $K_{n+m}$ minus a $4$-cycle. 
Hence, the second part of \cref{thm:tight} implies that---perhaps counterintuitively---the set of graphs for which a Gioan-type theorem holds is not closed under adding edges. 
From the proof of \cref{thm:tight} it follows that \cref{thm:main} cannot be extended to any graph that contains a $K_5$ minus a four-cycle $C_4$ or a $K_{3,2}$ minus two edges incident to the same vertex of the smaller partition class, as an induced subgraph.

We present two different proofs of \cref{thm:main}. 
Our first proof 
uses a similar approach as the
proof of Gioan's Theorem
by Schaefer~\cite{schaefer21}.
His 
proof heavily relies on a (plane) spanning star as a basis for transforming one drawing into the other. 
While plane spanning stars exist in any simple drawing of~$K_n$, also minus a non-perfect matching, 
this is in general not the case for complete multipartite graphs.
However, any simple drawing of a complete multipartite graph~$G$ contains a
plane spanning tree~\cite{agpvw-sssd-22}. 
We show that for drawings of $G$ with the same \ERS, such a plane spanning tree can be used for transforming one drawing into the other.
The resulting proof is shorter and probably more elegant than the second proof. 
But it does not directly yield a polynomial time transformation algorithm, as it is still
an open question~\cite{agpvw-sssd-22} whether a plane spanning tree can be found in polynomial time.

Our second proof 
yields a polynomial time algorithm for the transformation.
It uses a similar approach as the  
proof of Gioan's Theorem by Arroyo, McQuillan, Richter, and Salazar~\cite{gioan_proof}. 
Several ingredients 
of their proof 
are known properties of drawings of complete graphs or follow directly from such properties, 
while it was unknown whether analogous statements hold for drawings of other graphs. 
Hence, for our proof we discover a number of useful, fundamental properties of simple drawings of complete multipartite graphs.
For example, we establish a Carath{\'e}odory-type theorem for them.

The classic Carath{\'e}odory Theorem states that if a point~$p \in \R^2$ lies in the convex hull of a set~$A\subset\R^2$ of~$n\ge 3$ points, then there exists a triangle spanned by points of~$A$ that contains~$p$. 
In the terminology of drawings, if a point~$p$ lies in a bounded \cell of a straight-line drawing~$D$ of~$K_n$ in~$\R^2$, then there 
exists a $3$-cycle~$C$ in~$D$ so that~$p$ lies in the bounded \cell of~$C$. 
This statement has been generalized to simple (not necessarily straight-line) drawings of~$K_n$~\cite{balko_fulek_kyncl_15, bfsss-tdmctcg-20}. 
However, it clearly does not generalize to arbitrary (non-complete) graphs; consider for example a simple drawing of a path with self-intersections that forms a bounded cell. 
A natural question is, for which classes of graphs this statement, or a variation of it, holds. 
We show that it holds for complete multipartite graphs if in addition to $3$-cycles---which might not exist in those graphs---we also allow $4$-cycles to contain~$p$.

\begin{restatable}[Carath{\'e}odory-type theorem for simple drawings of complete multipartite graphs]{theorem}{propcaratheodory}\label{prop:caratheodory}
	Let~$D$ be a simple drawing of a complete multipartite graph $G$ in the plane. For every point~$p$ in a bounded \cell of~$D$,
	there exists a cycle~$C$ of length three or four in~$D$ such that~$p$ is contained in a bounded \cell of~$C$. 
	This statement is tight in the sense that it may not hold for~$G$ minus one~edge.
\end{restatable}

\subparagraph{Number of triangle flips.}
Schaefer~\cite[Remark 3.3]{schaefer21} showed that for~$K_n$,
polynomially many triangle flips are sufficient and gave an upper bound of~$O(n^{20})$ for the number of required flips.
Using a different approach in our second proof of~\cref{thm:main}, we show an upper bound of~$O(n^{16})$ triangle flips for complete multipartite graphs on $n$ vertices.
We further present drawings which, regardless of the approach, require at least~$\Omega(n^{6})$ triangle~flips.

\subparagraph{Motivation and related work.}

Originally, rotation systems were invented to investigate embeddings of graphs on higher-genus surfaces~\cite{MR898434}. 
Nowadays they are widely used to represent drawings of graphs in the plane and to derive their structural properties. 
Gioan's Theorem implies that for simple drawings of complete graphs, the set of crossing pairs of edges determines the drawing's \ERS. 
Conversely,
for drawings of complete graphs, the rotation system determines which pairs of edges cross~\cite{kyncl2009,PT04}. 
These relations are crucial in the study 
of simple drawings of complete graphs, their generation and enumeration~\cite{aafhpprsv-agdsc-15,kyncl2009,kyncl2013}. 

For non-complete graphs, the literature on 
rotation systems for simple drawings is rather sparse. Besides the recent work of Schaefer~\cite{schaefer21}, 
we are only aware of work by Cardinal and Felsner~\cite{cardinalfelsner}, who investigate the realization of complete bipartite graphs as outer drawings. The main reason why there are no further results on rotation systems beyond drawings of complete graphs is the lack of known properties in these cases. Our work contributes towards the generalization of rotation systems to drawings of wider graph classes, not only by the main statement but also due to the structural results obtained along the way.

We note that rotation systems of drawings also play a role in a wider context.
For example, they are crucial in a recent breakthrough result devising an algorithm for the subpolynomial approximation of the crossing number for non-simple drawings of general graphs \cite{ChuzhoyTan2022}. 

The study of triangle flips has a long history in several different contexts.
In addition to the mentioned work on Gioan's Theorem~\cite{gioan_proof,gioan,gioan_final,schaefer21}, this in particular includes work on arrangements of pseudolines~\cite{fps-aapl-2022,ringel55,roudneff88,SnoeyinkHershberger1991}, knot theory~\cite{ab-tkc-27,ito13, kauffman89,l-pubrm-15,r-ebk-27,t-rmck-83, yamada89}, as well as on transforming curves on compact oriented surfaces~\cite{celmsstt-tcslm-18}.

\subparagraph{Outline.}
In \cref{sec:def},
we mainly state definitions, introduce notation, and give a characterization of complete multipartite graphs.
In \cref{sec:caratheodory,ssec:tight_pf} we sketch the proofs of the Carath{\'e}odory-type \cref{prop:caratheodory} and \cref{thm:tight}, respectively. 
\cref{sec:main_proof} is devoted to proving 
\cref{thm:main}, where the first proof is given nearly fully, and the second one is shortly sketched to explain the algorithm.
In \cref{sec:numberofflips} we present bounds on the required number of triangle flips derived from the second proof.
We conclude the paper with open questions in \Cref{sec:conclusion}.

\section{Definitions and preliminaries}
\label{sec:def}

A graph~$G=(V,E)$ is \emph{multipartite} if its vertex set~$V$ can be partitioned into~$k$ nonempty subsets~$V_1,\ldots,V_k$, for some~$k\in\N$, such that each~$V_i$, for~$i\in\{1,\ldots,k\}$, induces an independent set in~$G$, that is, no two vertices in~$V_i$ are adjacent. A \emph{complete multipartite graph} $G=(V,E)$ contains \emph{all} edges outside of the independent sets, that is, 
we have $E=\{v_iv_j\colon v_i\in V_i \ \wedge \ v_j\in V_j \ \wedge \ 1\le i<j\le k\}$. For a multiset~$\{n_1,\ldots,n_k\}$ of natural numbers, there is a unique (up to isomorphism) complete multipartite graph~\kmult with~$|V_j|=n_j$, for all~$j\in\{1,\ldots,k\}$. Note that both the empty graph on $n$ vertices (with~$k=1$ and~$n_1=n$) and the complete graph~$K_n$ (with~$k=n$ and~$n_1=\cdots=n_k=1$) are complete multipartite graphs. We also have the following useful characterization, whose proof is an easy graph-theoretic exercise. For completeness, we include it in \cref{sec:lem:claim}. 

\begin{restatable}{lemma}{theclaim}\label{lem:claim}
  A graph~$G=(V,E)$ is complete multipartite if and only if for every edge~$uv\in E$ and every vertex~$w\in V\setminus\{u,v\}$ we have~$uw\in E$ or~$vw\in E$ (or both).
\end{restatable}

\enlargethispage{3ex}
\vspace{-2ex}\subparagraph{Drawings.}  A \emph{drawing}~$\gamma$ of a graph~$G=(V,E)$ is a geometric representation of~$G$ by points and curves on an oriented surface~$\Su$. More precisely, every vertex~$v$ of~$G$ is mapped to a point~$\gamma_v$ on~$\Su$ and every edge~$uv$ of~$G$ is mapped to a 
simple (that is, continuous and not self-intersecting)
curve~$\gamma_{uv}$ on~$\Su$ with endpoints~$\gamma_u$ and~$\gamma_v$, such that: 

\vspace{-1.5ex}\begin{enumerate}
	\item Any two vertices are mapped to distinct points ($\gamma_u=\gamma_v\Longrightarrow u=v$, for all $u,v\in V$).
	\item No vertex is mapped to the relative interior of an edge ($\gamma_{uv}\cap\gamma_w=\emptyset$, for all $uv\in E$ and $w\in V\setminus\{u,v\}$).
	\item Every pair of curves~$\gamma_e,\gamma_f$, for $e\ne f$, intersects in at most finitely many points, each of which is either a common endpoint or a proper, transversal crossing.
\end{enumerate}

\vspace{-1ex} \noindent In this paper, we consider drawings on the sphere~$\Su^2$, except for a few places---specified explicitly---where we consider drawings in the plane~$\R^2$. All our graphs and drawings are labeled. Hence, we often identify vertices and edges with their geometric representation in a drawing. Any subgraph~$H$ of~$G$ induces a \emph{subdrawing}~$\gamma[H]$ that is obtained by restricting~$\gamma$ to the vertices and edges of~$H$. 
For a graph~$F$, an \emph{$F$-subdrawing} of~$\gamma$ is a subdrawing~$\gamma[H]$ that is induced by some subgraph~$H$ of~$G$ that is isomorphic to~$F$.
A drawing partitions $\Su$ into vertices (\emph{endpoints}) and \emph{crossings} of the curves~$\{\gamma_e\colon e\in E\}$, \emph{edge fragments} (the connected components of the curves~$\{\gamma_e\colon e\in E\}$ after removing all vertices and crossings), and \emph{\cells} (the connected components of~$\Su$ after removing all vertices, crossings, and edge fragments). For a cell~$C$ we denote by~$\partial C$ the \emph{boundary} of~$C$. A cell that is bounded by exactly three edge fragments is called a \emph{tricell}.

The class of drawings of a graph is vast and for many purposes too rich to be directly useful. To begin with, it is not clear in general how to represent a drawing using a finite amount of space. Two natural approaches to address this concern are to (1)~further restrict the class of drawings or (2)~study drawings on a much coarser level, up to some notion of isomorphism. In this work, we use a combination of both of these approaches.

\subparagraph{Simple drawings.} An example for the first approach are \emph{straight-line drawings} in the Euclidean plane (also known as \emph{geometric graphs}), where the geometry of an edge is uniquely determined by the location of its endpoints; see
the \emph{Handbook of Discrete and Computational Geometry}~\cite[Chapter 10]{Handbook3rdedition} and references therein. In this work, we consider a
more general class of drawings, which appear in the literature as \emph{simple drawings}~\cite{dlm-sgdbp-19},
\emph{good drawings}~\cite{AMRS18,EG_1973}, \emph{topological graphs}~\cite{PST03}, \emph{simple topological graphs}~\cite{kyncl2009}, and even just as \emph{drawings}~\cite{h-etdcg-98}.
In a simple drawing,  
every pair of edges has at most one point in common,
either a common endpoint or a proper crossing. 
Additionally, we may assume that no three edges meet at a common point.
Simple drawings are 
a combinatorial/topological generalization of straight-line drawings. If the graph~$G$ has $n$ vertices, then every simple drawing of~$G$ has $O(n^4)$ crossings, edge fragments, and \cells. 
Simple drawings are also important for crossing minimization because 
all crossing-minimal drawings 
are simple~\cite{crossing_minimal}.

\subparagraph{Strong isomorphism.} An example for the second approach is the notion of strong isomorphism for drawings, defined as follows. Two drawings~$\gamma$ and~$\eta$ of a graph~$G=(V,E)$ are \emph{strongly isomorphic}, denoted by~$\gamma\cong\eta$, if there exists an orientation-preserving homeomorphism\footnote{Strong isomorphism can also be defined for unlabeled drawings; then a mapping for the vertex sets is needed. The homeomorphism is sometimes not required to be orientation-preserving; then, e.g., mirror-images of drawings are also considered to be strongly isomorphic.} of~$\Su$ that maps~$\gamma$ to~$\eta$, that is, $\gamma_v\mapsto\eta_v$, for all $v\in V$, and $\gamma_e\mapsto\eta_e$, 
for all $e\in E$. 
A combinatorial formulation, which is equivalent for connected drawings, can be obtained as follows~\cite{kyncl2009}: 

\vspace{-1.5ex}\begin{enumerate}
	\item\label{prop:weak} The same pairs of edges cross. (This is called \emph{weak isomorphism}.)
	\item\label{prop:order} The order of crossings along each edge is the same.
	\item\label{prop:rs} At each vertex and crossing the \emph{rotation}, that is, the clockwise circular order of incident edges, is the same (see next paragraph for more details).
\end{enumerate}

\vspace{-1ex}\noindent The notion of strong isomorphism encapsulates basically everything that can be said about a drawing from a topological or combinatorial point of view: the order of edges around vertices and \cells, which pairs of edges cross, and in which order the crossings appear along an edge. For our purposes, we consider strongly isomorphic drawings to be equivalent.

\begin{figure}[bt] 
	\centering
	\includegraphics[page=1]{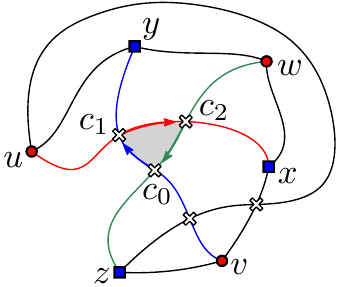}
	\hspace{3cm}
	\includegraphics[page=2]{iso}
	\caption{Two drawings of $K_{3,3}$ that have same \ERS but are not strongly isomorphic (because $ux$ crosses $vy$ and $wz$ in different order).
		The shaded tricell is an invertible triangle with different \parities in the two drawings.
		\label{fig:iso}}
\end{figure}

\enlargethispage{7ex}

\subparagraph{Extended rotation systems.} A coarser notion of equivalence can be obtained by requiring two drawings to have the same \emph{rotation system}, which is the collection of the rotations of all vertices. 
Property~\ref{prop:rs} 
in the above-mentioned combinatorial description uses a 
slightly stronger notion of equivalence, where also the 
rotations at crossings are the same in both drawings. 
More formally, the \emph{rotation of a crossing} ${\crossing}$ is the clockwise cyclic order of the four vertices of the crossing edge pair which is induced by the cyclic order of edge fragments around ${\crossing}$. 
(In other words, the rotation of a crossing~$\crossing$ is the rotation of an additional degree-4 vertex~$v_{\crossing}$ obtained by splitting the crossing edge pair at~$\crossing$ and replacing ${\crossing}$ by~$v_{\crossing}$.) 
The \emph{extended rotation system} (ERS) of a drawing is the collection of rotations of all vertices and crossings. 
Any two strongly isomorphic drawings have the same \ERS~\cite{kyncl2009}. But the converse is not true in general, as the example in~\cref{fig:iso} demonstrates.

\subparagraph{Crossing triangles.} 
In fact, the only difference between the two drawings in~\cref{fig:iso} with respect to strong isomorphism stems from
the tricell formed by the triple~$ux,vy,wz$ of pairwise crossing edges, which is shaded gray in the figure: In the left drawing, this \cell lies to the right of the oriented edge~$ux$, whereas in the right drawing, it lies to the left of~$ux$.
Given a simple drawing, a tricell~$\Delta$ in the subdrawing of three pairwise crossing edges~$e_1,e_2,e_3$ is called a \emph{crossing triangle}; the three edges~$e_1,e_2,e_3$ are said to \emph{span}~$\Delta$. 
Note that every edge triple in a simple drawing 
spans at most one crossing triangle. 
The following lemma shows that the crossing triangles are well-defined for complete multipartite graphs. We do not use it later on and it also follows from the proof of~\cref{thm:main}. However, for completeness, we include a direct (and much shorter) proof in \cref{sec:lem:triangles}. 

\begin{restatable}{lemma}{lemcrossing}\label{lem:triangles}
	In every simple drawing of a complete multipartite graph, the set of edge triples that span crossing triangles is uniquely determined by the \ERS. 
\end{restatable}

\vspace{-2ex}
\subparagraph{Invertible triangles and triangle flips.}
To formally define the triangle flip operation, globally fix an orientation $\pi$ of the edges of the abstract graph~$G$. 
This orientation can be arbitrary, but once we fix the graph, we also fix its orientation. 
With this orientation $\pi$, we can assign every crossing triangle a \parity as follows.  
The \emph{\parity} of a crossing triangle~$\Delta$ in a drawing is the parity (odd or even) of the number of bounding edges of~$\Delta$ such that~$\Delta$ lies to the left of the edge (when going along the edge according to its orientation). 
See \cref{fig:iso} for two drawings with even (left) and odd (right) parity of the crossing triangle.
A crossing triangle~$\Delta$ in a drawing $\gamma$ is \emph{invertible} if there exists another simple drawing $\gamma'\neq\gamma$ of the same graph $G$ with the same edge orientation $\pi$ and with the same
\ERS in which~$\Delta$ appears with the opposite \parity. 
In \Cref{prop:flippable}, we show that any invertible 
triangle in a drawing of a complete multipartite graph is empty in the sense that it does not contain any vertices.

Locally redrawing the edges of an empty crossing triangle and thereby changing its \parity is an 
elementary operation to transform a given drawing, say, the one in~\cref{fig:iso}(left), into a new drawing, such as the one in~\cref{fig:iso}(right). Up to strong isomorphism, there is a unique way for the redrawing. 
This operation is referred to as \emph{triangle flip}~\cite{gioan_proof}, \emph{triangle mutation}~\cite{gioan}, \emph{slide move}~\cite{schaefer21}, \emph{homotopy move}~\cite{celmsstt-tcslm-18, ito13}, or \emph{Reidemeister move of Type~3},
where the latter name has been extensively used\footnote{albeit in the context of knots also an above/below relationship among the curves is relevant} 
in knot theory~\cite{ab-tkc-27, kauffman89, l-pubrm-15, r-ebk-27,t-rmck-83, yamada89}.

\enlargethispage{3ex} 
\vspace{-0.1ex}
\subparagraph{Triangle flip graphs.} Based on the triangle flip as an elementary operation, we can define a meta graph whose vertices are drawings and whose edges correspond to triangle flips. We fix a graph~$G$ and consider all simple drawings of~$G$ on~$\Su$ up to strong isomorphism; these are the vertices of the triangle flip graph~$\TG$. Any two such drawings~$\gamma,\eta$ are connected by an edge in~$\TG$ if~$\eta$ can be obtained from~$\gamma$ by a single triangle flip. As triangle flips are reversible, edges are symmetric. So we consider~$\TG$ as an undirected graph.

\begin{figure}[htbp]
	\centering 
	\includegraphics[page=3]{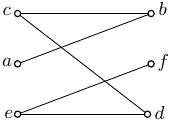}
	\hspace{3cm}
	\includegraphics[page=4]{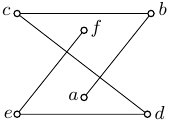}
	\caption{%
          Two drawings of a path with the same \ERS, but the order of crossings along the edge~$cd$ differs, thus, the drawings are not strongly isomorphic. Neither drawing contains any tricell to flip.\label{fig:no_tri}}
	\vspace{-3ex}
\end{figure}

Observe that a triangle flip does not change the rotation of any vertex or crossing, only the order of crossings along the edges changes. Therefore only drawings that have the same \ERS can be in the same component of~$\TG$. 
In general, the flip graph~$\TG$ may be disconnected. Consider, for instance, the two drawings of a path depicted in  in \cref{fig:no_tri}. As neither drawing contains any crossing triangle, both are isolated vertices in~$\TG$.

\section{A Carath{\'e}odory-type theorem for complete multipartite graphs} 
\label{sec:caratheodory}

This section is devoted to a proof outline of the Carath{\'e}odory-type \cref{prop:caratheodory}. 
The corresponding statement for simple drawings of~$K_n$, which is a direct generalization of the classic theorem for convex sets in~$\R^2$, was shown by Balko, Fulek, and Kyn\v{c}l~\cite{balko_fulek_kyncl_15}.
A simpler proof was given later by Bergold, Felsner, Scheucher, Schr{\"o}der, and Steiner~\cite{bfsss-tdmctcg-20}, 
whose proof idea we follow. 
The full proof of \cref{prop:caratheodory} can be found in Appendix~\ref{sec:appcaratheodory}. We give a proof outline here.

\begin{proof}[Sketch of Proof] 
If $G$ is empty or a star $K_{1,n}$, then the statement is vacuously true. So we assume that~$G$ is neither, and thus every pair of distinct vertices~$u,v\in V$ with~$uv\notin E$ has at least two distinct common neighbors. By studying a minimal counter-example we prove \cref{prop:caratheodory} by contradiction. To that aim, we consider a simple drawing $D$ of~$G$ and a point~$p$, such that the following holds: 

\vspace{-1.5ex}\begin{enumerate}
	\item The point $p$ is in a bounded \cell of~$D$.
	\item The point $p$ is not contained in a bounded \cell of any induced $C_i$-subdrawing of~$D$, for~$i\in\{3,4\}$.
	\item When removing any vertex from~$D$, the point $p$ lies in the unbounded \cell.
\end{enumerate}

Let $a$ be a vertex of $G$, and let $O$ be the smallest set of edges incident to $a$ such that removal of all edges of~$O$ from~$D$ puts~$p$ into the unbounded \cell of the resulting drawing~$D^-$. Then in~$D^-$ one can draw a simple curve~$P$ from~$p$ to the interior of the unbounded \cell of~$D$ so that~$P$ does not intersect any vertex or edge of~$D^-$. Subject to this constraint, we select~$P$ to minimize the number of crossings with edges of~$D$.  We show that we can assume every edge in $O$ crosses $P$ exactly once. 
Finally we consider an edge $ab \in O$, which crosses $P$ in a point $p_{ab}$, and analyze two cases depending on whether $ab$ crosses another edge between $a$ and $p_{ab}$ or not. We show that in both cases, $p$ is contained in a  bounded \cell of an induced $C_i$-subdrawing of~$D$, for~$i\in\{3,4\}$. 

\begin{figure}[bhtp]
	\centering
	\includegraphics[page=21]{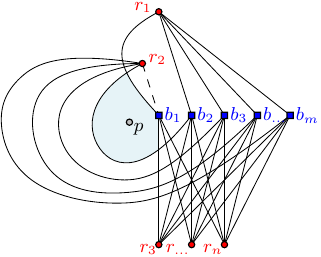}
	\caption {Drawing of $K_{m,n}$ minus one edge ($r_2b_1$, drawn dashed), based on~\Cref{fig:necessary}. 
		The point $p$ lies in a bounded cell, but in no $C_i$, for~$i\in\{3,4,5\}$.}
	\label{fig:minus1_necessary}
\end{figure}

To see that the theorem may not hold if we remove one edge from~$G$, consider the simple drawing of~$K_{m,n}$, $m,n\ge 2$, depicted in~\Cref{fig:minus1_necessary}.
When removing the edge $b_1r_2$, the point~$p$ still lies in a bounded cell, but any cycle that encloses~$p$ has at least six vertices. 
\end{proof}

\newpage
\section{Theorem~\ref{thm:main} is essentially tight}\label{ssec:tight_pf}

\cref{thm:tight} implies that \Cref{thm:main} is essentially tight: The removal or addition of very few edges may yield a graph for which the theorem does not hold. 
This implies that the class of graphs for which this Gioan-type theorem holds is not 
closed under the operation of taking (non-induced) subgraphs or supergraphs. We sketch the proof of \cref{thm:tight} by depicting the drawings we use to show tightness. 
The full proof can be found in Appendix~\ref{sec:apptight}. 

Each of \Cref{fig:necessary,fig:necessary:indep,fig:complete_minus_4,fig:Kmn_plus} contains two simple drawings of a graph with the same \ERS.
In all of them, the crossing order along $b_1r_1$ differs between the two drawings.
This order cannot be changed via triangle flips because the edges crossing $b_1r_1$ in different orders are pairwise non-crossing. 
\Cref{fig:necessary,fig:necessary:indep} cover the case of $K_{m,n}$ minus two adjacent or disjoint edges, 
\Cref{fig:complete_minus_4} is an extension of \Cref{fig:necessary} to $K_m$ minus a $4$-cycle, and \Cref{fig:Kmn_plus} shows subdrawings of \Cref{fig:complete_minus_4} that form a $K_{m-1,n+1}$ plus one edge.

\begin{figure}[htb]
	\centering
	\includegraphics[page=18]{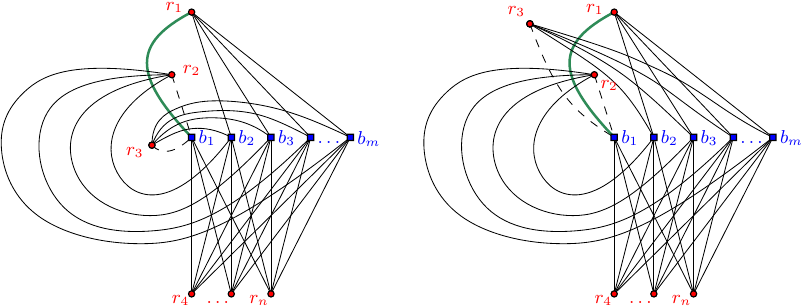}
	\caption {Two drawings of $K_{m,n}$ minus two adjacent edges $b_1r_2$ and $b_1r_3$ (drawn as dashed lines) that have the same \ERS but cannot be transformed into each other via triangle flips.
	}
	\label{fig:necessary}
\end{figure}

\begin{figure}[htb]
	\centering
	\includegraphics[page=10]{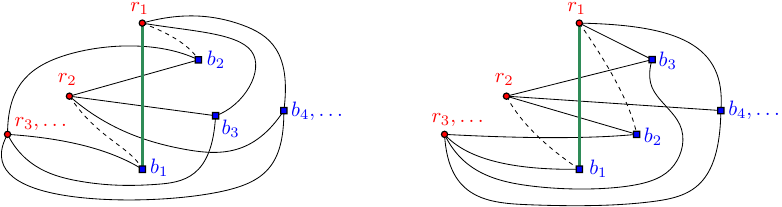}
	\caption {Two drawings of $K_{m,n}$ minus two independent edges $b_2r_1$ and $b_1r_2$ (drawn dashed) that have the same \ERS but cannot be transformed into each other via triangle flips.}
	\label{fig:necessary:indep}
\end{figure}

\begin{figure}[htb]
	\centering
	\includegraphics[page=20]{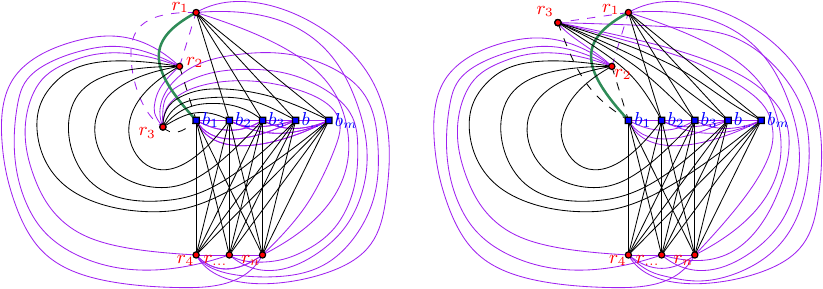}
	\caption {Two drawings of $K_{m}$ minus a $4$-cycle (drawn dashed) 
		that have the same \ERS, but cannot be transformed into each other via triangle flips. 
	}
	\label{fig:complete_minus_4}
\end{figure}

\begin{figure}[htb]
	\centering
	\includegraphics[page=19]{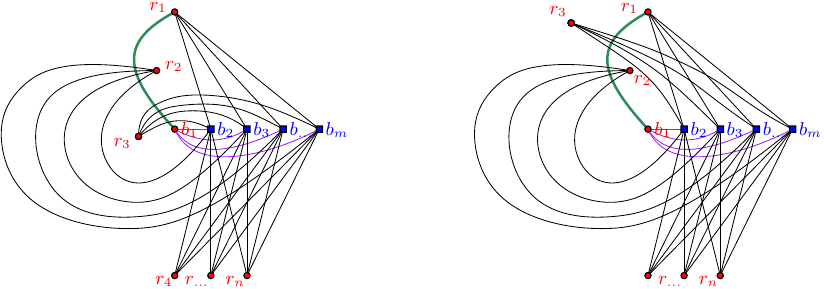}
	\caption {Two drawings of $K_{m-1,n+1}$ plus one edge ($b_1r_1$) that cannot be transformed into each other via triangle flips.}
	\label{fig:Kmn_plus}
\end{figure}

We remark that also two simple drawings with the same \ERS that cannot be transformed into each other via triangle flips exist for any graph that contains
(1) a $K_{5}$ minus a $4$-cycle, 
or (2) a $K_{2,3}$ minus two edges sharing a vertex in the bipartition class of cardinality two 
(where the list of induced subgraphs is not exhaustive).
This can be shown by choosing appropriate subdrawings in the construction from \cref{fig:complete_minus_4}.  

\newpage
\section{A Gioan-type theorem for complete multipartite graphs} 
\label{sec:main_proof}
In this section, we present our two proofs of~\cref{thm:main} and include a short algorithmic discussion of the second one. 

\subsection{First proof of Theorem~\ref{thm:main}}\label{ssec:newproof}
For our first proof of \Cref{thm:main}, we use the same general approach as Schaefer~\cite{schaefer21}.
To closely follow the lines of Schaefer, we also use homeomorphisms in this proof.

\begin{proof}
	Let $G$ be a complete multipartite graph, and let $D_1$ and $D_2$ be two simple drawings of~$G$ on~$\Su^2$ with the same \ERS.
	Let $R= \{r_1, r_2,\dots, r_n\}$ be a maximal independent set in~$G$ 
	and let $B= \{b_1, b_2,\dots, b_m\}$ denote the set of the remaining vertices. 
	Note that the graph on the vertex set $R\cup B$ together with all edges with an endpoint in $R$ and one in $B$ forms a complete bipartite graph $K_{n,m}$, and the set $R$ is an independent set in $G$ while $B$ might not necessarily be an independent set.

	By~\cite{agpvw-sssd-22}, the subdrawing of $D_1$ spanned by this $K_{n,m}$ 
	contains a spanning tree $T$ which is drawn crossing-free in this subdrawing and hence also in $D_1$.
	As $D_1$ and $D_2$ have the same crossing edge pairs, $T$ is drawn crossing-free in $D_2$ as well.
	Since the rotation systems of $D_1$ and $D_2$ are the same by assumption,
	the drawings of $T$ in $D_1$ and $D_2$ are homeomorphic.
	Thus there exists a drawing $D:\cong~D_1$ with the following properties.
\begin{enumerate}
	\item The drawing of $T$ is the same for both drawings $D$ and $D_2$, 
	implying that also the vertex locations are the same in both drawings.
	\item Considering the set of the  
	vertices and edges of $D$ and $D_2$ together as the \emph{combined drawing} of $D$ and~$D_2$, we denote the
	cyclical order of edges in $D$ and $D_2$ emanating from 
	a vertex as \emph{combined rotation} at that vertex. 
	For each edge $e$ of~$G$ (not in $T$) and each vertex $v$ of $e$, 
	the two drawings of $e$ are consecutive in the combined rotation at~$v$. 
	\item  For each edge $e$ of~$G$, the two drawings of $e$ are either identical or have only 
	finitely many points in common (two
	are its endpoints and the others
	are proper crossings). 
\end{enumerate}

	Our goal is to change $D$ via 
	triangle flips 
	(and orientation-preserving homeomorphisms) until we obtain~$D = D_2$. 
	Since the vertex locations in both drawings are the same, 
	we can speak about two drawings of an edge, one in $D$, and one in $D_2$, being the same or not.
	As in Schaefer's proof, we iteratively reduce the number of edges that are drawn differently in $D$ and $D_2$. 
	Let $E_=$ be the set of edges whose drawings in $D$ and $D_2$ are the same.
	Initially, $E_=$ contains at least all edges of $T$.
	If $E_=$ contains all edges of $G$ then we are done. 

	So suppose that this is not the case and consider an edge $e$ that is drawn differently in $D$ and $D_2$.
	Let $e_1$ and $e_2$ denote the curves representing $e$ in $D$ and $D_2$, respectively.
	Since $D$ and $D_2$ have the same \ERS, $e_1$ and $e_2$ cross the same edges of $T$ and they do so with the same crossing rotations.
	Moreover, the following lemma implies that they also cross those edges in the same order.
	The lemma can be proven relying on Lemma~\ref{lem:claim} and using a case distinction for drawings with six vertices. 

	\begin{restatable}{lemma}{orderOfCrossings}\label{lem:order_of_crossings}
	Let $D$ be a simple drawing of a complete multipartite graph $G$ on~$\Su^2$ and let $vw$ be an edge of $G$.
	Then for any pair of adjacent or disjoint edges crossed by $vw$, the \ERS of $D$ 
	determines the order in which $vw$ crosses them.
	\end{restatable}

Hence $e_1$ and $e_2$ are equivalent with respect to the drawing of $T$ (which is the same in $D$ and $D_2$), 
that is, $e_1$ has the same sequence of directed crossings with $T$ as~$e_2$.
	Let $\Gamma=e_1\cup e_2$ be the (not necessarily simple) closed curve formed by $e_1$ and $e_2$.
	A \emph{lens} in~$\Gamma$ is a cell of $\Gamma$ whose boundary is formed by 
	exactly two edge fragments of~$\Gamma$, where one is from~$e_1$ and one is from~$e_2$. 
	Next, consider the drawing $D_T$ of $T$ plus the drawings $e_1$ and $e_2$ of $e$.
	A lens of $\Gamma$ is called \emph{empty} if it contains no vertices of $T$ (and hence also no vertices of $G$) in its interior.
	With the next lemma, we show that $\Gamma$ forms an \emph{empty lens}. 
	This lemma is a special case of a result of Hass and Scott on intersecting curves on surfaces~\cite[Lemma 3.1]{HS85_intersections}, which is also known as the \emph{bigon criterion}~\cite[Section 1.2.4]{Farb12_primer}. Schaefer~\cite[Lemma 3.2]{schaefer21} gives an elementary proof in the planar (or spherical) case when the plane spanning tree $T$ is a star. However, he only uses that the star is a spanning subdrawing that is crossing-free and that $e_1$ and $e_2$ are equivalent with respect to the star. Thus, we can follow the proof line by line to obtain the result for any plane spanning tree $T$. 

	\begin{restatable}[\cite{Farb12_primer,HS85_intersections,schaefer21}]{lemma}{treeEmptyLens}\label{lem:tree_empty_lens}
		Let $D_1$ and $D_2$ be two simple drawings of a graph on~$\Su^2$ that
		contain the same crossing-free drawing $D_T$ of a spanning tree $T$ as a subdrawing.
		Let $e$ be an edge for which the drawings $e_1$ and $e_2$ differ,  
		but are equivalent with respect to $D_T$. Then $\Gamma=e_1\cup e_2$ forms an empty lens. 
	\end{restatable}

	Let $L$ be an empty lens of~$\Gamma$, which is formed by the edge fragments $\gamma_1$ of $e_1$ and $\gamma_2$ of $e_2$,
	respectively. 
	Each of the two points of $\gamma_1 \cap \gamma_2$ is either an endpoint or a crossing between $e_1$ and $e_2$.
	Recall that, in the combined drawing of $D$ and $D_2$, $e_1$ and $e_2$ are 
	consecutive in the combined rotation at each of their endpoints.
	Hence, independent of whether the points of $\gamma_1 \cap \gamma_2$ are crossings or endpoints, 
	$\gamma_2$ is what Schaefer calls a ``\emph{homotopic detour} of $\gamma_1$ on $e_1$''. 
	We next need his detour lemma, which we restate here using slightly different terminology (and for drawings on the sphere instead of in the plane). 

	\begin{lemma}[detour lemma~{\cite[Lemma 2.1]{schaefer21}}]\label{lem:detour_lemma}
		Let $\gamma_2$ be a homotopic detour of the arc $\gamma_1$ on the edge $e_1$ in a simple drawing of a graph. 
		Let $F$ be the set of edges which cross $\gamma_2$ at least twice.
		Then we can apply a sequence of triangle flips and homeomorphisms of the sphere~$\Su^2$ so
		that in the resulting drawing, $\gamma_1$ is routed arbitrarily close to $\gamma_2$, without intersecting it.
		The triangle flips and homeomorphisms only affect a small open neighborhood of the
		region bounded by $\gamma_1 \cup \gamma_2$, and only edges in $F$ and the $\gamma_1$ part of $e_1$ are redrawn.
	\end{lemma}

	Note that the set $F$ of edges that are affected by the transformation is disjoint from $E_=$,
	because any edge of $E_=$ is identical in $D$ and $D_2$ and hence intersects $\gamma_2$ at most once.

	If at least one of the points of $\gamma_1 \cap \gamma_2$ is a crossing, then after applying the detour lemma,
	we can redraw $e_1$ (via a homeomorphism) to have at least one fewer crossing with $e_2$ and repeat the process of
	applying 
	\cref{lem:tree_empty_lens,lem:detour_lemma}
	with the redrawn edge.
	
	If none of the points of $\gamma_1 \cap \gamma_2$ is a crossing, then $e_1 \cup e_2$ is a simple closed curve and 
	$\gamma_1=e_2$ is a homotopic detour of $\gamma_2=e_1$. Hence, after one final application of \cref{lem:detour_lemma},
	we can redraw $e_1$ to be identical to $e_2$. With this step, $e_2$ is added to $E_=$ and we have reduced the number 
	of edges differing between $D$ and $D_2$ by one.

	Repeating this process for the remaining differing edges we obtain two identical drawings.
	Omitting the homeomorphisms, the process yields a sequence of triangle flips for transforming $D_1$ into $D_2$ (up to strong isomorphism), which completes the proof of the theorem.
\end{proof} 

\subsection[Second Proof of Theorem~\ref{thm:main}]{Second Proof of \cref{thm:main}}\label{ssec:main_pf_outline}

Our second proof of \Cref{thm:main}, which we briefly outline here,
uses the same general framework as the proof of Gioan's Theorem by Arroyo, McQuillan, Richter, and Salazar~\cite{gioan_proof}.
We present only a brief outline here.
The full proof, starting with a more detailed outline, can be found in \cref{app:main_proof}. 

\begin{proof}[Sketch of Proof] 
We consider
two simple drawings~$D_1$ and~$D_2$ of a complete (multipartite)
graph~$G=(V,E)$ with the same \ERS, and one of them, say~$D:=D_1$, is
iteratively transformed to become ``more similar'' to the
other. Similarity is measured using a subgraph~$X$ of~$G$ for which we
demand as an invariant that the induced subdrawings~$D[X]$
and~$D_2[X]$ are strongly isomorphic. 
In each iteration, we will add one edge to~$X$ and then perform a sequence of
triangle flips in~$D$ so as to reestablish the invariant.

Initially, we establish the invariant in the following way.
	As in the first proof, we consider an independent set $R\subseteq V$ of
	vertices such that~$G$ contains a complete bipartite subgraph
	between~$R$ and~$B:=V\setminus R$.
If~$G$ is complete, then~$R$
contains a single vertex only; in general, it may contain several
vertices. We then pick one vertex~$r_0\in R$ and start by taking~$X$
to be the maximal induced substar of~$G$ centered at~$r_0$ (which
includes all vertices of~$B$).
Then the invariant
holds because both drawings have the same rotation system by
assumption.

We then consider the (possibly) remaining vertices of~$R$ in an
arbitrary order. Let~$r\in R$ be the next vertex to be
considered. First, we show that the position of~$r$ in the
induced---strongly isomorphic, by the invariant---subdrawings~$D[X]$
and~$D_2[X]$ is consistent, that is, the vertex~$r$ lies in the same
(according to isomorphism) face of these drawings
(\cref{prop:rot_cell}, whose proof uses the Carath{\'e}odory-type \cref{prop:caratheodory}).

We add the edges incident to~$r$ one by one to~$X$. When adding an
edge~$rb$ to~$X$ to obtain~$X'=X\cup \{rb\}$, the drawings~$D[X']$
and~$D_2[X']$ may not be strongly isomorphic because the edge~$rb$ may
cross other edges in a different order in both drawings. We consider a
sort of overlay~$O$ of both drawings~$D[X']$ and~$D_2[X']$, in which
the two versions of~$rb$ together form a closed curve~\curveC with 
$O(|V(X')|^4)$ self-crossings (\cref{prop:existence:ett}), where $|V(X')|$ is the number of vertices of $X'$. 
In~\curveC, we can identify a nice substructure, which we refer to as a \emph{free lens}, and show that it always exists (\cref{prop:empty:lens:e1:e2}).
A lens in~\curveC is \emph{free} if it does not contain any vertex of~$O$; it may contain edge crossings, though. 
Each such edge crossing corresponds to an invertible triangle
in~$D$. Invertible triangles are empty of vertices
(\cref{prop:flippable}), not only of the vertices in~$X$ but also of
the (possibly) not yet considered vertices of~$R$. Hence,
the edges of $D$ that cross an invertible triangle~$\Delta$ behave similarly to a
collection of pseudolines inside~$\Delta$, except that not all pairs need
to cross.
Let $m$ 
be the number of edges that cross~$\Delta$.
Using a classic sweeping algorithm by Hershberger and Snoeyink~\cite[Lemma 3.1]{SnoeyinkHershberger1991},
all $m$ 
edges can be ``swept'' out of $\Delta$ via triangle flips in $D$, where the total number of flips is bounded by $O(m^3)$.
After these flips, $\Delta$ has become a crossing triangle and can be flipped in~$D$.
Processing all invertible
triangles inside a selected free lens in this fashion effectively
destroys this lens. And after iteratively destroying all free lenses,
the resulting drawing~$D[X']$ is strongly isomorphic to~$D_2[X']$.

After all vertices in~$R$ and the complete bipartite subgraph of~$G$
between~$R$ and~$B$ have been added to~$X$, we add the remaining edges
(the ones with both endpoints in~$B$) in exactly the same fashion as described
above.
\end{proof}

While the outline of the above proof mostly follows the one for $K_n$~\cite{gioan_proof}, its core challenges lie in the proofs of several statements, whose analogues are known for $K_n$ but not for complete multipartite graphs. 
These include in particular the proofs of Lemmata 11, 13, and 14 (while the proof of Lemma 12 is quite straightforward). 
We discuss these challenges at the end of \Cref{ssec:main_pf}. 

\subparagraph*{Algorithmic complexity.} The above proof yields an algorithm that can be implemented using standard computational geometry data structures.
Its runtime is polynomial in the size of the input and the number of performed triangle flips.

\section{On the number of triangle flips}\label{sec:numberofflips}

The \emph{flip distance} between two different drawings of a complete multipartite graph with the same \ERS is the minimum number of triangle flips that are required to transform one drawing into the other.
This section is devoted to obtain bounds on the flip distance.
 
For an upper bound, 
Schaefer~\cite[Remark~3.3]{schaefer21} showed that any two simple drawings of~$K_n$ with the same rotation system can be transformed into each other with at most~$O(n^{20})$ triangle flips.
Using our second proof of \cref{thm:main}, we can obtain an upper bound of $O(n^{16})$ on the flip distance between two simple drawings of any complete multipartite graph with $n$ vertices and the same \ERS (and thus also for such drawings of $K_n$). 

\begin{restatable}{theorem}{thmbound}\label{thm:bound}
	Let $D_1$ and $D_2$ be two simple drawings of a complete multipartite graph $G$ 
	on~$\Su^2$ with $n$ vertices and 
	with the same \ERS. Then $D_1$ can be transformed into~$D_2$ via a sequence of $O(n^{16})$ triangle flips, obtained via the algorithm in the second proof of \Cref{thm:main}.
\end{restatable}

\begin{proof}
	We analyze the number of flips performed through the second proof of \cref{thm:main}.
	Recall that in this proof, we iteratively consider the edges of~$G$. 
	We perform flips in a drawing~$D$ (initially set to $D_1$) so that the subdrawings of $D$ and $D_2$ 
	induced by the already considered edges become (strongly) isomorphic.

	When considering a new edge $e$, we imagine to add both versions of it (the one from~$D$ and the one from $D_2$) to the already isomorphic subdrawing $X$ of $D$ and $D_2$.
	By \cref{prop:existence:ett}, this can be done in such a way that in the combined drawing, the two copies of $e$ have $O(|V(X)|^4)=O(n^4)$ crossings, 
	where $|V(X)|$ is the number of vertices of $X$. 

	Let $C$ be the closed curve formed by the two copies of $e$.
	In order to transform $D$ to make the drawing of $e$ in $D$ isomorphic to the one in $D_2$, we iteratively resolve a free lens of $C$. 
	At every iteration, we reduce the number of crossings of $C$, except for the very last iteration (i.e, for the very last lens). 
	Hence, the number of lenses we need to resolve when processing $e$ is bounded by $O(n^4)$ as well.
	To resolve a free lens, we need to flip all inverted triangles in this lens that have $e$ as an edge, of which there are at most $O(n^4)$ many. 
	For one inverted triangle $\Delta$ intersected by $m=O(n^2)$ edges, this can be done with $O(m^3)=O(n^6)$ flips.
	Hence resolving one free lens can be achieved with $O(n^4)\cdot O(n^6)=O(n^{10})$ flips.

	Repeating this for all lenses of $C$ and for each of the $O(n^2)$ edges of $G$, we obtain an upper bound of 
	$O(n^2) \cdot O(n^4) \cdot O(n^{10})= O(n^{16})$ for the total number of triangle flips.
\end{proof}

\begin{figure}[htb]
	\centering
	\includegraphics[page=2]{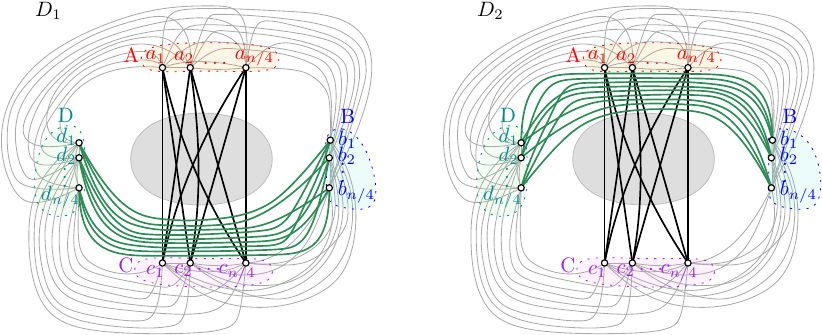}
	\caption {Two simple drawings of $K_n$ with the same \ERS whose flip distance is~$\Omega(n^6)$. %
	}
	\label{fig:lowerbound}
\end{figure}

\begin{restatable}{theorem}{distanceLower}\label{thm:distance_lower}
	Let $G$ be a multipartite graph $G$ with $n$ vertices that contains two vertex-disjoint subgraphs each forming a $K_{m,m}$ for some $m=\Theta(n)$. 
	Then $G$ admits two drawings $D_1$ and $D_2$ with the same \ERS that have flip distance $\Omega(n^6)$.
\end{restatable}
\begin{proof}[Proof idea]
To transform the two drawings of~$K_n$ in \cref{fig:lowerbound} into each other, 
each of the $\Theta(n^2)$ edges $b_id_j$ needs to be moved over the $\Theta(n^4)$ crossings formed by edges $a_kc_\ell$, yielding the $\Omega(n^6)$ lower bound.
	An according example of two drawings of a $K_{m,m}$ can be obtained by disregarding all edges $a_ib_j$ and $c_id_j$.
	A detailed proof can be found in Appendix~\ref{sec:appnumberofflips_lower}. 
\end{proof}

\section{Conclusion \& open questions}
\label{sec:conclusion}

We have shown that Gioan's Theorem holds for complete multipartite graphs (\cref{thm:main}), extending previous results~\cite{gioan_proof,gioan,gioan_final,schaefer21}. 
Further, we have shown that the class of graphs for which an analogue statement holds is not closed under addition or removal of edges (\cref{thm:tight}). 
We also provide several obstructions such that Gioan's Theorem does not hold for any graph that contains any of these obstructions as a substructure.  
However, the list of obstructions is probably incomplete.
A full characterization of graphs for which a Gioan-type statement for drawings with the same \ERS holds remains open.

\begin{quest}
	Can we completely characterize all graphs for which a Gioan-type 
	theorem holds for drawings with the same \ERS? 
\end{quest}

Further, having the same \ERS is not the only necessary condition for a Gioan-type statement to hold. 
Another example of such a condition is that incident or disjoint edges must have the same crossing orders over all drawings. 
The constructions in the proof of \cref{thm:tight} rely on violating this condition. 

\begin{quest}
	Can we characterize all graphs for which a Gioan-type
	theorem holds for classes of drawings which fulfill (subsets of) obviously necessary conditions? 
\end{quest}

In \cref{sec:caratheodory}, we have proven a Carath{\'e}odory-type theorem for simple drawings of complete multipartite graphs with the same \ERS (\cref{prop:caratheodory}).
It would be interesting to know for which further classes of graphs a similar statement is true.

Naturally, we would also like to narrow or even close the gap between the lower bound of~$\Omega(n^6)$ and the upper bound of $O(n^{16})$  for the flip distance, obtained in \cref{sec:numberofflips}.

\begin{quest}
	What is the worst case flip distance between two simple drawings of a complete multipartite graph on~$n$ vertices with a given \ERS? 
\end{quest}

\bibliography{gioan_multipartite_arxiv}

\begin{thebibliography}{10}

\bibitem{aafhpprsv-agdsc-15}
Bernado~M. \'Abrego, Oswin Aichholzer, Silvia Fern\'andez-Merchant, Thomas
  Hackl, J\"urgen Pammer, Alexander Pilz, Pedro Ramos, Gelasio Salazar, and
  Birgit Vogtenhuber.
\newblock {{All Good Drawings of Small Complete Graphs}}.
\newblock In {\em Proc. $31^{st}$ European Workshop on Computational Geometry
  EuroCG '15}, pages 57--60, Ljubljana, Slovenia, 2015.
\newblock URL:
  \url{http://www.csun.edu/~sf70713/publications/all_good_drawings_2015.pdf}.

\bibitem{agpvw-sssd-22}
Oswin Aichholzer, Alfredo Garc{\'i}a, Irene Parada, Birgit Vogtenhuber, and
  Alexandra Weinberger.
\newblock Shooting stars in simple drawings of {$K_{m,n}$}.
\newblock In Patrizio Angelini and Reinhard von Hanxleden, editors, {\em Graph
  Drawing and Network Visualization}, pages 49--57, Cham, 2023. Springer
  International Publishing.
\newblock \href {https://doi.org/10.1007/978-3-031-22203-0_5}
  {\path{doi:10.1007/978-3-031-22203-0_5}}.

\bibitem{ab-tkc-27}
James~W. Alexander and Garland~B. Briggs.
\newblock On types of knotted curves.
\newblock {\em Ann. Math.}, 28(1/4):562--586, 1927.
\newblock \href {https://doi.org/10.2307/1968399} {\path{doi:10.2307/1968399}}.

\bibitem{gioan_proof}
Alan Arroyo, Dan McQuillan, R.~Bruce Richter, and Gelasio Salazar.
\newblock Drawings of {$K_n$} with the same rotation scheme are the same up to
  triangle-flips ({G}ioan’s theorem).
\newblock {\em Australas. J. Comb.}, 67(2):131--144, 2017.
\newblock URL: \url{https://ajc.maths.uq.edu.au/pdf/67/ajc_v67_p131.pdf}.

\bibitem{AMRS18}
Alan Arroyo, Dan McQuillan, R.~Bruce Richter, and Gelasio Salazar.
\newblock Levi's lemma, pseudolinear drawings of {$K_n$}, and empty triangles.
\newblock {\em J. Graph Theory}, 87(4):443--459, 2018.
\newblock \href {https://doi.org/10.1002/jgt.22167}
  {\path{doi:10.1002/jgt.22167}}.

\bibitem{balko_fulek_kyncl_15}
Martin Balko, Radoslav Fulek, and Jan Kyn\v{c}l.
\newblock Crossing numbers and combinatorial characterization of monotone
  drawings of {$K_n$}.
\newblock {\em Discrete Comput. Geom.}, 53:107--143, 2015.
\newblock \href {https://doi.org/10.1007/s00454-014-9644-z}
  {\path{doi:10.1007/s00454-014-9644-z}}.

\bibitem{bfsss-tdmctcg-20}
Helena Bergold, Stefan Felsner, Manfred Scheucher, Felix Schröder, and Raphael
  Steiner.
\newblock Topological drawings meet classical theorems from convex geometry.
\newblock In {\em Proc. 28th Internat. Sympos. Graph Drawing}, volume 12590 of
  {\em Lecture Notes Comput. Sci.}, pages 281--294. Springer-Verlag, 2020.
\newblock \href {https://doi.org/10.1007/978-3-030-68766-3_22}
  {\path{doi:10.1007/978-3-030-68766-3_22}}.

\bibitem{cardinalfelsner}
Jean Cardinal and Stefan Felsner.
\newblock Topological drawings of complete bipartite graphs.
\newblock {\em J. Comput. Geom}, 9(1):213--246, 2018.
\newblock \href {https://doi.org/10.20382/jocg.v9i1a7}
  {\path{doi:10.20382/jocg.v9i1a7}}.

\bibitem{celmsstt-tcslm-18}
Hsien{-}Chih Chang, Jeff Erickson, David Letscher, Arnaud de~Mesmay, Saul
  Schleimer, Eric Sedgwick, Dylan Thurston, and Stephan Tillmann.
\newblock Tightening curves on surfaces via local moves.
\newblock In {\em Proc. 29th ACM-SIAM Sympos. Discrete Algorithms}, pages
  121--135, 2018.
\newblock \href {https://doi.org/10.1137/1.9781611975031.8}
  {\path{doi:10.1137/1.9781611975031.8}}.

\bibitem{ChuzhoyTan2022}
Julia Chuzhoy and Zihan Tan.
\newblock A subpolynomial approximation algorithm for graph crossing number in
  low-degree graphs.
\newblock In {\em Proceedings of the 54th Annual ACM SIGACT Symposium on Theory
  of Computing}, STOC 2022, page 303–316, New York, NY, USA, 2022.
  Association for Computing Machinery.
\newblock \href {https://doi.org/10.1145/3519935.3519984}
  {\path{doi:10.1145/3519935.3519984}}.

\bibitem{dlm-sgdbp-19}
Walter Didimo, Giuseppe Liotta, and Fabrizio Montecchiani.
\newblock A survey on graph drawing beyond planarity.
\newblock {\em {ACM} Comput. Surv.}, 52(1):4:1--4:37, 2019.
\newblock \href {https://doi.org/10.1145/3301281} {\path{doi:10.1145/3301281}}.

\bibitem{EG_1973}
Paul Erd\H{o}s and Richard~K. Guy.
\newblock Crossing number problems.
\newblock {\em Am. Math. Mon.}, 88:52--58, 1973.
\newblock \href {https://doi.org/10.2307/2319261} {\path{doi:10.2307/2319261}}.

\bibitem{Farb12_primer}
Benson Farb and Dan Margalit.
\newblock {\em A primer on mapping class groups}, volume~49 of {\em Princeton
  Mathematical Series}.
\newblock Princeton University Press, Princeton, NJ, 2012.

\bibitem{fps-aapl-2022}
Stefan Felsner, Alexander Pilz, and Patrick Schnider.
\newblock Arrangements of approaching pseudo-lines.
\newblock {\em Discrete Comput. Geom.}, 67:380--402, 03 2022.
\newblock \href {https://doi.org/10.1007/s00454-021-00361-w}
  {\path{doi:10.1007/s00454-021-00361-w}}.

\bibitem{gioan}
Emeric Gioan.
\newblock Complete graph drawings up to triangle mutations.
\newblock In {\em Proc. 31st Internat. Workshop Graph-Theoret. Concepts Comput.
  Sci.}, volume 3787 of {\em Lecture Notes Comput. Sci.}, pages 139--150.
  Springer, 2005.
\newblock \href {https://doi.org/10.1007/11604686_13}
  {\path{doi:10.1007/11604686_13}}.

\bibitem{gioan_final}
Emeric Gioan.
\newblock Complete graph drawings up to triangle mutations.
\newblock {\em Discrete Comput. Geom.}, 67:985--1022, 2022.
\newblock \href {https://doi.org/10.1007/s00454-021-00339-8}
  {\path{doi:10.1007/s00454-021-00339-8}}.

\bibitem{MR898434}
Jonathan~L. Gross and Thomas~W. Tucker.
\newblock {\em Topological graph theory}.
\newblock Wiley-Interscience Series in Discrete Mathematics and Optimization.
  John Wiley \& Sons Inc., New York, 1987.
\newblock A Wiley-Interscience Publication.

\bibitem{h-etdcg-98}
Heiko Harborth.
\newblock Empty triangles in drawings of the complete graph.
\newblock {\em Discrete Math.}, 191:109--111, 1998.
\newblock \href {https://doi.org/10.1016/S0012-365X(98)00098-3}
  {\path{doi:10.1016/S0012-365X(98)00098-3}}.

\bibitem{HS85_intersections}
Joel Hass and Peter Scott.
\newblock Intersections of curves on surfaces.
\newblock {\em Isr. J. Math.}, 51(1-2):90--120, 1985.

\bibitem{ito13}
Noboru Ito and Yusuke Takimura.
\newblock (1,2) and weak (1,3) homotopies on knot projections.
\newblock {\em J. Knot Theory Ramif.}, 22(14), 2013.
\newblock \href {https://doi.org/10.1142/S0218216513500855}
  {\path{doi:10.1142/S0218216513500855}}.

\bibitem{kauffman89}
Louis~H. Kauffman.
\newblock Invariants of graphs in three-space.
\newblock {\em Trans. Am. Math. Soc.}, 311(2):697--710, 1989.
\newblock \href {https://doi.org/10.1090/S0002-9947-1989-0946218-0}
  {\path{doi:10.1090/S0002-9947-1989-0946218-0}}.

\bibitem{kyncl2009}
Jan Kyn\v{c}l.
\newblock Enumeration of simple complete topological graphs.
\newblock {\em Eur. J. Comb.}, 30:1676--1685, 2009.
\newblock \href {https://doi.org/10.1016/j.ejc.2009.03.005}
  {\path{doi:10.1016/j.ejc.2009.03.005}}.

\bibitem{crossings_to_extended_rot}
Jan Kyn\v{c}l.
\newblock Simple realizability of complete abstract topological graphs in {P}.
\newblock {\em Discrete Comput. Geom.}, 45:383--399, 2011.
\newblock \href {https://doi.org/10.1007/s00454-010-9320-x}
  {\path{doi:10.1007/s00454-010-9320-x}}.

\bibitem{kyncl2013}
Jan Kyn\v{c}l.
\newblock Improved enumeration of simple topological graphs.
\newblock {\em Discrete Comput. Geom.}, 50:727--770, 2013.
\newblock \href {https://doi.org/10.1007/s00454-013-9535-8}
  {\path{doi:10.1007/s00454-013-9535-8}}.

\bibitem{l-pubrm-15}
Marc Lackenby.
\newblock A polynomial upper bound on {Reidemeister} moves.
\newblock {\em Ann. Math.}, 82(2):491--564, 2015.
\newblock \href {https://doi.org/10.4007/annals.2015.182.2.3}
  {\path{doi:10.4007/annals.2015.182.2.3}}.

\bibitem{PST03}
J{\'{a}}nos Pach, J{\'{o}}zsef Solymosi, and G{\'{e}}za T{\'{o}}th.
\newblock Unavoidable configurations in complete topological graphs.
\newblock {\em Discrete Comput. Geom.}, 30(2):311--320, 2003.
\newblock \href {https://doi.org/10.1007/s00454-003-0012-9}
  {\path{doi:10.1007/s00454-003-0012-9}}.

\bibitem{PT04}
J\'{a}nos Pach and G\'{e}za T\'{o}th.
\newblock How many ways can one draw a graph?
\newblock {\em Combinatorica}, 26(5):559--576, 2006.
\newblock \href {https://doi.org/10.1007/s00493-006-0032-z}
  {\path{doi:10.1007/s00493-006-0032-z}}.

\bibitem{r-ebk-27}
Kurt Reidemeister.
\newblock Elementare {Begr{\"u}ndung} der {Knotentheorie}.
\newblock {\em Abh. Math. Sem. Univ. Hamburg}, 5:24--32, 1927.
\newblock \href {https://doi.org/10.1007/BF02952507}
  {\path{doi:10.1007/BF02952507}}.

\bibitem{ringel55}
Gerhard Ringel.
\newblock Teilungen der {Ebene} durch {Geraden} oder topologische {Geraden}.
\newblock {\em Math. Z.}, 64:79--102, 1956.
\newblock \href {https://doi.org/10.1007/BF01166556}
  {\path{doi:10.1007/BF01166556}}.

\bibitem{roudneff88}
Jean-Pierre Roudneff.
\newblock Tverberg-type theorems for pseudoconfigurations of points in the
  plane.
\newblock {\em Eur. J. Comb.}, 9(2):189--198, 1988.
\newblock \href {https://doi.org/10.1016/S0195-6698(88)80046-5}
  {\path{doi:10.1016/S0195-6698(88)80046-5}}.

\bibitem{schaefer21}
Marcus Schaefer.
\newblock Taking a detour; or, {Gioan’s} theorem, and pseudolinear drawings
  of complete graphs.
\newblock {\em Discrete Comput. Geom.}, 66:12--31, 2021.
\newblock \href {https://doi.org/10.1007/s00454-021-00296-2}
  {\path{doi:10.1007/s00454-021-00296-2}}.

\bibitem{SnoeyinkHershberger1991}
Jack {Snoeyink} and John {Hershberger}.
\newblock Sweeping arrangements of curves.
\newblock In {\em Discrete and Computational Geometry: Papers from the DIMACS
  Special Year}, volume~6 of {\em DIMACS}, pages 309--349. AMS, 1991.
\newblock \href {https://doi.org/10.1090/dimacs/006/21}
  {\path{doi:10.1090/dimacs/006/21}}.

\bibitem{crossing_minimal}
László~A. Székely.
\newblock A successful concept for measuring non-planarity of graphs: the
  crossing number.
\newblock {\em Discr. Math.}, 276(1):331--352, 2004.
\newblock 6th International Conference on Graph Theory.
\newblock \href {https://doi.org/10.1016/S0012-365X(03)00317-0}
  {\path{doi:10.1016/S0012-365X(03)00317-0}}.

\bibitem{Handbook3rdedition}
Csaba~D. T{\'o}th, Joseph O'Rourke, and Jacob~E. Goodman, editors.
\newblock {\em Handbook of Discrete and Computational Geometry, Third Edition}.
\newblock Chapman and Hall/CRC, 2017.
\newblock URL: \url{https://www.routledge.com/9781498711395}.

\bibitem{t-rmck-83}
Bruce Trace.
\newblock On the {Reidemeister} moves of a classical knot.
\newblock {\em Proc. Amer. Math. Soc.}, 89(4):722--724, 1983.
\newblock \href {https://doi.org/10.1090/S0002-9939-1983-0719004-4}
  {\path{doi:10.1090/S0002-9939-1983-0719004-4}}.

\bibitem{yamada89}
Sh\^{u}ji Yamada.
\newblock An invariant of spatial graphs.
\newblock {\em J. Graph Theory}, 13(5):537--551, 1989.
\newblock \href {https://doi.org/10.1002/jgt.3190130503}
  {\path{doi:10.1002/jgt.3190130503}}.

\end{thebibliography}

\newpage
\appendix

\section{Missing proofs of Section~\ref{sec:def}}
\subsection{Characterizing complete multipartite graphs (Proof of Lemma~\ref{lem:claim})}\label{sec:lem:claim}

\theclaim*
\begin{proof}
	  ``$\Rightarrow$'': If~$uv\in E$, then~$u$ and~$v$ are in different sets of the partition. So~$w$ can be in the same partition set with at most one of~$u$ or~$v$. Without loss of generality assume that~$u$ and~$w$ are in different partition sets. Then~$uw\in E$.
	
	  ``$\Leftarrow$'': Define the non-adjacency relation~$\Bumpeq$ on~$V\times V$ by~$u\Bumpeq v\iff uv\notin E$. We claim that~$\Bumpeq$ is an equivalence relation. Reflexivity and symmetry are obvious. For transitivity suppose that~$u\Bumpeq v$ and~$v\Bumpeq w$. Then~$uv\notin E$ and~$vw\notin E$. If~$v\in\{u,w\}$, then~$u\Bumpeq w$ is immediate. So suppose that~$v\in V\setminus\{u,w\}$. We want to show that~$u\Bumpeq w$, that is, $uw\notin E$. Suppose to the contrary that~$uw\in E$. Then by the assumption in the statement of the lemma, applied to the edge~$uw$ and the vertex~$v$, we have~$vu\in E$ or~$vw\in E$, in contradiction to~$u\Bumpeq v$ and~$v\Bumpeq w$. Therefore, we have~$uw\notin E$ and~$u\Bumpeq w$, which proves that~$\Bumpeq$ is transitive.
	
		Let~$V_1,\ldots,V_k$ denote the partition of~$V$ into equivalence classes according to~$\Bumpeq$. Clearly, by definition of~$\Bumpeq$, each~$V_i$, for~$i\in\{1,\ldots,k\}$, is an independent set in~$G$, and for each pair~$u\in V_i$ and~$v\in V_j$ with~$1\le i<j\le k$ we have~$u\not\Bumpeq v$, that is, $uv\in E$. In other words, the graph~$G$ is complete multipartite with vertex partition~$V_1, \ldots V_k$. 
\end{proof}

\subsection{\ERS determines crossing triangles (Proof of Lemma~\ref{lem:triangles})}\label{sec:lem:triangles}

\lemcrossing*

\begin{proof}
	Consider a multipartite graph~$G=(V,E)$ on~$n$ vertices. If there is no simple drawing of~$G$ that contains a crossing triangle, then the statement of the lemma holds. In particular, this is the case for~$n\le 5$ because six vertices are required to span a crossing triangle. So we may suppose that~$n\ge 6$, and that there exists a drawing~$\Gamma$ of~$G$ that contains a crossing triangle~$T$. Let $ab,cd,ef\in E$ be the three edges that define~$T$. Three pairwise crossing edges in a simple drawing can only form two possible arrangements, up to labeling and (strong) isomorphism: Type~1 (\cref{fig:tri:1}) corresponds to a crossing triangle (that is, all vertices are on the boundary of one \cell) and Type~2 (\cref{fig:tri:2}) has two vertices on the boundary of one \cell and four vertices on the boundary of the other \cell. 
	
	\begin{figure}[thbp]
		\begin{minipage}[t]{.32\linewidth}
			\centering\includegraphics[page=2]{triangles}
			\subcaption{Type~1}\label{fig:tri:1} 
		\end{minipage}
		\begin{minipage}[t]{.32\linewidth}
			\centering\includegraphics[page=1]{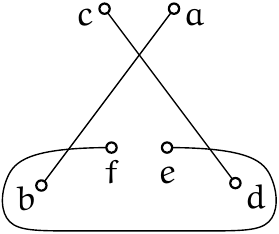}
			\subcaption{Type~2}\label{fig:tri:2}
		\end{minipage}
		\begin{minipage}[t]{.32\linewidth}
			\centering\includegraphics[page=3]{triangles}
			\subcaption{Type~2}\label{fig:tri:3}
		\end{minipage}
		\vspace{0.5ex}
		\caption{Possible arrangements formed by three pairwise crossing edges.\label{fig:tri}}
	\end{figure}
	
	So suppose for the sake of contradiction that there exists another drawing~$\Gamma'$ of~$G$ with the same \ERS as~$\Gamma$ in which the edges $ab$, $cd$, and $ef$ form an arrangement of Type~2. As the Type~1 arrangement is completely symmetric, we may fix its labeling without loss of generality as indicated in \cref{fig:tri:1} and select $ef$ to be the edge in~$\Gamma'$ that has both vertices on the same \cell of the Type~2 arrangement. Then the remaining vertices are uniquely determined by the \ERS; the two possible cases are depicted in \cref{fig:tri:2,fig:tri:3}. We will consider both cases in order. In each case we add some of the remaining edges of~$G$, which we know to exist by \cref{lem:claim}, to arrive at a contradiction in all cases. 
	
	\begin{case}[In~$\Gamma'$, the three edges form the arrangement depicted in \cref{fig:tri:2}]\label{triangles:case:1} 
		By \cref{lem:claim} we have $ae\in E$ or $af\in E$.

		\begin{subcase}[$ae\in E$]\label{triangles:case:1.1}
			See \cref{fig:tri:4,fig:tri:5} for an illustration.
			Then in~$\Gamma'$ the edge~$ae$ crosses~$cd$. Consider the subdrawing $\Gamma'[\{a,b,e,f\}]$: Vertex~$c$ is inside the tricell that is bounded by the edge~$ae$ and vertex~$d$ is not. There are two different ways to draw the edge~$ae$ in~$\Gamma$ so that it crosses~$cd$, see the red and blue curve in \cref{fig:tri:4}: In one case (red curve), both vertices~$c$ and~$d$ are in the tricell of~$\Gamma[\{a,b,e,f\}]$ that is bounded by the edge~$ae$; in the other case (blue curve), no vertex is inside this tricell. In either case this yields a contradiction to \cref{prop:Ksmall}.
		\end{subcase}
		
		\begin{figure}[thbp]
			\begin{minipage}[t]{.24\linewidth}
				\centering\includegraphics[page=4]{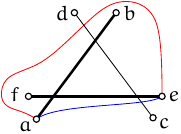}
				\subcaption{$\Gamma$: $ae\in E$}\label{fig:tri:4} 
			\end{minipage}\hfill
			\begin{minipage}[t]{.24\linewidth}
				\centering\includegraphics[page=5]{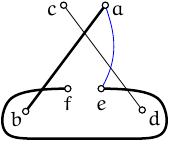}
				\subcaption{$\Gamma'$: $ae\in E$}\label{fig:tri:5}
			\end{minipage}\hfill
			\begin{minipage}[t]{.24\linewidth}
				\centering\includegraphics[page=6]{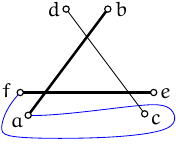}
				\subcaption{$\Gamma$: $af\in E$}\label{fig:tri:6}
			\end{minipage}\hfill
			\begin{minipage}[t]{.24\linewidth}
				\centering\includegraphics[page=7]{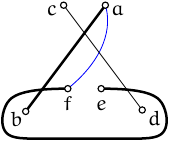}
				\subcaption{$\Gamma'$: $af\in E$}\label{fig:tri:7}
			\end{minipage}
			\vspace{0.5ex}
			\caption{%
				\cref{triangles:case:1} in the proof of \cref{lem:triangles}.\label{fig:tricase}}
		\end{figure}
		
		\begin{subcase}[$af\in E$]\label{triangles:case:1.2}
			See \cref{fig:tri:6,fig:tri:7} for an illustration. 
			Then in~$\Gamma'$ the edge~$af$ crosses~$cd$, and there is only one way to draw the edge~$af$ in~$\Gamma$ so that it crosses~$cd$ with the same rotation, see the blue curve in \cref{fig:tri:6}. Observe that in $\Gamma[\{a,b,e,f\}]$, the vertex~$c$ is contained in the tricell that is bounded by the edge~$af$, whereas in $\Gamma'[\{a,b,e,f\}]$ the tricell that is bounded by the edge~$af$ does not contain any vertex. This yields a contradiction to \cref{prop:Ksmall}.
		\end{subcase}\noqed
	\end{case}
	
	\begin{case}[In~$\Gamma'$, the three edges form the arrangement depicted in \cref{fig:tri:3}]\label{triangles:case:2}  
		By \cref{lem:claim} we have $de\in E$ or $df\in E$.
		
		\begin{subcase}[$de\in E$]\label{triangles:case:2.1}
			See \cref{fig:tri:8,fig:tri:9} for an illustration. 
			Then in~$\Gamma'$ the edge~$de$ crosses~$ab$. Consider the subdrawing $\Gamma'[\{c,d,e,f\}]$: Vertex~$b$ is inside the tricell that is bounded by the edge~$de$ and vertex~$a$ is not. There are two different ways to draw the edge~$de$ in~$\Gamma$ so that it crosses~$cd$ with the same rotation, see the red and blue curve in \cref{fig:tri:4}: In one case (red curve), both vertices~$a$ and~$b$ are in the tricell of~$\Gamma[\{c,d,e,f\}]$ that is bounded by the edge~$de$; in the other case (blue curve), no vertex is inside this tricell. In either case this yields a contradiction to \cref{prop:Ksmall}. 
		\end{subcase}
		
		\begin{figure}[thbp]
			\begin{minipage}[t]{.24\linewidth}
				\centering\includegraphics[page=8]{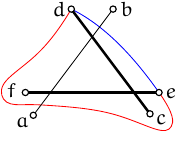}
				\subcaption{$\Gamma$: $de\in E$}\label{fig:tri:8} 
			\end{minipage}\hfill
			\begin{minipage}[t]{.24\linewidth}
				\centering\includegraphics[page=9]{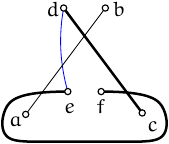}
				\subcaption{$\Gamma'$: $de\in E$}\label{fig:tri:9}
			\end{minipage}\hfill
			\begin{minipage}[t]{.24\linewidth}
				\centering\includegraphics[page=10]{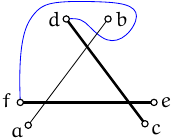}
				\subcaption{$\Gamma$: $df\in E$}\label{fig:tri:10}
			\end{minipage}\hfill
			\begin{minipage}[t]{.24\linewidth}
				\centering\includegraphics[page=11]{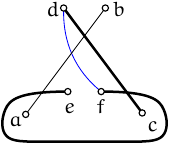}
				\subcaption{$\Gamma'$: $df\in E$}\label{fig:tri:11}
			\end{minipage}
			\vspace{0.5ex}
			\caption{\cref{triangles:case:2} in the proof of \cref{lem:triangles}.\label{fig:tricase2}}
		\end{figure}
		
		\begin{subcase}[$df\in E$]\label{triangles:case:2.2}
			See \cref{fig:tri:10,fig:tri:11} for an illustration. 
			Then in~$\Gamma'$ the edge~$df$ crosses~$ab$, and there is only one way to draw the edge~$df$ in~$\Gamma$ so that it crosses~$cd$ with the same rotation, see the blue curve in \cref{fig:tri:10}. Observe that in $\Gamma[\{c,d,e,f\}]$, the vertex~$b$ is contained in the tricell that is bounded by the edge~$df$, whereas this tricell does not contain any vertex in $\Gamma'[\{c,d,e,f\}]$. This yields a contradiction to \cref{prop:Ksmall}.
		\end{subcase}\noqed
	\end{case}
	This completes the proof of \cref{lem:triangles}.
\end{proof}

\section{Missing proof of Section~\ref{sec:caratheodory}: A Carath{\'e}odory-type theorem} 
\label{sec:appcaratheodory}
In this section, we present the proof of \cref{prop:caratheodory}. 
As stated before, the corresponding result has been shown for simple drawings of~$K_n$ by Balko, Fulek, and Kyn\v{c}l~\cite{balko_fulek_kyncl_15}, and by Bergold, Felsner, Scheucher, Schr{\"o}der, and Steiner~\cite{bfsss-tdmctcg-20}. We follow the proof idea of the latter, 
but the adaptation to the multipartite setting---specifically, the proof of the explicit \emph{claim} in the proof below---is nontrivial.

\propcaratheodory*
\begin{proof}
	If $G$ is empty or a star $K_{1,n}$, then the statement is vacuously true, since $D$ contains no bounded cell and hence the set of possible choices for $p$ is empty.
	So we may assume that $G$ is neither empty nor a star, and hence every pair of distinct vertices~$u,v\in V$ with~$uv\notin E$ has at least two distinct common neighbors.
	
	Suppose for the sake of contradiction that there exists a simple drawing~$D$ of~$G$ such that there is a point $p$ in a bounded \cell of~$D$ but $p$ is not contained in a bounded \cell of any induced $C_i$-subdrawing of~$D$, for~$i\in\{3,4\}$. We choose $D$ to be minimal with respect to the number of vertices. That is, if we remove any vertex (and all its incident edges) from~$D$, then~$p$ lies in the unbounded \cell.
	
	Let~$a$ be any vertex of the graph, and let~$O$ be a smallest set of edges incident to~$a$ so that removal of all edges of~$O$ from~$D$ puts~$p$ into the unbounded \cell of the resulting drawing~$D^-$. Then in~$D^-$ one can draw a simple curve~$P$ from~$p$ to \emph{infinity} (any point in the interior of the unbounded \cell of~$D$) so that~$P$ does not intersect any vertex or edge of~$D^-$. Subject to this constraint, we select~$P$ to minimize the number of crossings with edges of~$D$. Observe that by the minimality of~$O$, every edge~$o\in O$ is crossed at least once by~$P$, and adding~$o$ to~$D^-$ puts~$p$ into a bounded \cell of the resulting drawing. Also note that the edges in~$O$ are pairwise non-crossing because~$D$ is a simple drawing and all edges in~$O$ are incident to the common vertex~$a$.
	
	\begin{claim*}\label{claim:P_crossings}
		We may assume that every edge in~$O$ crosses~$P$ exactly once.
	\end{claim*}
	\begin{proof}
		Suppose that there is an edge~$ab\in O$ that crosses~$P$ at least twice. Trace the edge~$ab$ from~$b$ to~$a$ and denote by~$\chi_i$, for~$i=1,\ldots,k$, the $i$-th crossing with~$P$ encountered along the way. By assumption we have~$k\ge 2$.  Now we have two curves between~$\chi_1$ and~$\chi_2$, one along the edge~$ab$ and another one along~$P$. Denote the former curve by~$\gamma_{ab}$ and the latter one by~$\gamma_P$. Together they form a closed Jordan curve~\caratheodoryC$=\gamma_{ab}\cup \gamma_P$. Denote the bounded region of~\caratheodoryC by~$R$.
		
		We claim that there is an edge~$uw$ in~$D$ such that~$u\in R$ and~$w\notin R$.
		To see this, observe that $P'=(P\setminus \gamma_p) \cup \gamma'_{ab}$ is a curve from $p$ to \emph{infinity}, where $\gamma'_{ab}$ is a close copy of $\gamma_{ab}$ such that $\gamma'_{ab}\cap ab=\{\chi_1,\chi_2\}$.
		If no edge of~$D$ crosses~$\gamma_{ab}$, then~$P'$ would have fewer crossings than~$P$ (at least one), in contradiction to the minimality of~$P$. Hence, there is an edge~$uw$ of~$D$ that crosses~$\gamma_{ab}$. As~$D$ is simple, the crossing edges~$ab$ and~$uw$ do not share an endpoint and cross exactly once. As only edges incident to~$a$ may cross~$P$, the edge~$uw$ does not cross~$P$. Thus, the edge~$uw$ has exactly one crossing with~\caratheodoryC, and so exactly one of its endpoints is in~$R$. Without loss of generality we take~$u\in R$ and~$w\not\in R$. 
		We distinguish two cases, depending on whether or not $b \in R$.
		
		\begin{case}[$b\in R$]
			Then $bw$ is not an edge of~$G$, as any simple curve from~$b$ to~$w$ crosses~\caratheodoryC, but an edge~$bw$ can neither cross~$ab$ (as an incident edge) nor~$P$ (as not being incident to~$a$). Hence both $aw$ and $bu$ are edges of~$G$ (by \cref{lem:claim} applied to $w$ with $ab$ and to $b$ with $uw$, respectively), implying that $abuw$ forms a $C_4$ in $G$.
			As~$\chi_1$ and~$\chi_2$ are the first two crossings of $ab$ with~$P$ when traversing~$ab$ from~$b$ to~$a$, we have~$p\in R$.
			Further, as neither~$bu$ nor~$uw$ crosses~$P$ (not being adjacent to~$a$), 
			we conclude that~$p$ lies in a bounded \cell of the $C_4$-subdrawing induced by~$abuw$, 
			which completes the proof in this case.
		\end{case}
		\begin{case}[$b\notin R$]
			Then, for analogous reasons as in the first case, $bu$ is not an edge of~$G$ and both $bw$ and $au$ are edges of~$G$, 
			implying that $abwu$ forms a $C_4$ in $G$.
			As the edge~$bw$ is not incident to~$a$, it must not cross~$P$. 
			Thus, the closed curve that, starting from~$b$, follows the edge~$ba$ up to its crossing with~$uw$, then follows~$uw$ to~$w$, and then returns to~$b$ via the edge~$bw$, crosses~$P$ exactly once, at~$\chi_1$, and hence contains~$p$ in its interior.
			Therefore, the point~$p$ lies in a bounded \cell of the $C_4$-subdrawing induced by~$abwu$, 
			which concludes the proof of this case.
		\end{case} 
		Altogether, this completes the proof of the claim. \emptyqed
	\end{proof}
	
	For an edge~$ab\in O$ denote by~$p_{ab}$ the unique (by the claim) point in~$ab\cap P$. We conclude by again considering two cases (numbered 3 and 4 for convenience).
	
	\begin{case}[There exist edges~$ab\in O$ and~$cd$ in~$D$ such that~$cd$ crosses~$ab$ between~$a$ and~$p_{ab}$] 
		Then by \cref{lem:claim} at least one of~$c$ or~$d$ is adjacent to~$b$ in~$G$. Without loss of generality assume that~$bc$ is an edge. Consider the closed curve~\caratC that starts at the crossing~$ab\cap cd$, follows~$cd$ to~$c$, then follows~$cb$ to~$b$, and then follows~$ab$ back to~$ab\cap cd$. The path~$P$ crosses~\caratC exactly once, namely on the part along~$ab$. (It crosses there by assumption, only once along~$ab$ by the claim, and not at all along the edges~$bc$ and~$cd$ because these are not incident to~$a$.) Thus, $p$ is bounded by \caratC (or, equivalently, $p$ and \emph{infinity} are on different sides of~\caratC).
		If $a$ and $d$ are adjacent in $G$, it follows that $p$ lies in a bounded \cell of the $C_4$-subdrawing induced by~$abcd$. If $a$ and $d$ are not adjacent in $G$, then, by Lemma~\ref{lem:claim}, the vertex $a$ must be adjacent to $c$ in $G$, and $d$ must be incident to~$b$. Since $ab$ crosses~$cd$, the sides of the 3-cycles $abc$ and $abd$ containing $p$ intersect exactly in the region bounded by~\caratC, which contains $p$ and consequently does not contain \emph{infinity}. Hence, \emph{infinity} can be on the same side of $p$ in at most one of the 3-cycles $abc$ and $abd$. Thus, $p$ lies in a bounded \cell in (at least) one of those 3-cycles, which concludes the proof in this case.
	\end{case}
	
	\begin{case}[In~$D$, every edge~$ab\in O$ is uncrossed between~$a$ and~$p_{ab}$]
		Let~$ab$ be the first edge from~$O$ that is crossed by~$P$. Then we can reroute~$P$ to not cross~$ab$ at~$p_{ab}$ but instead follow~$ab$ to~$a$, without crossing any edge of~$D$. Let~$ac$ be the next edge incident to~$a$ along the same \cell. We claim that~$ac\notin O$. To see this, suppose for the sake of contradiction that~$ac\in O$. Then we continue to route~$P$ from~$a$ along~$ac$ to~$p_{ac}$, without crossing any edge of~$D$, and from there along its original track to \emph{infinity}. The resulting path from~$p$ to \emph{infinity} crosses strictly fewer edges from~$D$ than~$P$, in contradiction to the crossing-minimality of~$P$. Thus, the claim holds and~$ac\notin O$; in particular, $P\cap ac=\emptyset$. If~$bc\in E$, then~$p$ lies in the bounded \cell of the $3$-cycle~$abc$ because~$P$ crosses~$abc$ exactly once, at~$p_{ab}$. 
		Otherwise, we have~$bc\notin E$. 
		As stated at the very beginning of the proof, every pair of non-adjacent distinct vertices has at least two distinct common neighbors, 
		and so $b$ and~$c$ have at least two distinct common neighbors in~$G$. One of these neighbors is~$a$; let~$d\ne a$ denote another common neighbor of~$b$ and~$c$. As~$P$ crosses exactly one edge, namely~$ab$, of the $C_4$-subdrawing induced by~$abcd$, it follows that~$p$ lies in a bounded \cell of this subdrawing.
	\end{case}
	
	In summary, in every case we have established a~$C_i$-subdrawing of~$D$, for~$i\in\{3,4\}$ so that~$p$ lies in a bounded \cell of the subdrawing.
	This concludes the main part of the proof.
	
	\begin{figure}[htb]
		\centering
		\includegraphics[page=6]{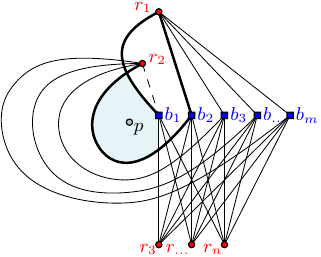}
		\caption {Drawing of $K_{m,n}$ minus one edge ($r_2b_1$, drawn dashed), based on~\Cref{fig:appnecessary}. 
			The point $p$ lies in a bounded cell, but in no $C_i$, for~$i\in\{3,4,5\}$.}
		\label{fig:appminus1_necessary}
	\end{figure}
	
	To see that the theorem may not hold if we remove one edge from~$G$, consider the simple drawing of~$K_{m,n}$, for~$m,n\ge 2$, depicted in~\Cref{fig:appminus1_necessary}. The only missing edge is~$r_2b_1$. The point~$p$ lies in a bounded cell. To form a cycle that encloses~$p$, we need the edge~$r_1b_1$ and the vertex~$r_2$, which, in a cycle, has two incident edges connecting to $b_i$ and $b_j$, for~$2 \leq i < j$. Thus, any cycle that encloses~$p$ has at least six vertices.
\end{proof}

\section{Missing proof of Section~\ref{ssec:tight_pf}: Tightness of Theorem~\ref{thm:main}}\label{sec:apptight}
This section is devoted to the proof of \cref{thm:tight}, which is sketched in Section~\ref{ssec:tight_pf}. The figures of Section~\ref{ssec:tight_pf} are repeated here for better readability. 
For convenience, we also repeat the statement of the theorem before its proof.

\thmTight*

\begin{proof}
	Let us first consider~$K_{m,n}$ minus two edges. The case where the two missing edges are adjacent is depicted in \cref{fig:appnecessary}. It shows two drawings of~$K_{m,n}$ minus two edges $b_1r_2$ and $b_1r_3$ that have the same \ERS. 
	First note that the two drawings are not strongly isomorphic because the order in which~$r_1b_1$ crosses the other edges differs in both drawings. We claim that the two drawings cannot be transformed into each other via triangle flips. The edge~$r_1b_1$ only crosses edges incident to~$r_2$ or~$r_3$. The subdrawing of all edges incident to~$r_2$ or~$r_3$ is plane. Thus, there is no crossing triangle that involves the edge~$r_1b_1$. Consequently, no triangle flip can change the order of crossings along~$r_1b_1$. (The \ERS~determines the set of crossing edge pairs, and this set remains invariant under triangle flips. Therefore, the edge~$r_1b_1$ can never be part of a crossing triangle.)
	
	\begin{figure}[htb]
		\centering
		\includegraphics[page=5]{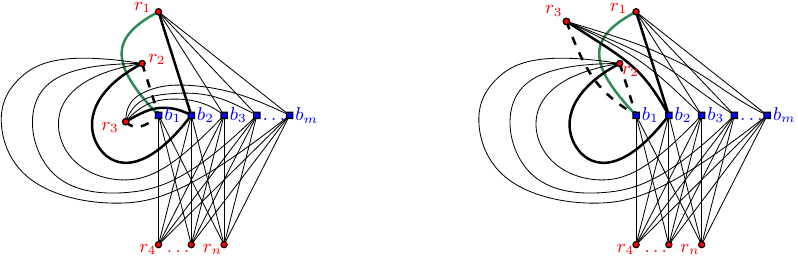}
		\caption {Two drawings of $K_{m,n}$ minus two adjacent edges $b_1r_2$ and $b_1r_3$ (drawn as dashed lines) that have the same \ERS but cannot be transformed into each other via triangle flips. The drawing for the smallest case, $K_{2,3}$, is shown bold. No triangle flip can change the crossing order along the green edge $r_1b_1$, which is different in the two drawings.}
		\label{fig:appnecessary}
	\end{figure}
	
	It remains to consider the case where the two missing edges are independent, which is depicted in \cref{fig:appnecessary:indep}. The vertices labeled~$r_3,\ldots$ and~$b_4,\ldots$ stand for an arbitrarily large (possibly empty) cluster of red and blue vertices, respectively, that are connected to the remaining vertices in the same way (topologically). The edge~$b_1r_1$ crosses~$b_2r_2$ and~$b_3r_3$ in a different order in both drawings. Among those edges that cross~$b_1r_1$ no two cross. So there is no crossing triangle that involves~$b_1r_1$ and therefore no sequence of triangle flips can change the order of crossings along~$b_1r_1$. Thus, there is no way to transform one drawing into the other using triangle flips.
	
	\begin{figure}[htbp]
		\centering
		\includegraphics[page=10]{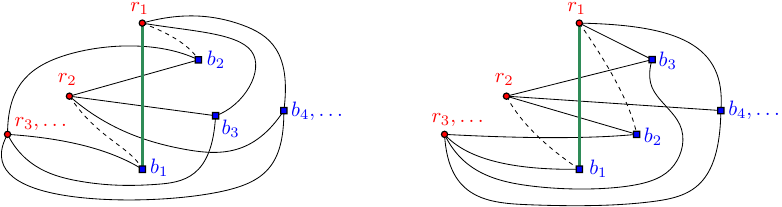}
		\caption {Two drawings of $K_{m,n}$ minus two independent edges $b_2r_1$ and $b_1r_2$ (shown dashed) that have the same \ERS but cannot be transformed into each other via triangle flips.}
		\label{fig:appnecessary:indep}
	\end{figure}

		\begin{figure}[htb]
		\centering
		\includegraphics[page=8]{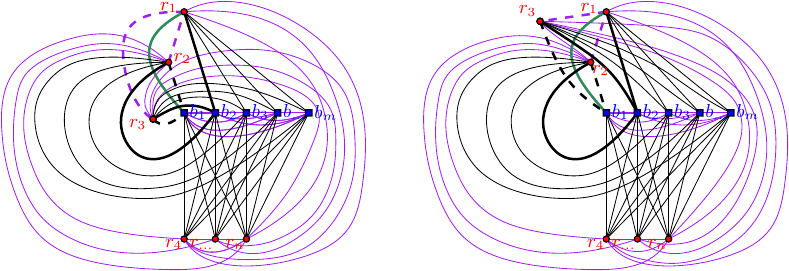}
		\caption {Two drawings of $K_{m}$ minus four edges building a $C_4$ that have the same \ERS, but cannot be transformed into each other via triangle flips. The missing edges are $b_1r_2$, $r_2r_1$, $r_1r_3$, and $r_3r_1$ and drawn as dashed lines. (The drawing for $K_{5}$ is marked in bold lines.) No triangle flip can change the crossing order in which the green edge $r_1b_1$ crosses the black edges incident to $r_1$ and $r_2$.}
		\label{fig:appcomplete_minus_4}
	\end{figure}

\begin{figure}[htb]
	\centering
	\includegraphics[page=17]{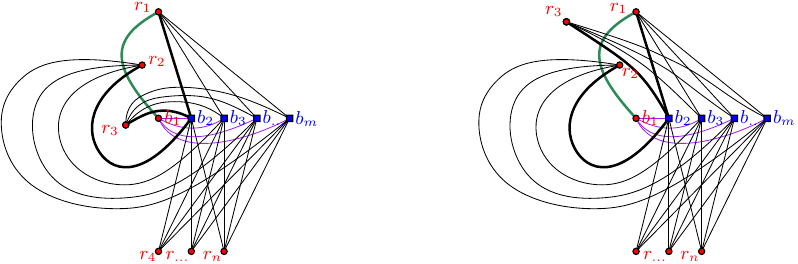}
	\caption {Two drawings of $K_{m-1,n+1}$ plus one edge ($b_1r_1$) that cannot be transformed into each other via triangle flips.}
	\label{fig:addKmn_plus}
\end{figure}

	To obtain the statement for~$K_n$, we extend the drawings in \cref{fig:appnecessary} to drawings of a complete graph minus $4$ edges that form a $C_4$ such that both drawings have the same \ERS. The extension is shown in \cref{fig:appcomplete_minus_4}, where the original edges are drawn black and the edges of the extension are drawn purple. This drawing contains the drawing of \cref{fig:appnecessary} as a subdrawing. As before, the edge~$r_1b_1$ crosses the original (black) edges incident to~$r_2$ and~$r_3$ in a different order, and there are no crossing triangles that involve the edge~$r_1b_1$ and the edges incident to~$r_2$ and~$r_3$. Thus, there is no way to transform one drawing into the other using triangle flips.

	Finally, the statement for~$K_{m,n}$ plus one edge in a partition class of size at least four can be obtained from the construction depicted in \cref{fig:appcomplete_minus_4} in the following way: color the four vertices $r_1$, $r_2$, $r_3$, and $b_1$ of the missing~$C_4$ with one color, the vertex $b_2$ with the other color, and the remaining vertices arbitrarily, see for an example Figure~\ref{fig:addKmn_plus}. Then disregarding all edges between two vertices of the same color except for the edge $r_1b_1$ yields two different drawings of~$K_{m,n}$ plus one edge. With the same reasoning as for $K_n$ minus $C_4$, those two drawings cannot be transformed into each other via triangle flips.
\end{proof}

\section{Missing proofs of Section~\ref{sec:main_proof}}\label{app:main_proof}

\subsection{Proof of Lemma~\ref{lem:order_of_crossings}}\label{sec:orderofcrossings}

In this section, we proof Lemma~\ref{lem:order_of_crossings}, which is part of the first proof of \cref{thm:main}.
For convenience, we repeat the lemma before its proof.

\orderOfCrossings*

\begin{proof}
	We prove the statement by showing that if in two different drawings of a complete multipartite graph $G$, 
	an edge $vw$ crosses a pair of adjacent or disjoint edges in different orders, then the according
	drawings must have different \ERSs. We distinguish two cases, depending on whether the two edges crossed by $vw$ are adjacent or disjoint. 
	
	\begin{case}[The edge $vw$ crosses two adjacent edges $ab$ and $ac$ in different orders]\label{case:crossingorder:adj}
		We distinguish two subcases, depending on the topology of the drawing of $vw$, $ab$, and $ac$; 
		see \cref{fig:crossingorder_adjacent}.
		
		\begin{figure}[htbp]
			\begin{minipage}[b]{.48\linewidth}
				\centering\includegraphics[page=2]{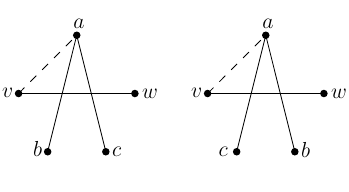}
				\subcaption{\cref{case:crossingorder:adj:case_1}}\label{fig:crossingorder_adjacent_case1} 
			\end{minipage}
			\hfill
			\begin{minipage}[b]{.48\linewidth}
				\centering\includegraphics[page=4]{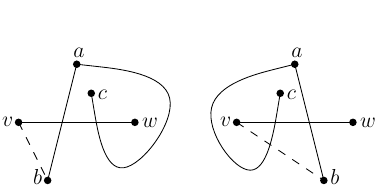}
				\subcaption{\cref{case:crossingorder:adj:case_2}}\label{fig:crossingorder_adjacent_case2}
			\end{minipage}
			\caption{\cref{case:crossingorder:adj}: the two topologically different cases how $vw$ can intersect $ab$ and $ac$ (up to relabelling). \label{fig:crossingorder_adjacent}}
		\end{figure}
		
		\begin{subcase}[$ab$ and $ac$ cross $vw$ from the same side]\label{case:crossingorder:adj:case_1}
			By \cref{lem:claim}, at least one of $va$ and $wa$ is an edge of $G$. 
			Assume w.l.o.g.\ that $va$ is in $G$ (the other choice is symmetric).
			As $va$ must not cross any of $vw$, $ab$, and $ac$, the drawing of $va$ is unique in each drawing. 
			The resulting drawings have different rotations at $a$ and hence different \ERSs. 
		\end{subcase}
		\begin{subcase}[$ab$ and $ac$ cross $vw$ from different sides]\label{case:crossingorder:adj:case_2} 
			By \cref{lem:claim}, at least one of $vb$ and $wb$ is an edge of $G$. 
			Assume w.l.o.g.\ that $vb$ is in $G$ (the other choice is symmetric).
			As $vb$ must not cross any of $vw$ and $ab$, the drawing of $vb$ is again unique in each drawing. 
			However, in one of the drawings $vb$ does not cross any edge, while in the other one it must cross $ac$,
			again implying different \ERSs. 
		\end{subcase}\noqed
	\end{case}
	
	\begin{case}[The edge $vw$ crosses two disjoint edges $ab$ and $cd$ in different orders]\label{case:crossingorder:disj}
		The only possibility for the topology of the drawing of $vw$, $ab$, and $ac$ in this case is depicted in \cref{fig:crossingorder_disjoint}.
		We will again use other existing edges for our reasoning.
		
		\begin{figure}[htbp]
			\centering\includegraphics[page=5]{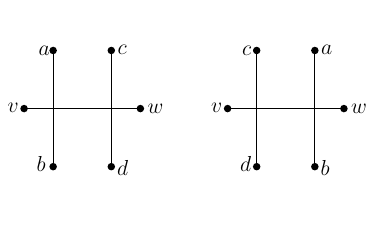}
			\caption{\cref{case:crossingorder:disj}: Depiction of how $vw$ intersects $ab$ and $cd$ (up to relabelling). \label{fig:crossingorder_disjoint}}
		\end{figure}
		
		Consider the two edges $vw$ and $ab$. By \cref{lem:claim}, each of the four vertices must have an edge to at least 
		one of the vertices of the other edge. Hence, out of the four edges $va$, $wa$, $vb$, and $wb$, either $va$ and $wb$ or 
		$vb$ and $wa$ are in $G$. Assume w.l.o.g.\ that $va$ and $wb$ are in $G$ (the other choice is symmetric).
		Note that each of $va$ and $wb$ might cross $cd$, but none of them can cross any of $vw$ and $ab$.
		Further, by \cref{case:crossingorder:adj}, we know that the crossing order of
		any of the edges with two adjacent edges must be the same in both drawings.
		
		\begin{subcase}[$va$ crosses $cd$ between $c$ and the crossing of $vw$ and $cd$]\label{case:crossingorder:disj:case_1}
			See \cref{fig:crossingorder_disjoint_case1}. 
			Then in each of the drawings, the crossing rotation of $va$ and $cd$ is fixed, 
			as the other rotation would require a second crossing 
			(which is not allowed by the simplicity of the drawing).
			As these fixed crossing rotations are different in the two drawings, the \ERSs are different as well.
		\end{subcase}
		
		\begin{subcase}[$va$ crosses $cd$ between $d$ and the crossing of $vw$ and $cd$]\label{case:crossingorder:disj:case_2}
			See \cref{fig:crossingorder_disjoint_case2}. 
			Then it follows from the left drawing that $wb$ must cross $cd$. 
			However, in the right drawing $wb$ must not cross $cd$.
		\end{subcase}
		
		\begin{figure}[htbp]
			\begin{minipage}[t]{.48\linewidth}
				\centering\includegraphics[page=6]{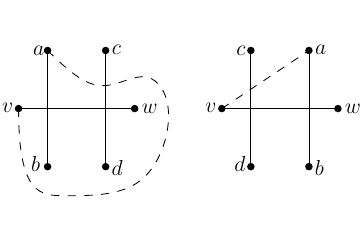}
				\subcaption{\cref{case:crossingorder:disj:case_1}}\label{fig:crossingorder_disjoint_case1} 
			\end{minipage}
			\hfill
			\begin{minipage}[t]{.48\linewidth}
				\centering\includegraphics[page=14]{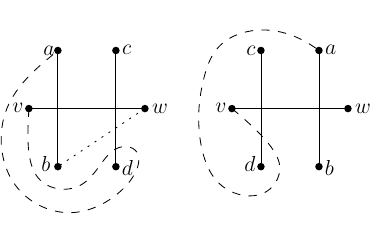}
				\subcaption{\cref{case:crossingorder:disj:case_2}}\label{fig:crossingorder_disjoint_case2}
			\end{minipage}
			\caption{\cref{case:crossingorder:disj}: The two cases where $va$ crosses $cd$. \label{fig:crossingorder_adjacent_details}}
		\end{figure}
		
		Note that, if $va$ does not cross $cd$ but $wb$ crosses $cd$, \cref{fig:crossingorder_disjoint_case1,fig:crossingorder_disjoint_case2}
		apply analogously. Hence the only missing case is that none of the edges crosses $cd$; see \cref{fig:crossingorder_disjoint_noncrossing}. 
		
		\begin{figure}[htbp]
			\centering\includegraphics[page=8]{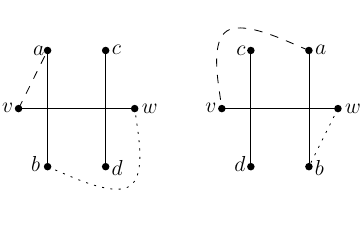}
			\caption{\cref{case:crossingorder:disj:case_3}: The situation where none of $va$ and $wb$ crosses $cd$. \label{fig:crossingorder_disjoint_noncrossing}}
		\end{figure}
		
		\begin{subcase}[None of $va$ and $wb$ crosses $cd$]\label{case:crossingorder:disj:case_3}
			Consider the two edges $vw$ and $cd$. 
			By \cref{lem:claim}, each of the four vertices must have an edge to at least one of the vertices of the other edge. 
			Hence, out of the four edges $vc$, $wc$, $vd$, and $wd$, either $vc$ and $wd$ or $vd$ and $wc$ are in $G$, which yields two subcases. 
			
			\begin{subsubcase}[$vc$ and $wd$ are edges of $G$]\label{case:crossingorder:disj:case_31}
				We can assume that none of the edges $vc$ and $wd$ crosses $ab$, 
				as otherwise the two drawings have different \ERSs by \cref{case:crossingorder:disj:case_1} and \cref{case:crossingorder:disj:case_2}.
				(independent of how $va$ and $wb$ are drawn). 
				Hence, for the two drawings to have the same crossings, $vc$ must be added without crossing any edge (due to the right drawing),
				which gives a unique way to add it to each of the drawings. 
				These unique ways force different rotations at $v$ and hence again different \ERSs of the two drawings; 
				see \cref{fig:crossingorder_disjoint_noncrossing_case1}.
			\end{subsubcase}
			
			\begin{figure}[htbp]
				\begin{minipage}[t]{.48\linewidth}
					\centering\includegraphics[page=9]{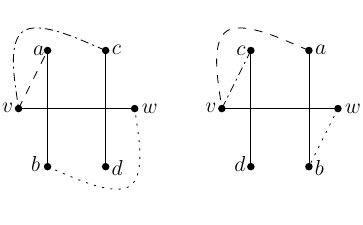}
					\subcaption{\cref{case:crossingorder:disj:case_31}}\label{fig:crossingorder_disjoint_noncrossing_case1} 
				\end{minipage}
				\hfill
				\begin{minipage}[t]{.48\linewidth}
					\centering\includegraphics[page=13]{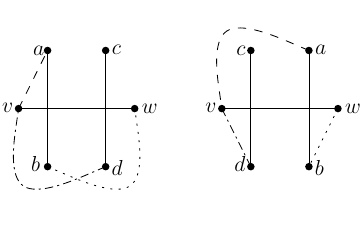}
					\subcaption{\cref{case:crossingorder:disj:case_32}}\label{fig:crossingorder_disjoint_noncrossing_case2}
				\end{minipage}
				\caption{The two possibilities in \cref{case:crossingorder:disj:case_3}\label{fig:crossingorder_disjoint_noncrossing_details}}
			\end{figure}
			
			\begin{subsubcase}[$vd$ and $wc$ are edges of $G$]\label{case:crossingorder:disj:case_32}
				As in the previous case, we can assume that none of the edges $vd$ and $wc$ crosses $ab$ 
				by \cref{case:crossingorder:disj:case_1} and \cref{case:crossingorder:disj:case_2}.
				Hence in the left drawing, $vd$ must cross $wb$, while in the right drawing, $vd$ cannot cross any of the edges, 
				again implying different \ERSs of the two drawings; 
				see \cref{fig:crossingorder_disjoint_noncrossing_case2}.
			\end{subsubcase}\noqed
		\end{subcase}\noqed
	\end{case}
	
	As the statement holds in each of the cases this completes the proof of the lemma.
\end{proof}

\subsection[Second Proof of Theorem~\ref{thm:main}]{Second Proof of \cref{thm:main}}\label{ssec:main_pf}
For our second proof of \Cref{thm:main}, we use the same general proof framework as \cite{gioan_proof} where we iteratively transform one of the drawings in order to increase the strongly isomorphic parts of both drawings. 
However, many steps of the proof in~\cite{gioan_proof} rely on known properties of drawings of complete graphs which do not hold for complete multipartite graphs in general (or were not known to do so prior to the work at hand).
In the following, we present the big picture of the proof,
while deferring the proofs of some key lemmata to subsequent sections. 
We then 
highlight the novel contributions of the proof 
and discuss the key differences to \cite{gioan_proof}.

\thmmain*

\begin{proof}
	Let $G$ be a complete multipartite graph, and let $D_1$ and $D_2$ be two simple drawings of~$G$ on~$\Su^2$ with the same \ERS.
	Let $R= \{r_1, r_2,\dots, r_n\}$ be a maximal independent set in~$G$, and let $B= \{b_1, b_2,\dots, b_m\}$ denote the set of the remaining vertices. 
	We call the vertices in $R$ and $B$ the red and blue vertices, respectively. Note that the graph on vertex set $R\cup B$ together with all edges with a red and a blue endpoint forms a complete bipartite graph $K_{n,m}$, and the set $R$ is an independent set in $G$ while $B$ might not necessarily be an independent set.
	Let $D:\cong~D_1$. 
	We change $D$ by performing triangle flips, until we obtain~$D\cong D_2$. 
	We iteratively consider the vertices~$r_1,\ldots,r_n$. For each vertex $r_i$, we iteratively consider the incident edges~$r_ib_1,\ldots,r_ib_m$. 
	After that the remaining edges, which connect two vertices in~$B$, are considered in some arbitrary but fixed order.
	Formally, we consider the edges of $G$ in an order $(a_1, \dots, a_t)$, where for every $1\le i \le t$, $a_i=r_jb_k$ or $a_i = b_kb_{\ell}$ for some $1\le j \le n$ and $1 \le k, \ell \le m$, and where

	\vspace{-1ex}\begin{itemize} 
		\item the edge $r_ib_j$ precedes $r_ib_k$ for all $1 \leq i \leq n$ and $1 \leq j < k \leq m$,
		\item the edge $r_ib_j$ precedes $r_kb_{\ell}$ for all $1 \leq i < k \leq n$ and $1 \leq j, \ell \leq m$, and
		\item the edge $r_ib_j$ precedes $b_kb_{\ell}$ for all $1 \leq i \leq n$ and $1 \leq j, k, \ell \leq m$.
	\end{itemize}

	For $1\le i\le t$, we denote by~$X_{i}$ the subgraph of~$G$ induced by the edge set $\{a_1, \dots, a_i\}$.
	When considering an edge~$a_i$, the goal is to establish~$D[X_{i}]\cong D_2[X_{i}]$,
	where $D[X_{i}]$ and $D_2[X_{i}]$ are the corresponding subdrawings of $D$ and $D_2$, respectively.
	
	For the base case~$i\le m$, observe that $D[K_{1,i}]\cong D_2[K_{1,i}]$ because there is only one simple drawing of~$K_{1,i}$ (our graphs are labeled but the \ERS is given).
	
	For the general case~$i > m$, assume that~$D[X_{i-1}]\cong D_2[X_{i-1}]$.
	Observe that, since~$i > m$, all vertices of~$B$ are already present in~$X_{i-1}$.
	In the case when a vertex $r_j$ is introduced for the first time in $X_i$, (i.e., $r_j$ is present in $X_i$ but not in $X_{i-1}$), we first argue that the position of vertex~$r_j$ is consistent between~$D[X_{i-1}]$ and~$D_2[X_{i-1}]$.
	To show this, we use the following lemma, whose proof relies on \cref{prop:caratheodory} (Carath{\'e}odory's Theorem) and is deferred to \cref{sec:cell_determined}. 
	
	\begin{restatable}{lemma}{proprotcell}\label{prop:rot_cell}
		Let $F$ be a simple drawing of a complete multipartite graph on~$\Su^2$.
		For any vertex~$v$ in $F$, the \ERS of $F$ uniquely determines which \cell of $F':=F \setminus \{v\}$ contains~$v$. 
	\end{restatable}
	
	Since $D[X_{i-1}]\cong D_2[X_{i-1}]$, the two drawings topologically have the same \cells. 
	By the order of the edges, if $r_j$ is introduced for the first time in $X_i$, then $X_{i-1}$ is a complete bipartite graph between the vertices in $B$ and the vertices in $\{r_1, \dots, r_{j-1}\}$.
	Further, $X_{i-1+m}$ is then a complete bipartite graph between the vertices in $B$ and the vertices in $\{r_1, \dots, r_{j}\}$.
	Hence, as $D$ and $D_2$ have the same \ERS, by
	\cref{prop:rot_cell} applied to~$F=D[X_{i-1+m}]$ and to $D_2[X_{i-1+m}]$, both times with~$v=r_i$, 
	we conclude that~$r_i$ lies in the same \cell in~$D[X_{i-1}]$ and~$D_2[X_{i-1}]$. 
	
	Now consider the edge $a_i = xb_j$, where $x$ can be in either $R$ or $B$. 
	The aim is to use a sequence of triangle flips to transform~$D$ such that $D[X_{i}]\cong D_2[X_{i}]$. Let~$e_1$ denote the curve that represents~$xb_j$ in~$D$.
	We imagine adding another copy~$\ett$ of~$xb_j$ to~$D$, which corresponds to the curve $e_2$ that represents the edge~$xb_j$ in~$D_2$ and serves as a ``target'' curve which we aim to transform~$e_1$ into. The following lemma, which is proven in \cref{sec:proof_existence_ett}, guarantees the existence of a suitable target curve.
	
	\begin{restatable}{lemma}{propexistenceett}\label{prop:existence:ett}
		There exists a simple curve~$\ett$ such that~$D[X_{i-1}]\cup\ett\cong D_2[X_{i}]$ and~$e_1$ and~$\ett$ have 
		$O(|V(X_{i-1})|^4)$ 
		intersections in~$D[X_{i}]\cup\ett$, where $|V(X_{i-1})|$ is the number of vertices of $X_{i-1}$.
	\end{restatable}
	
	Now fix such a curve~$\ett$. Then~$\Gamma=e_1\cup\ett$ forms a (not necessarily simple) closed curve. 
	A \emph{lens} in~$\Gamma$ is a cell whose boundary is formed by 
	exactly two edge fragments of~$\Gamma$, one from~$e_1$ and one from~$\ett$. 
	With the next lemma, we show that 
	there is a lens in~$\Gamma$ which we can use as a starting point for transforming~$e_1$ into~$\ett$ via triangle flips in $D$. 
	Its proof is presented in \Cref{sec:adding_vertex}.
	
	\begin{restatable}{lemma}{propemptylens}\label{prop:empty:lens:e1:e2}
		In~$\Gamma$ there is a \emph{free lens}, i.e., a lens that does not contain any vertex of~$D[X_{i}]$.
	\end{restatable}
	
	Now consider a free lens~$L$ that exists by \cref{prop:empty:lens:e1:e2}. 
	While~$L$ does not contain any vertex of~$D[X_{i-1}]$, it may contain crossings of~$D[X_{i-1}]$. 
	As a next step, we aim to transform~$D$ using triangle flips such that~$L$ does not contain any crossings of~$D[X_{i-1}]$. 
	Let~$\chi\in L$ be a crossing of two edges~$a',a''$ in~$D[X_{i-1}]$. 
	As $x$ and $b_j$ are the only vertices of $e_1 \cup \ett$, it follows that each of~$a',a''$ crosses~$\partial L$ twice; as both~$D$ and~$D_2$ are simple drawings, one of these crossings is with~$e_1$ and the other is with~$\ett$. 
	Thus,~$a'$, $a''$, and $e_1$ form a crossing triangle~$\Delta_{e_1}$ in $D$.
	Moreover, the corresponding crossing triangle in~$D_2$ has the opposite \parity, and hence $\Delta_{e_1}$ is invertible. 
	By the following lemma, whose proof is presented in \cref{sec:flippable}, the triangle~$\Delta_{e_1}$ is empty of \emph{all} vertices of~$D$ if $D$ is a complete bipartite graph (we already knew this for the vertices already in~$X_{i}$, but not for vertices in $R$ that are not present in $D[X_{i}]$).
	
	\begin{restatable}[Invertible triangles are empty]{lemma}{propflippable}\label{prop:flippable}
		Let $D$ be a simple drawing of a complete bipartite graph~$G$,
		and let $\Delta$ be an invertible triangle in $D$. Then all vertices of $D$ lie outside~$\Delta$. 
	\end{restatable}
	Thus, while $D[X_{i}]$ contains only edges between $R$ and $B$, the crossing triangle~$\Delta_{e_1}$ has to be empty of all vertices of~$D$. If $D[X_{i}]$ also contains edges between two vertices of $B$, then all vertices of $D$ are in $D[X_{i}]$ and thus $\Delta_{e_1}$ is empty of all vertices of $D$ by \cref{prop:empty:lens:e1:e2}.
	
	We claim that also all edges that cross~$\Delta_{e_1}$ can be ``swept'' out of~$\Delta_{e_1}$. 
	To see this, consider the set~$\Xi$ of all edges of~$D$ that cross~$\Delta_{e_1}$. Note that every edge from~$\Xi$ crosses exactly two sides (that is, edges) of~$\Delta_{e_1}$. For two sides~$a,b$ of~$\Delta_{e_1}$ let~$\Xi_{a,b}\subseteq\Xi$ denote the set of edges that cross both~$a$ and~$b$. Pick some pair~$a,b$ for which~$\Xi_{a,b}\ne\emptyset$ (if none exists, we are done), and let~$\eta$ be the edge from~$\Xi_{a,b}$ that crosses~$a$ closest to the common endpoint of~$a$ and~$b$. Let~$R_\eta$ denote the closed triangular region that is bounded by~$a$, $b$, and~$\eta$ and whose interior lies inside~$\Delta_{e_1}$, and let~$\Xi_{\eta}$ denote the set of edges from~$\Xi$ that cross the interior of~$R_\eta$.
	Note that by the choice of~$\eta$, every edge in~$\Xi_{\eta}$ crosses each of~$\eta$ and~$\partial R_\eta\setminus\eta$ exactly once. 
	Hence, locally in $R_\eta$, the edges in~$\Xi_{\eta} \cup \{a\}$ behave similarly to a collection of pseudolines, except that not every pair may cross
	(also called an \emph{arrangement of bi-infinite curves having the one-intersection property} by Hershberger and Snoeyink~\cite{SnoeyinkHershberger1991}).
	We will use the following lemma for sweeping $R_{\eta}$ with~$\eta$. 

	\begin{lemma}[{\cite[Lemma~3.1]{SnoeyinkHershberger1991}}]
		Any arrangement of bi-infinite curves having the one-intersection property can be swept starting with any curve from the arrangement using three operations: 
		passing a triangle, passing the first ray, and taking the first ray.
	\end{lemma}

	Note that the operation \emph{passing a triangle} is a triangle flip. 
	The other two operations are only applied if the arrangement contains either lines that do not intersect the sweeping curve or lines that intersect nothing but the sweeping curve.
	Thus, the other two operations are not applied, since all edges in~$\Xi_{\eta} \cup \{a\}$ cross each of $b$ and ${\eta}$. 
	Since our sweeping curve $\eta$ intersects all edges in~$\Xi_{\eta} \cup \{a\}$, %
	we can sweep~$\eta$ through~$R_\eta$ via
	a sequence of triangle flips, in this way sweeping~$\eta$ once over each crossing among the edges from~$\Xi_{\eta}\cup\{a,b\}$. 
	After the last flip in this process, which is for the crossing triangle~$\eta a b$, the edge~$\eta$ does not cross~$\Delta_{e_1}$ anymore. 
	Hence, after repeating this process $|\Xi|$ times---each time for an appropriate choice of $\eta$---the crossing triangle~$\Delta_{e_1}$ is empty and, therefore, can be flipped as well. 
	We remark that the number of flips to empty $\Delta_{e_1}$ with this process is bounded from above by the number of crossings between edges in $\Xi$ and $\Delta_{e_1}$ times the number of edges in $\Xi$.
	Hence at most $O(|\Xi|^3)$ flips are required.
	
	Processing all remaining crossings inside~$L$ in the described fashion, we establish that in the resulting drawing, the lens~$L$ does not contain any vertex or crossing of~$D[X_{i-1}]$. 
	In other words, locally around~$L$, the edge~$e_1$ is topologically identical to~$\ett$ with respect to~$D[X_{i-1}]$.
	Thus, we can adapt~$\ett$ by replacing its edge part on~$\partial L$ with a close copy of the edge part of~$e_1$ on~$\partial L$, effectively removing the lens~$L$ from~$\Gamma$. 
	As a result, the edges~$e_1$ and~$\ett$ have fewer crossings than before in~$D$, and the parameters~$D$ and $\Gamma=e_1 \cup \ett$ again meet the conditions of \cref{prop:empty:lens:e1:e2}. 
	Repeatedly applying this procedure, 
	we eventually obtain a drawing~$D[X_{i}]\cup \ett$ where~$e_1$ and~$\ett$ do not cross and hence $\Gamma$ forms a simple closed curve. 
	By~\cref{prop:empty:lens:e1:e2}, one of the two cells bounded by $\Gamma$ contains no vertices of~$D[X_{i}]$. 
	So after one last round of transformations as described above, we obtain a drawing~$D[X_{i}]\cup \ett$ in which all vertices and crossings lie on one side of~$\Gamma$.
	Hence we have obtained $D[X_{i}]\cong D_2[X_{i}]$.
	Processing all vertices~$r_i$, for $i=2,\ldots,n$, and in turn handling all edges incident to~$r_i$ eventually yields a drawing~$D\cong D_2$.  
\end{proof}

\subparagraph*{Key differences to the proof~\cite{gioan_proof} for complete graphs.} 

The key differences to the proof in~\cite{gioan_proof} are concentrated in the statements of \cref{prop:rot_cell}, \ref{prop:empty:lens:e1:e2}, and \ref{prop:flippable}.

Firstly, when extending the `isomorphic subgraph' by a vertex $v$, we require that~$v$ lies in the same cell in both drawings induced by the subgraph, as stated in \cref{prop:rot_cell}. For the complete graph this statement is an easy corollary of the known fact that each \cell is the intersection of $3$-cycles, and that the position of a fourth vertex with respect to a $3$-cycle is determined by the rotation system. However, multipartite graphs may not have any $3$-cycles. We circumvent this issue by also allowing $4$-cycles---which, in contrast to triangles, may self-cross---and developing a corresponding Carath{\'e}odory-type theorem for simple drawings of complete multipartite graphs (\cref{prop:caratheodory}). 

Secondly, for extending the `isomorphic subgraph' $X_{i-1}$ by an edge~$a_i$, which is drawn with different crossing orders as $e_1$ in $D[X_{i}]$ and as $e_2$ in $D_2[X_{i}]$, we think of $e_2$ drawn virtually into $D[X_{i}]$ as $\ett$ and aim to untangle $e_1$ and $\ett$ via triangle flips.
For this we require that the curve formed by both edge incarnations encloses at least one `free lens'; see \cref{prop:empty:lens:e1:e2}. 
The corresponding statement for the complete graph follows from the facts that the drawings are simple and that each pair of vertices is adjacent. 
However, in a graph where not all pairs of vertices are adjacent, there are fewer restrictions on how the edges~$e_1$ and~$e_2$ can be drawn, which allows for a possibly complicated interaction between them.  
As a result, more deliberate considerations and new structural insights are needed in order to prove this property. 
We consider \cref{prop:empty:lens:e1:e2} as the core lemma for our proof.

Finally, another essential element of the proof of Gioan's Theorem for the complete graph~\cite{gioan_proof} is that for any invertible triangle~$T$ all vertices lie on the same side of~$T$. For the complete graph, this statement again can be shown using the fact that all vertices are pairwise adjacent. This does not hold for general multipartite graphs, and therefore in order to prove the according statement of \cref{prop:flippable}, we need a different approach. 

\subsection{\ERS determines positions of vertices (Proof of Lemma~\ref{prop:rot_cell})}\label{sec:cell_determined}

To show \cref{prop:rot_cell}, we first consider the crucial base case of subgraphs on up to four vertices.

\begin{restatable}{lemma}{propKsmall}
	\label{prop:Ksmall}
	Let $D$ be a simple drawing of a complete multipartite graph on~$\Su^2$, 
	let $D'$ be a subdrawing of $D$ on at most four vertices, and let $v$ be a vertex in $D \setminus D'$. Then the \ERS of $D$ uniquely determines which \cell of $D'$ contains~$v$. 
\end{restatable}
\begin{proof}
	Let $G=(V,E)$ denote the (complete multipartite) graph 
	represented by~$D'$. If~$G$ has no edges or if~$G$ is a tree, then the statement is vacuously true because then~$D'$ has only one \cell. Hence, we may suppose that~$G$ is either~$K_{1,1,1}=K_3=C_3$, $K_{2,2}=C_4$, $K_{1,1,2}$ or~$K_4$. 

	First consider the case that~$G$ has exactly three vertices. Then~$G=K_3$. Let~$e\in E$. Then by \cref{lem:claim} we have~$vu$ in $D$, for some endvertex~$u$ of~$e$. We trace the edge~$uv$ in~$D' \cup D[uv]$ starting from~$u$. The given \ERS tells us which of the two cells incident to~$u$ the edge~$uv$ enters when leaving~$u$. As~$uv$ is adjacent to two of the three edges of~$G$, it may only cross the third edge and only once. The \ERS tells us whether or not this crossing exists and thereby the cell of~$D'$ that contains~$v$, as claimed.

	So we may assume that~$G$ has exactly four vertices. Then~$G$ is one of~$K_{2,2}$, $K_{1,1,2}$, or~$K_4$. Up to (strong) isomorphism, each of these three graphs admits exactly two different drawings on~$\Su^2$, one without crossings and one with exactly one crossing, as depicted in \cref{fig:four_vertices}. We pick an edge~$e\in E$ as follows: If~$G=K_{1,1,2}$, then let~$e$ be the edge between the two degree three vertices; otherwise, select~$e\in E$ arbitrarily. By \cref{lem:claim} we have~$vu$ in $D$, for some endvertex~$u$ of~$e$. We trace the edge~$uv$ in~$D' \cup D[uv]$ starting from~$u$.  The given \ERS tells us which cell incident to~$u$ the edge~$uv$ enters when leaving~$u$. 
	It also tells us which other edges to cross and the rotations of all those crossings. 
	We claim that this information suffices to uniquely determine the cell of~$D'$ that contains~$v$. 

	Note that~$uv$ may cross each edge of~$D'$ at most once and each edge incident to~$u$ not at all. Thus, tracing~$uv$ corresponds to a trail\footnote{A \emph{trail} in a graph is a walk that uses every edge at most once. In contrast to a path, vertices may be visited several times.} in the dual, more precisely, in the graph~$D^*$ that is obtained from the dual of~$D'$ by removing the edges that are dual to an edge incident to~$u$. Again, see \cref{fig:four_vertices} for an illustration, where the graphs~$D^*$ are shown in pink.

	\begin{figure}[htb]
		\centering\includegraphics[page=4]{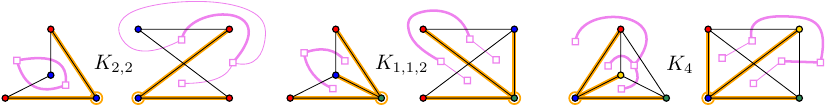}
		\caption {The three graphs and their two drawings each to consider in \cref{prop:Ksmall}. Their respective duals are drawn with square vertices in pink. The vertex~$u$ is marked with an orange circle, and its incident edges, which must not be crossed by the edge~$uv$, are also marked in orange. The dual edges shown bold are in 1-to-1 correspondence to edges of~$G$.}
		\label{fig:four_vertices}
	\end{figure}

	Consider first the case~$G\in\{K_{1,1,2},K_4\}$, and observe that then the graph~$D^*$ is always a tree. Moreover, most of the dual edges in these four drawings are in 1-to-1 correspondence to edges of~$G$; we call these edges \emph{bold}, and they are also shown bold in \cref{fig:four_vertices}. The only non-bold edges are those that are incident to the leaves of~$D^*$ in the drawings~$D'$ with one crossing, as both of these edges cross the same edge of~$G$. As~$D^*$ is a tree, whenever the dual trail that we trace encounters a bold edge that corresponds to an edge of~$G$ that must be crossed, it must take this edge. As all non-bold edges are incident to leaves of~$D^*$ (where there is only one option to continue, anyway) and no two non-bold edges are adjacent in~$D^*$ (so that there is never a choice between the two), it follows that the trail traced by~$uv$ in~$D^*$ is uniquely determined by the list of crossed edges in these cases.

	It remains to consider the case~$G=K_{2,2}$. The drawing without crossings has only two cells, and every crossing switches between them. So the target cell is uniquely determined by the number of crossings along~$uv$ (which is zero, one or two). 
	The drawing with one crossing is more interesting because here we need the crossing rotation. Note that our trace starts in a cell incident to~$u$, for which there are two options. The tricell incident to~$u$ is a leaf of~$D^*$ and using the incident edge rules out the other non-bold edge of~$D^*$. Therefore, the trace is uniquely determined if it starts into this tricell. Otherwise, the edge~$uv$ enters the other cell incident to~$u$, which is bounded by six edge-fragments. If the bold edge must be used, things are clear because it cannot be combined with the non-bold edge that is incident to a leaf of~$D^*$. Otherwise, the bold edge must not be used and then we can select the correct non-bold edge to take---if any---by considering the crossing rotation.
\end{proof}

With \cref{prop:Ksmall} and \cref{prop:caratheodory}, we can now prove \cref{prop:rot_cell}. The idea to the next proof is similar to the independently developed proof of Lemma 3.4 in Gioan's full proof~\cite{gioan_final}.

\proprotcell*
\begin{proof}  
	If~$F'$ has only one cell, then the statement is vacuously true. Hence, we may suppose that~$F'$ is neither empty nor a star. 
	Let~$c_1$ and~$c_2$ be any two distinct (hence disjoint) \cells of~$F'$. To prove the lemma it suffices to show that at least one of~$c_1$ or~$c_2$ cannot contain~$v$. 

	By marking one cell as the unbounded cell, we may consider~$F'$ as a drawing~$\Gamma'$ in the plane~$\R^2$. We select~$c_1$ to take the role of the unbounded cell in~$\Gamma'$, which makes~$c_2$ a bounded \cell in $\Gamma'$. Then \cref{prop:caratheodory} implies that for any point~$p\in c_2$ there is a $C_i$-subdrawing~$C$~in~$\Gamma'$, for some~$i\in\{3,4\}$, such that~$p$ lies in a bounded \cell of~$C$. As~$c_2$ is a \cell of~$\Gamma'$, it follows that~$C$ does not depend on the choice of~$p$, that is, the whole \cell~$c_2$ is contained in a bounded \cell of~$C$. In contrast, the \cell~$c_1$ is contained in the unbounded \cell of~$C$ by construction. As by \cref{prop:Ksmall} the \cell of~$C$ (in~$F'$ and thus also in~$\Gamma'$) that contains~$v$ is uniquely determined, at most one of~$c_1$ or~$c_2$ may contain~$v$. 
\end{proof}

\subsection{A new curve can be drawn nicely (Proof of Lemma~\ref{prop:existence:ett})}\label{sec:proof_existence_ett}

We use the notation introduced in~\cref{sec:main_proof}. 
Recall that we have two drawings $D$ and $D_2$ of the given graph~$G$, 
where we assume $D[X_{i-1}] \cong D_2[X_{i-1}]$ for some $i > m$.
We consider the curves~$e_1$ and~$e_2$ of the edge~$xb_j$ in $D$ and $D_2$, respectively.

\propexistenceett*

\begin{proof}
	The first part of the statement 
	is obvious, given that~$D[X_{i-1}]\cong D_2[X_{i-1}]$. 
	To see the second part of the statement, consider a cell~$c$ of~$D[X_{i-1}]$. 
	The curve~$e_1$ in~$D[X_{i}]$ intersects~$c$ in a (possibly empty) finite set~$J_1$ of pairwise disjoint edge fragments
	that connect points on~$\partial c$ (or inside~$c$ if the endpoint of~$e_1$ lies inside~$c$). 
	Analogously, the curve~$\ett$ in~$D[X_{i-1}]\cup\ett$ intersects~$c$ in a (possibly empty) finite set~$J_2$ of pairwise disjoint edge fragments 
	that connect points on~$\partial c$  (or inside~$c$ if the endpoint of~$\ett$ lies inside~$c$). 
	Therefore, we can choose~$\ett$ so that each edge fragment in~$J_2$ crosses each edge fragment in~$J_1$ at most once (and we also choose them this way). 
	Handling the arcs in each cell of~$D[X_{i-1}]$ accordingly results in a curve~$\ett$ that intersects~$e_1$ finitely many times.

	In total, $e_1$ and $\ett$ each have $O(|X_{i-1}|^2)$ edge fragments, since they intersect every edge of $D[X_{i-1}]$ at most once.
	Further, every edge fragment of~$e_1$ crosses every edge fragment of~$\ett$ at most once.
	Hence the number of crossings between $\ett$ and $e_1$ is bounded by $O(|X_{i-1}|^4)$ as claimed.
\end{proof}

\subsection{An empty lens exists (Proof of Lemma~\ref{prop:empty:lens:e1:e2})}\label{sec:adding_vertex}

This section is devoted to the proof of \cref{prop:empty:lens:e1:e2}. It is the key lemma in our proof of Gioan's Theorem for complete multipartite graphs, as it provides a means to untangle the possibly complicated relationship between the two geometric representations of the edge~$xb_j$ under consideration, one of which stems from each of the two drawings. We start by setting up some terminology.

Given two simple drawings~$H_1,H_2$ of~$G$ 
with the same \ERS, we consider the subdrawings induced by the edge set~$X_{i}$ as defined in \cref{ssec:main_pf}. Assuming~$H:=H_1[X_{i-1}]\cong H_2[X_{i-1}]$, we consider a virtual copy~$\ett$ of the edge~$xb_j$ in~$H_1[X_{i}]$, which already has a curve~$e_1$ that represents the edge~$xb_j$. The purpose of~$\ett$ is to mimic the role of~$xb_j$ in~$H_2$, and set a target state for the transformation of~$H_1$ so as to obtain~$H_1[X_{i}]\cong H_2[X_{i}]$. We argued in the proof of \cref{thm:main} (using \cref{prop:existence:ett}) that we can find a simple curve~$\ett$ such that (1)~$H\cup\ett\cong H_2[X_{i}]$ and (2)~$e_1$ and~$\ett$ have finitely many intersections in~$H^+=\Ha[X_{i}]\cup\ett$.

\begin{figure}[hbtp]
	\begin{minipage}[t]{.32\linewidth}
		\centering\includegraphics[page=1]{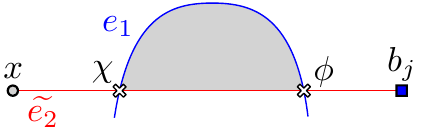}
	\end{minipage}
	\hfill
	\begin{minipage}[t]{.32\linewidth}
		\centering\includegraphics[page=2]{lenses}
	\end{minipage}
	\hfill
	\begin{minipage}[t]{.32\linewidth}
		\centering\includegraphics[page=3]{lenses}
	\end{minipage}
	\caption{Different types of bigons induced by an edge fragment~$\chi\phi$ of~$e_1$ in~$\Gamma$: A lens (left), a quasi-lens (middle), and an anti-lens (right).\label{fig:lenses}}
\end{figure}

\begin{figure}[htbp]
		\centering\includegraphics[page=4]{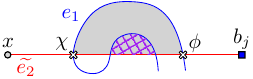}
	\caption{A quasi-lens (shaded grey) containing one lens (marked with a purple grid).\label{fig:quasi_lense}}
\end{figure}

\subparagraph{Quasi-lenses.} Denote by~$\Gamma=H^+[e_1\cup\ett]$ 
the subdrawing induced by the closed curve~$e_1\cup\ett$. Recall that a lens of a drawing is a cell whose boundary consists of exactly two edge fragments. (For a lens of~$\Gamma$ one fragment must be from~$e_1$ and the other from~$\ett$.) We slightly generalize this concept and define a \emph{quasi-lens} of~$\Gamma$ to be an open region of~$\Su^2$ that (1)~is bounded by a simple closed curve that is formed by a single edge fragment~$e$ of~$\Gamma$ together with the part of the other edge ($e_1$ if~$e\subseteq\ett$ and~$\ett$ if~$e\subseteq e_1$) between the endpoints of~$e$ and that (2)~does not contain either of~$b_j$ or~$x$. Note that, in particular, a lens of~$\Gamma$ is also a quasi-lens, but not necessarily the other way around, as a quasi-lens need not be a cell of~$\Gamma$. See \cref{fig:lenses} for an illustration.
Furthermore, each edge fragment of~$\Gamma$ induces at most one quasi-lens---unless~$b_j$ and $x$ are the two endpoints of the fragment, in which case it induces two quasi-lenses (that are, in fact, lenses). We can think of lenses as (inclusion-)minimal quasi-lenses, as the following simple lemma illustrates. 
Note that the quasi-lens of \cref{fig:lenses} contains a lens, as marked in \cref{fig:quasi_lense}. We will show that this always has to be the case. 

\begin{lemma}\label{obs:lensInside}
	Every quasi-lens of~$\Gamma$ contains a lens.
\end{lemma}
\begin{proof}
	Consider a quasi-lens~$Q$ of~$\Gamma$, and suppose without loss of generality that~$Q$ is induced by a fragment~$uv$ of~$e_1$. Let~$e'$ denote the part of~$\ett$ between~$u$ and~$v$. We proceed by induction on the number of edge fragments that form~$e'$. If this number is one, that is, if~$e'$ is a fragment of~$\ett$, then~$Q$ is a lens and the statement holds. Otherwise, the edge~$e_1$ crosses~$e'$ at some point~$\chi$. Trace~$e_1$ starting from~$\chi$ into~$Q$. As~$e_1$ is a simple curve and as neither endpoint of~$e_1$ lies in~$Q$ by definition (of quasi-lens), this trace must leave~$Q$ eventually at some point~$\psi\in\partial Q\cap\ett$, so that~$\chi\psi$ is a fragment of~$e_1$. Then~$\chi\psi$ induces a quasi-lens~$Q'\subsetneq Q$ that has strictly fewer fragments of~$\ett$ on its boundary than~$Q$. By the inductive hypothesis, the quasi-lens~$Q'$ contains a lens, and, therefore, so does~$Q$. 
\end{proof}

As~$\Gamma$ is a subdrawing of~$H^+$, we can classify the cells and quasi-lenses of~$\Gamma$ according to what parts of~$H^+$ they contain. Specifically, a cell or quasi-lens of~$\Gamma$ is

\vspace{-1ex}
\begin{itemize} 
	\item \emph{stabbed} if it contains\footnote{As all cells and quasi-lenses are open sets,  ``contains'' is equivalent to ``contains in its interior''.} at least one vertex of~$H$; 
	\item \emph{free} if it is not stabbed (that is, it does not contain any vertex of~$H$); 
	\item \emph{normal} if it contains at least one crossing but not any vertex of~$H$; 
	\item \emph{empty} if it contains at least one edge fragment but not any vertex or crossing of~$H$; 
	\item \emph{redundant} if it does not contain any vertex or edge fragment of~$H$; 
	\item \emph{essential} if it is not redundant (that is, it contains some vertex or edge fragment of~$H$). 
\end{itemize}

\subparagraph{Anti-lenses.} By definition a quasi-lens does not contain~$x$ or~$b_j$. But we also need to work with regions that contain one of these two points. We define an \emph{anti-lens} of~$\Gamma$ to be an open region of~$\Su^2$ that (1)~is bounded by a simple closed curve that is formed by a single edge fragment~$e$ of~$\Gamma$ together with the part of the other edge ($e_1$ if~$e\subseteq\ett$ and~$\ett$ if~$e\subseteq e_1$) between the endpoints of~$e$ and that (2)~contains exactly one of~$b_j$ or~$x$. 
Observe that an an anti-lens is induced by two crossings between $e_1$ and~$\ett$ that are consecutive along $e_1$ or~$\ett$ and have the same crossing rotation.
Equivalently, 
an edge fragment of~$e_1$ or~$\ett$ in~$\Gamma$ induces an anti-lens if and only if it connects to the two different sides of~$\ett$ or~$e_1$, respectively, at its endpoints. 
In particular, for an anti-lens~$A$ that is induced by an edge fragment~$f$ of~$c\in\{e_1,\ett\}$ in~$\Gamma$, one of the two ``stubs'' that comprise~$c\setminus f$ starts inside~$A$ and the other outside of~$A$ (if we consider them to be directed from~$A$ to~$\{x,b_j\}$). 
Hence we refer to them as the \emph{inside} and the \emph{outside stub} of~$A$. We have the following analogue of \cref{obs:lensInside} for anti-lenses.

\begin{lemma}\label{lem:antilensinside}
	Every anti-lens~$A$ of~$\Gamma$ contains a quasi-lens that is induced by an edge fragment of the inside stub of~$A$.
\end{lemma}
\begin{proof}
	Consider an anti-lens~$A$ and suppose without loss of generality that~$A$ is induced by a fragment~$\chi\phi$ of~$e_1$ and that the inside stub~$s$ of~$A$ starts at~$\phi$. Let~$\psi$ be the first intersection of~$s$ with~$\ett$ (which exists because~$s$ ends on~$\ett$). Let~$f$ denote the edge fragment~$\phi\psi$ of~$s$. Note that~$f\setminus\{\phi,\psi\}\subset A$. We prove the statement by induction on the number of intersections between~$s$ and~$\ett$.
	
	If there is exactly one such intersection (in which case $\psi\in\{x,b_j\}$) or if~$f$ attaches to the same side of~$\ett$ at~$\phi$ and~$\psi$, then~$f$ induces a quasi-lens~$Q\subset A$ in~$\Gamma$. 
	
	Otherwise, we know that $f$ attaches to different sides of~$\ett$ at~$\phi$ and~$\psi$. Then~$f$ induces an anti-lens~$A'\subset A$, whose inside stub~$s'$ has at least one fewer intersection with~$\ett$ than~$s$ because~$\phi\in s\setminus s'$. By the inductive hypothesis there exists a quasi-lens~$Q\subset A'\subset A$ in~$\Gamma$.
\end{proof}

\subparagraph*{Free Lenses.}
We are now ready to prove \cref{prop:empty:lens:e1:e2}, based on another lemma whose proof we will then develop over the remainder of the section. 

\propemptylens*
\begin{proof}
	As a base case suppose that~$e_1$ and~$\ett$ do not cross. Then~$\Gamma$ has only two cells, both of which are lenses. Both lenses have the same boundary, which is~$e_1\cup\ett$, and both~$x$ and~$b_j$ lie on this boundary. 
	In~$\Hc$ all blue vertices have edges to all red vertices in $X_i$. 
	As~$i > m$, there must be at least one red vertex in~$\Hc$ (for instance, $r_1$). As~$\Ha[X_{i-1}]\cong\Hb[X_{i-1}]$ and~$\Ha[X_{i-1}]\cup\ett\cong\Hb[X_{i}]$, any edge of~$\Hc$ either crosses both~$e_1$ and~$\ett$, or it crosses neither. It follows that all vertices of~$\Hc$ lie in the same lens of~$\Gamma$, and so the other lens is free.
	
	Otherwise, there are at least two lenses in~$\Gamma$. In particular, the first edge fragment of~$\ett$ forms a quasi-lens, which either is or contains a lens~$L$ with~$b_j\notin\partial L$. By \cref{lem:normal} this suffices to guarantee a free lens in~$\Gamma$.
\end{proof}

\begin{restatable}{lemma}{lemnormal}\label{lem:normal}
	If there is a lens~$L$ in~$\Gamma$ for which~$b_j\notin\partial L$, then there is a free lens in~$\Gamma$. 
\end{restatable}

To prove \cref{lem:normal}, we combine a number of observations concerning the structure of~$\Gamma$. A first such observation characterizes the placement of the red vertices of $H$. 

\begin{lemma}\label{lem:allredpoints}
	All neighbors of~$b_j$ in~$H$ lie in a single cell $U$ of $\Gamma$ with $b_j\in\partial U$.
\end{lemma}
\begin{proof} 
	Consider a neighbor~$y$ of~$b_j$ in~$H$. The edge~$yb_j$ is adjacent to~$xb_j$ and, therefore, crosses neither~$e_1$ nor~$\ett$ in~$H^+$. It follows that~$y$ lies in a cell~$C$ of~$\Gamma$ with $b_j\in\partial C$. As~$H_1$ and~$H_2$ have the same \ERS, the edges~$e_1$ and~$\ett$ are consecutive in the rotation around~$b_j$ in~$H^+$. It follows that all neighbors of~$b_j$ in~$H^+$ lie in the same cell of~$\Gamma$.
\end{proof}

Unfortunately, the situation for the blue vertices is not symmetric because they need not be connected to~$x$ or~$b_j$ in~$H$. In the following, we consider both curves~$e_1$ and~$\ett$ to be oriented from~$x$ to~$b_j$. This induces an orientation on the edge fragments around each cell of~$\Gamma$. A (quasi-)lens~$Q$ is \emph{cyclic} if the edge fragments on~$\partial Q$ form an oriented cycle; otherwise, $Q$ is \emph{acyclic}. As a next step, we want to show that all essential quasi-lenses in~$\Gamma$ are acyclic.

\begin{lemma}\label{obs:lensvertex}
	Every quasi-lens that has $x$ or $b_j$ on its boundary is acyclic.
\end{lemma}
\begin{proof}
	All edge fragments that are incident to~$x$ are oriented away
	from~$x$. Similarly, all edge fragments that are incident to~$b_j$
	are oriented towards~$b_j$.
\end{proof}

\begin{lemma}\label{obs:qquasishell}
	There is a free lens in~$\Gamma$ or every quasi-lens~$Q$ with $b_j\notin\partial Q$ is stabbed by a blue vertex of~$H$.
\end{lemma}
\begin{proof}
	Assume that all lenses in~$\Gamma$ are stabbed, which by \cref{obs:lensInside} implies that all quasi-lenses in~$\Gamma$ are stabbed. Let~$Q$ be a quasi-lens in~$\Gamma$. By definition (of quasi-lens)~$Q$ does not contain~$b_j$, and we have $b_j\notin\partial Q$ by assumption. Thus, by \cref{lem:allredpoints} we have~$Q\cap R=\emptyset$. Therefore, $Q$ is stabbed by a blue vertex, as claimed.
\end{proof}

\begin{restatable}{lemma}{lemacyclic}\label{lem:acyclic}
	There is a free lens in~$\Gamma$ or every essential quasi-lens of~$\Gamma$ is acyclic.
\end{restatable}
\begin{proof}
	Suppose for the sake of a contradiction that all lenses in~$\Gamma$ are stabbed and there exists an essential cyclic quasi-lens~$Q$. 
	Without loss of generality, we make the following two assumptions about the quasi-lens~$Q$:

	\vspace{-1.5ex}\begin{enumerate}
		\item\label{assume1} $Q$~is induced by a fragment~$\beta\xi$ of~$e_1$ in~$\Gamma$ such that~$\beta$ is closer to~$b_j$ along~$e_1$ (and closer to~$x$ along~$\ett$).
		\item\label{assume2} No fragment of~$e_1$ induces a cyclic quasi-lens along the part of~$\ett$ between~$\beta$ and~$b_j$ (otherwise, we would let such a quasi-lens take the role of~$Q$). 
	\end{enumerate} 

	\vspace{-1ex} \noindent For ease of illustration we imagine~$\ett$ as a horizontal line segment such that~$x$ lies to the left of~$b_j$ and~$e_1$ as a curve that ``meanders'' around~$\ett$. By \cref{obs:lensvertex} and \cref{obs:qquasishell} we know that $Q$ contains a blue vertex~$v$ of~$H$. Let~$r\in R\setminus\{x\}$. Such a vertex exists because~$i > m$ and thus $r_1\in R\setminus\{x\}$. As~$r$ lies in the exterior of~$Q$, the edge~$vr$ of~$H$ crosses~$\partial Q$. We prove the statement in two steps. First, we consider the case that~$Q$ is a lens. Then we address the extension to quasi-lenses.
	
	So suppose for now that~$Q$ is a lens.
	Then we may suppose without loss of generality that~$vr$ crosses~$\partial Q$ along~$e_1$ (otherwise, as~$Q$ is also induced by a fragment of~$\ett$, we may exchange the roles of~$e_1$ and~$\ett$). 
	As~$e_1\cong_H\ett$, the rotation of the crossing between $vr$ of~$H$ and $\ett$ is the same as the rotation of the crossing between $vr$ and~$e_1$. There are no further crossings between $vr$ and $\ett$ or~$e_1$.  
	The labeling of the endpoints of~$\ett$ also fixes which of the endpoints the two stubs of~$e_1$ connect to, which results in the following two cases. In Case~1 (\cref{fig:acyclic}(left)), the edge $vr$ intersects~$\ett$ between~$Q$ and~$b_j$; in Case~2 (\cref{fig:acyclic}(right)), it intersects~$\ett$ between~$Q$ and~$x$. If~$x$ is blue, then also~$rx$ is an edge of~$H$, and the roles of~$x$ and~$b_j$ in this proof are interchangeable. Therefore, we may opt to treat this situation as an instance of Case~2,  which conversely allows us to assume~$x\in R$ in Case~1 without loss of generality.
	
	In both cases we also add the edge~$rb_j$ to the picture. As~$rb_j$ is adjacent in~$H$ to all other edges considered so far, it does not cross any of them. We consider the two cases in order. Removal of~$\partial Q$ splits~$c\in\{e_1,\ett\}$ into two parts; denote the part that connects to~$p\in\{x,b_j\}$ as~\emph{$p$-stub} of~$c$ with respect to~$Q$. We consider a $p$-stub to be directed from~$Q$ to~$p$.
	
	\begin{figure}[htbp]
		\begin{minipage}[t]{.48\linewidth}
			\centering\includegraphics[page=1]{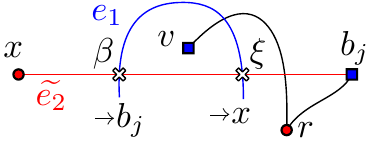}
		\end{minipage}
		\hfill
		\begin{minipage}[t]{.48\linewidth}
			\centering\includegraphics[page=2]{acyclic_kpt}
		\end{minipage}
		\caption{The two cases in the proof of \cref{lem:acyclic}: \cref{quasi:case:1} (left) and \cref{quasi:case:2} (right).\label{fig:acyclic}}
	\end{figure}
	
	\begin{case}[$x \in R$ and $vr$ intersects~$\ett$ between~$Q$ and~$b_j$]\label{quasi:case:1}
		Each edge of~$H$ is crossed at most once by~$e_1$ and~$\ett$. So there are no further crossings of~$e_1$ or~$\ett$ with~$vr$. Let~$c_{x}$ denote the first intersection of the~$x$-stub~$s_x^1$ of~$e_1$ w.r.t.~$Q$ with the~$x$-stub~$s_x^2$ of~$\ett$ w.r.t.~$Q$ (well-defined because~$x$ is such an intersection). Consider the simple closed curve~$C$ that goes from~$c_{x}$ along~$\ett$ to~$\xi$ and then along~$s_x^1$ back to~$c_{x}$. Now consider the~$b_j$-stub~$s_b^1$ of~$e_1$ w.r.t.~$Q$. As~$C$ cannot cross either of the edges~$vr$ and~$rb_j$, it follows that~$b_j$ and the starting fragment of~$s_b^1$ are on different sides of~$C$. Therefore, in order to reach~$b_j$, the stub~$s_b^1$ crosses~$C$. While~$s_b^1$ cannot cross~$C\cap e_1$, it can cross~$C\cap\ett$, possibly many times. However, it cannot cross~$\partial Q$ because~$Q$ is a lens, and so it crosses~$s_x^2$. Denote by~$\mathcal{R}_C$ the region of~$\Su$ enclosed by~$C$ that does not contain~$b_j$.

		We claim that there exists a quasi-lens~$Q'\subset\mathcal{R}_C$ that is induced by a fragment of~$s_b^1$ in~$\Gamma$.
		To see this, consider the first intersection~$c_b$ of~$s_b^1$ with~$s_x^2$. If~$s_b^1$ crosses~$s_x^2$ from left to right at~$c_b$, then the claim holds because the part of~$s_b^1$ up to~$c_b$ induces a cyclic quasi-lens with~$\ett$. Otherwise, the stub~$s_b^1$ crosses~$s_x^2$ from right to left at~$c_b$, that is, the initial edge fragment of~$s_b^1$ induces an anti-lens in~$\Gamma$ whose inside stub is formed by the part of~$s_b^1$ that starts at~$c_b$. In this case the claim follows by \cref{lem:antilensinside}.
		
		As~$Q'$ and~$b_j$ are separated by~$C$, it follows by \cref{obs:qquasishell} that~$Q'$ contains a blue vertex~$u$ of~$H$. See \cref{fig:acyclic:2}(left) for an illustration.
		Next we consider the edge~$ur$ of~$H$. It cannot cross the other edges that are incident to~$r$ in~$H$, in particular, the edges~$vr$ and~$rb_j$. But~$ur$ can (and actually has to) cross~$e_1$ and~$\ett$, but at most once each and with the same rotation. Observe that~$Q'$ and~$r$ are on different sides of~$C$. So when tracing~$ur$ starting from~$u$, the edge has to cross both~$\partial Q'$ and~$C$ in order to reach~$r$. As it may cross~$e_1$ at most once, $ur$~can leave~$C$ only by crossing~$s_x^2$; let us denote this crossing by~$p_u=ur\cap s_x^2$. 
		Consequently, the edge~$ur$ also crosses~$s_b^1$ with the same rotation.
		These two crossings can occur in either order, which determines the drawing of~$ur$ so that either way there is a simple closed curve~$C'$ that starts at~$p_u$, then reaches~$r$ along~$ur$, then continues along~$rv$ until the crossing with~$\ett$, from where it returns back to~$p_u$. Observe that~$x$ and~$v$ are on different sides of~$C'$. See \cref{fig:acyclic:2}(middle) for an illustration.
		
		\begin{figure}[htbp]
			\begin{minipage}[t]{.32\linewidth}
				\centering\includegraphics[page=3]{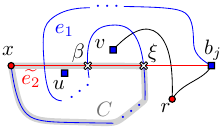}
			\end{minipage}
			\hfill
			\begin{minipage}[t]{.32\linewidth}
				\centering\includegraphics[page=4]{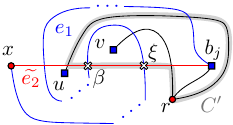}
			\end{minipage}
			\hfill
			\begin{minipage}[t]{.32\linewidth}
				\centering\includegraphics[page=5]{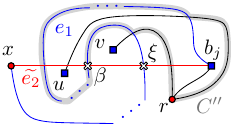}
			\end{minipage}
			\caption{Situations in \cref{quasi:case:1} of the proof of \cref{lem:acyclic}.\label{fig:acyclic:2}}
		\end{figure}
		
		Finally, we consider the edge~$vx$, which exists because we assume $x \in R$ for this case. Note that it may not be an edge of~$H$, but still we need to be able to draw it consistently (with the same crossings and crossing rotations) in both~$H\cup e_1$ and~$H\cup\ett$, so as to be able to extend these drawings to drawings of the complete multipartite graph with the same~\ERS. 
		So let us first consider the drawing of~$vx$ in~$H\cup\ett$. Note that~$vx$ must not cross the adjacent edges~$\ett$ and~$vr$. As~$vx$ crosses~$C'$ it follows that it crosses~$ur$ so that when passing through this crossing from~$v$ to~$x$, the vertex~$u$ is on the left side. So the drawing of~$vx$ in~$H\cup e_1$ must cross~$ur$ in the same manner. In~$H\cup e_1$, the edge~$vx$ must not cross the adjacent edges~$e_1$ and~$vr$. Consider the simple closed curve~$C''$ that starts from~$r$ along~$ur$ up to the crossing with~$e_1$, from there continues along~$e_1$ up to the crossing with~$vr$, and from there returns back to~$r$ along~$vr$. See \cref{fig:acyclic:2}(right) for an illustration. Observe that~$v$ is on one side of~$C''$ and the edge~$ur$ is on the other side and on~$\partial C''$. The only edge along~$\partial C''$ that~$vx$ may cross is~$ur$; but doing so would cross~$ur$ with the wrong crossing rotation (so that when passing through this crossing from~$v$ to~$x$, the vertex~$u$ is on the right side). Therefore, there is no way to draw~$vx$ in~$H\cup e_1$ as required, in contradiction to our assumption that both~$H\cup e_1$ and~$H\cup\ett$ can be extended to drawings of the complete multipartite graph with the same \ERS. So we conclude that this case cannot occur.
	\end{case}
	
	\begin{case}[$vr$ intersects~$\ett$ between~$Q$ and~$x$]\label{quasi:case:2}
		As both drawings~$H\cap e_1$ and~$H\cap\ett$ are simple, there are no further crossings of~$vr$ with $e_1$ or~$\ett$.
		Consider the closed curve~$C$ that goes from~$b_j$ along~$rb_j$ to~$r$ and then along~$rv$ until it reaches~$\partial Q$ from where it continues along~$\partial Q\cap e_1$ to~$\beta$, and then along~$\ett$ back to~$b_j$.
		Now consider the~$x$-stub~$s_x^1$ of~$e_1$ w.r.t.~$Q$. See \cref{fig:acyclic:3} for an illustration.
		As~$\ett$ cannot cross~$\partial Q$ nor the edge~$rb_j$ and it can only cross~$vr$ once, it follows that~$x$ and the initial fragment of~$s_x^1$ are on different sides of~$C$.
		Therefore, in order to reach~$x$, the stub~$s_x^1$ crosses~$C$.
		As~$s_x^1$ cannot cross~$vr$, $rb_j$, or~$\partial Q$ (the latter because~$Q$ is a lens), this crossing is along the~$b_j$-stub of~$\ett$ w.r.t.~$Q$, at a point~$\phi$. 
		Then the edge fragment of~$s_x^1$ that ends at~$\phi$ induces a cyclic quasi-lens~$Q'$ in~$\Gamma$ along the part of~$\ett$ between~$\beta$ and~$b_j$. This is a contradiction to our 
		Assumption~\ref{assume2}
		about~$Q$ and, therefore, this case cannot occur, either.
	\end{case}
	
	\begin{figure}[htbp]
		\centering\includegraphics[page=6]{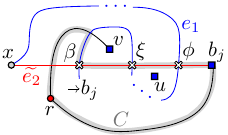}
		\caption{The situation in \cref{quasi:case:2} in the proof of \cref{lem:acyclic}.\label{fig:acyclic:3}}
	\end{figure}
	
	Since we derive a contradiction in both cases, we conclude that~$Q$ is not a lens, which completes the first part of the proof. It remains to consider the case that~$Q$ is a general quasi-lens that is not a lens.  
	Then by \cref{obs:lensInside} there is a lens~$L\subset Q$. See \cref{fig:quasimore}(left) for an illustration. As~$L$ lies along the part of~$\ett$ between~$\beta$ and~$b_j$, by Assumption~\ref{assume2} 
	it is acyclic. Furthermore, by the assumption of the lemma, the lens~$L$ is stabbed by a vertex~$v$ of~$H$. As neither~$x$ nor~$b_j$ lies on~$\partial L$, by \cref{obs:qquasishell} we know that~$v$ is blue and, therefore, connected to~$r$ by an edge in~$H$. As~$r\notin Q$, the edge~$vr$ crosses~$\partial Q$ exactly once. Given that~$vr$ also crosses both~$\partial L$ and~$e_1$ exactly once, we conclude that it crosses~$\partial Q\cap\ett$ at a point~$\mu$ between~$\beta$ and~$\xi$. Observe that~$vr$ cannot cross~$\partial L\cap e_1$ because such a crossing would have a different rotation than the crossing with~$\ett$ at~$\mu$. It follows that~$\mu\in\partial L$. Let~$\nu$ denote the vertex of~$L$ between~$\mu$ and~$\xi$. 
	
	\begin{figure}[htbp]
		\begin{minipage}[t]{.48\linewidth}
			\centering\includegraphics[page=11]{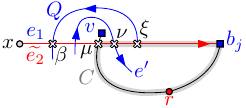}
		\end{minipage}
		\hfill
		\begin{minipage}[t]{.48\linewidth}
			\centering\includegraphics[page=12]{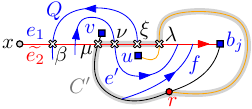}
		\end{minipage}
		\hfill
		\caption{Extending the statement to general quasi-lenses in \cref{lem:acyclic}.\label{fig:quasimore}}
	\end{figure}
	
	Consider the simple closed curve~$C$ that goes from~$\mu$ along~$\ett$ to~$b_j$, then along~$b_jr$ to~$r$, and then along~$rv$ back to~$\mu$. Let~$e'$ denote the part of~$e_1$ from~$\nu$ towards~$b_j$ up to its next intersection with~$C$ (which exists, possibly at~$b_j$). Note that~$e'$ cannot cross~$vr$ because such a crossing would have the wrong rotation compared to the crossing~$vr\cap\ett=\mu$. It cannot cross the adjacent edge~$rb_j$, either, nor the part of~$\ett$ between~$\mu$ and~$\nu$ because~$L$ is a lens. So~$e'$ intersects~$C\cap\ett$ somewhere between~$\nu$ and~$b_j$. Furthermore, given that~$vr$ crosses~$\ett$, it must cross~$e_1$ as well. Denote by~$f$ the part of~$e_1$ from~$e_1\cap vr$ towards~$b_j$ up to the next intersection with~$C$ (which exists, possibly at~$b_j$). 
	To have the correct rotation at the crossing of $e_1$ and~$vr$, the curve~$f$ starts on the same side of~$C$ as~$e'$. 
	Also, being a part of~$e_1$, the curve~$f$ has the same crossing restrictions as~$e'$ and, therefore, it ends on~$C\cap\ett$, somewhere between~$\nu$ and~$b_j$. As the intersections of~$e'$ and~$f$ with~$C\cap\ett$ are distinct, it follows that~$e'$ crosses~$\ett$ properly between~$\nu$ and~$b_j$. In other words, the fragment~$e'$ of~$e_1$ induces an acyclic quasi-lens~$Q'$ in~$\Gamma$.
	
	As $x\notin\partial Q'$ and $b_j\notin\partial Q'$, by \cref{obs:qquasishell} there is a blue vertex~$u\in Q'$. Consider the edge~$ur$ in~$H^+$: It starts inside~$Q'$ and its target~$r$ lies outside of~$Q'$ and on~$C$. However, it cannot reach~$r$ by staying on the same side of~$C$ because this would require crossing both~$e'$ and~$f$. Hence, the edge~$ur$ has to cross~$C$. But it cannot cross the adjacent edges~$vr$ and~$rb_j$, nor can it cross through~$Q$, as this would require crossing~$e_1$ and~$\ett$ with different rotations. So we conclude that~$ur$ crosses~$C$ along~$\ett$, at a point~$\lambda$ between~$\xi$ and~$b_j$. See \cref{fig:quasimore}~(right) for an illustration. 
	
	Let~$C'$ denote the simple closed curve that goes from~$r$ along~$rv$ to~$\mu$, then along~$\ett$ to~$\lambda$, and then along~$ur$ back to~$r$. Now consider the curves~$e'$ and~$f$. Both are part of~$e_1$, so one appears before the other along~$e_1$. 
	We assume that~$f$ appears before~$e'$ along~$e_1$.
	(The reasoning when $e'$ appears before~$f$ along~$e_1$ is analogous.)
	Note that both~$e'$ and~$f$ start at a crossing of~$e_1$ with~$C\cap C'$. 
	Thus, in order to connect~$f$ to~$e'$, the edge~$e_1$ must get from the endpoint of~$f$, which lies on~$C$ but strictly on one side of~$C'$, to the other side of~$C'$. In other words, somewhere between~$f$ and~$e'$, the edge~$e_1$ crosses~$C'$. Let~$\phi$ denote the first such crossing. Then~$\phi\notin vr$ because we already know the unique crossing~$e_1\cap vr$, which is the startpoint of~$f$. Also~$\phi\notin ur$ because then~$\phi$ would have the wrong rotation, compared to the crossing~$\lambda=ur\cap\ett$. It follows that~$\phi\in\ett\cap C'$. Let~$\sigma$ denote the crossing~$e_1\cap\ett$ that immediately precedes~$\phi$ along~$e_1$. As~$e_1$ cannot cross the adjacent edge~$rb_j$, the fragment~$\sigma\phi$ of~$e_1$ attaches to the same side of~$\ett$ at~$\phi$ and~$\sigma$. Thus, this fragment induces a quasi-lens~$Q''$ in~$\Gamma$. Moreover, by definition of~$\phi$ (as the first crossing of~$e_1$ with~$C'$ after~$f$) and the position of the endpoint of~$f$ (after all of~$C'\cap\ett$ along~$\ett$) we conclude that~$Q''$ is cyclic. But this is a contradiction to our choice of~$Q$, concretely to Assumption~\ref{assume2} above.
	Hence, such a cyclic quasi-lens~$Q$ does not exist to begin with, which concludes the proof of the lemma.
\end{proof}

After these preparations we are now ready to prove \cref{lem:normal}.
\lemnormal*
\begin{proof}
	Assume for the sake of a contradiction that every lens in~$\Gamma$ is stabbed. Then by \cref{obs:lensInside} every quasi-lens in~$\Gamma$ is stabbed and, therefore, essential. Let~$L$ be a lens in~$\Gamma$ with $b_j\notin\partial L$, which exists by assumption. As~$L$ is stabbed, it contains a vertex~$v$ of $H$. Moreover, as $b_j\notin\partial L$ we have~$v\in B$ by \cref{lem:allredpoints}. As~$L$ 
	is essential, by \cref{lem:acyclic} it is acyclic.
	
	Let $\chi_1, \ldots, \chi_q$ be the intersections of~$e_1$ and~$\ett$, ordered 
	along~$\ett$, with $\chi_1=x$ and $\chi_q=b_j$.
	As $b_j\notin\partial L$, the lens~$L$ is defined by a fragment of~$\ett$ between $\chi_k$ and $\chi_{k+1}$, for some $1 \leq k \leq q-2$. For ease of illustration we imagine~$\ett$ as a horizontal line segment such that~$x$ lies to the left of~$b_j$ and~$e_1$ as a curve that ``meanders'' around~$\ett$. We may assume without loss of generality that $\eblue$ crosses $\ered$ from \emph{left to right} at~$\chi_{k+1}$ (otherwise, flip the drawing vertically).
	Then, as~$L$ is acyclic, the edge~$e_1$ crosses~$\ett$ from right to left at~$\chi_{k}$.
	
	Let~$r\in R\setminus\{x\}$. Such a vertex exists because~$i > m$ and thus $r_1\in R\setminus\{x\}$. As~$L\cap R=\emptyset$, the edge~$vr$ in~$H$ crosses~$\partial L$ in exactly one of its two edge fragments.
	Assume without loss of generality that~$vr$ crosses~$\partial L$ along~$\eblue$ (otherwise, exchange the roles of $\eblue$ and $\ered$),
	and denote this crossing by~$\psi$. Then~$vr$ crosses~$\eblue$ from right to left. As~$vr$ crosses~$\ered$ in the same direction, it must ``go around''~$b_j$ or~$x$ and cross~$\ett$ between~$\chi_{k+1}$ and~$b_j$ before ending at~$r$; see \cref{fig:normalLensGeneral}. (In the figure, $\chi_k$ is depicted as a crossing, but we could also have~$\chi_k=x$.)
	
	\begin{figure}[htbp]
		\centering\includegraphics[page=1]{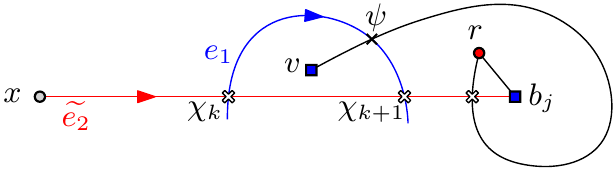}
		\caption{Base situation in the proof of \cref{lem:normal}\label{fig:normalLensGeneral}.}
	\end{figure}
	
	Let~$\chi_\ell$ denote the next consecutive intersection of~$e_1$ with~$\ett$ after~$\chi_{k+1}$ (possibly~$\chi_\ell=b_j$). We distinguish several cases depending on the relation of $k$ and $\ell$, and on the rotation of $\chi_\ell$. 
	
	\begin{case}[$\ell < k$ and $\eblue$ crosses $\ered$ from right to left at $\chi_\ell$]\label{ln:c1}
		Then the edge fragment of~$e_1$ between~$\chi_{k+1}$ and~$\chi_\ell$ induces a cyclic quasi-lens in~$\Gamma$, in contradiction to \cref{lem:acyclic}. Hence this case cannot occur.
	\end{case}
	
	\begin{case}[$\ell < k$ and $\eblue$ crosses $\ered$ from left to right at $\chi_\ell$]\label{ln:c2}
		Then the edge fragment~$\chi_{k+1}\chi_\ell$ of~$e_1$ forms an anti-lens~$A$ in~$\Gamma$ whose inside stub~$s_A$ is formed by the part of~$e_1$ from~$\chi_\ell$ to~$b_j$. See \cref{fig:normalLensCase12} for an illustration. By \cref{lem:antilensinside} there is a quasi-lens~$Q\subset A$ in~$\Gamma$ that is induced by an edge fragment of~$s_A$. As~$A$ separates~$Q$ from~$b_j$, by \cref{obs:qquasishell} there is a blue vertex~$u\in Q$. We claim that there is no way to add the edge~$ur$ to this drawing.
		
		\begin{figure}[htbp]
		\begin{center}
			\begin{minipage}[t]{.48\linewidth}
				\includegraphics[page=8]{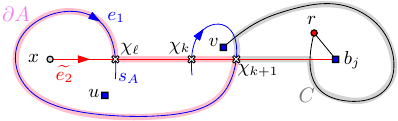}
			\end{minipage}
		\end{center}
		\caption{\cref{ln:c2} in the proof of \cref{lem:normal}: $\ell<k$.\label{fig:normalLensCase12}}
	\end{figure}

		To see this, observe first that the edge fragment~$\chi_k\chi_{k+1}$ of~$\ett$ is not crossed by~$s_A$ because~$L$ is a lens. In particular, the closures of~$Q$ and~$L$ are disjoint. As~$u\in Q$ and~$r\notin Q$, as a first step an edge from~$u$ to~$r$ has to leave~$Q$, by crossing~$\partial Q$ along one of~$e_1$ or~$\ett$. As~$\partial Q\cap\partial L=\emptyset$ and~$r\notin L$, this implies that~$ur$ cannot enter~$L$ (without crossing one of~$e_1$ or~$\ett$ a second time on its way to~$r$, which is forbidden). Next observe that~$u$ and~$r$ are also separated by the simple closed curve~$C$ that starts at~$\chi_{k+1}$ along~$\ett$ towards~$b_j$ up to the crossing with the edge~$rb$, then along~$rb$ towards~$b$ up to the crossing with~$\partial L$, and then along~$\partial L$ back to~$\chi_{k+1}$. So the edge~$ur$ has to also cross both~$\partial A$ and~$C$. The curve~$C$ consists of three parts: 

		\vspace{-1.5ex}\begin{enumerate}
			\item The part along~$\partial L$, which cannot be crossed by~$ur$, as argued above.
			\item The part along the edge~$br$, which is adjacent to~$ur$ and therefore cannot be crossed by~$ur$, either. 
			\item The part along~$\ett$, which is the only remaining option to be crossed by~$ur$. 
		\end{enumerate}
	
	\newpage
	\noindent The boundary~$\partial A$ consists of two parts: 

	\vspace{-1.5ex}\begin{enumerate}
		\item The part along~$\ett$, which~$ur$ cannot cross, given that it already crosses~$\ett$ along~$C$ (and the only common point~$\chi_{k+1}$ is part of~$\partial L$).
		\item The part along~$e_1$, which is the only remaining option to be crossed by~$ur$. 
	\end{enumerate}

	\noindent However, in addition the edge~$ur$ also crosses~$\partial Q$. As~$C\cap\partial Q=\emptyset$ and the only common point~$\partial A\cap s_A$ is~$\chi_\ell\in\ett$, there is no way to realize all these required crossings along~$ur$. So this case cannot occur, either. 
	\end{case}

	\begin{case}[either $k+1 < \ell < q$ and $\eblue$ crosses $\ered$ from left to right at $\chi_\ell$ or $\ell=q$ and hence~$\chi_\ell=b_j$]\label{ln:c3}
		Then there is no way to draw the edge~$vr$ without crossing one of~$e_1$ or~$\ett$ more than once, which is forbidden. See \cref{fig:normalLensCase3} for an illustration. So this case cannot occur, either.
	\end{case}
	
	\begin{figure}[htbp]
		\begin{minipage}[t]{.48\linewidth}
			\centering\includegraphics[page=4]{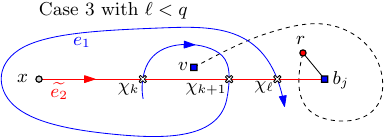}
		\end{minipage}
		\hfill
		\begin{minipage}[t]{.48\linewidth}
			\centering\includegraphics[page=5]{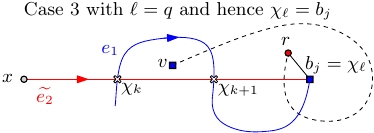}
		\end{minipage}
		\caption{\cref{ln:c3} in the proof of \cref{lem:normal}: $k+1 < \ell < q$ or $\ell=q$.\label{fig:normalLensCase3}}
	\end{figure}
	
	\begin{case}[$k+1 < \ell < q$ and $\eblue$ crosses $\ered$ from right to left at $\chi_\ell$]
		Then the edge fragment $\chi_{k+1}\chi_{\ell}$ of~$\eblue$ forms a quasi-lens~$Q$ with~$\ered$, which by \cref{obs:qquasishell} contains a blue vertex~$u$. 
		The edge~$ur$ must leave~$Q$, which involves crossing one of~$e_1$ or~$\ett$.
		
		Assume first that~$ur$ crosses~$\eblue$ along~$\partial Q$. Then~$ur$ crosses~$\eblue$ from left to right.
		Let~$F$ be the region that contains~$r$ and is bounded by the edge~$\eblue$ between~$\chi_{k+1}$ and~$\psi$, 
		by~$vr$ between $\psi$ and the crossing of~$br$ with~$\ered$, and by~$\ered$ between that crossing and~$\chi_{k+1}$; see the shaded area in \cref{fig:normalLensCase4}.
		The part of~$ur$ directly after crossing~$\eblue$ lies outside~$F$. To reach~$r$, the edge~$ur$ must enter~$F$. 
		However, it must not intersect~$vr$, it must not intersect~$\eblue$ again, and it must intersect~$\ered$ from left to right. 
		Hence $ur$ cannot be completed, a contradiction.
	
		So we may assume that~$ur$ crosses~$\ett$ from right to left along~$\partial Q$.
		Let~$\Delta\subset F$ be bounded by~$b_jr$, the part of~$vr$ between~$r$ and the crossing of~$vr$ and~$\ered$, and the part of~$\ered$ between that crossing and~$b_j$.
		The part of~$ur$ directly after crossing~$\ered$ lies in~$F$. 
		To reach~$r$, the edge~$ur$ must stay in~$F$ and it must not intersect~$\Delta$, leading to the drawing depicted in \cref{fig:normalLensCase4}(right). 

		\begin{figure}[htbp]
		\begin{minipage}[t]{.48\linewidth}
			\centering\includegraphics[page=6]{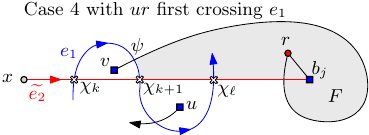}
		\end{minipage}
		\hfill
		\begin{minipage}[t]{.48\linewidth}
			\centering\includegraphics[page=7]{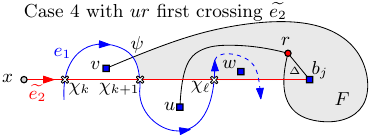}
		\end{minipage}
		\caption{Case 4 of \cref{lem:normal}\label{fig:normalLensCase4}}
		\end{figure}

	\noindent	Consider how~$e_1$ continues beyond~$\chi_\ell$: 

		\vspace{-1.5ex}\begin{enumerate}
			\item It cannot cross~$ur$ because the rotation of~$ur\cap e_1$ does not match with the rotation of~$ur\cap\ett$.
			\item It cannot cross~$rv$ because of the crossing~$rv\cap e_1=\psi$.
			\item It cannot cross~$\ett$ to the left of~$\chi_\ell$ because this would create a cyclic quasi-lens, in contradiction to \cref{lem:acyclic}.
			\item Thus, it crosses~$\ett$ to the right of~$\chi_\ell$, thereby inducing a (cyclic) quasi-lens~$Q'$.
		\end{enumerate}

\noindent	The edge~$wr$ of~$H$ must cross~$\partial Q'$ to reach~$r$, which means it crosses one of~$e_1$ or~$\ett$. So it also crosses the other, with the same crossing rotation. As~$wr$ cannot cross the adjacent edges~$vr$ and~$ur$, it has to leave~$F$ so as to cross~$\ett$. In order to reach~$r$, which lies inside~$F$, the edge~$wr$ has to enter~$F$ again. But this is impossible because:

	\vspace{-1.5ex}\begin{enumerate}
		\item The edge~$wr$ cannot cross~$\ett$ again.
		\item It cannot cross~$vr$.
		\item It cannot reach~$F$ through~$L$ without crossing~$e_1$ twice.
	\end{enumerate} 

\noindent Therefore, this case cannot occur, either.
	\end{case}

	To summarize, in each of the four cases we obtained a contradicition to the assumption that every lens in~$\Gamma$ is stabbed. Hence, there exists a free lens in~$\Gamma$.
\end{proof}

\subsection{Invertible triangles are empty (Proof of Lemma~\ref{prop:flippable})}\label{sec:flippable}

By definition of an invertible triangle, there exists a drawing $D'$ where the crossing triangle~$\Delta'$ formed by the same combinatorial edges as those for $\Delta$ has a different \parity than $\Delta$.
We call $\Delta'$ the \emph{inverted triangle} of $\Delta$ in $D'$.
\cref{prop:flippable} is an immediate corollary of two lemmata below (see the end of the section). 

\begin{lemma}\label{lem:point_inside_both_triangles}
	For $m,n\geq 3$ and $m+n \geq 7$, 
	let $D$ and $D'$ be two drawings of $K_{m,n}$ on the sphere with the same \ERS and let $\Delta$ and $\Delta'$ be crossing triangles in $D_1$ and $D_2$, respectively, such that $\Delta'$ is an inverted triangle of $\Delta$. Let $v$ be a vertex that in $D$ lies inside $\Delta$. Then in~$D'$, $v$~cannot lie inside $\Delta'$. 
\end{lemma}

\begin{figure}[htb]
	\centering
	\includegraphics[page=1]{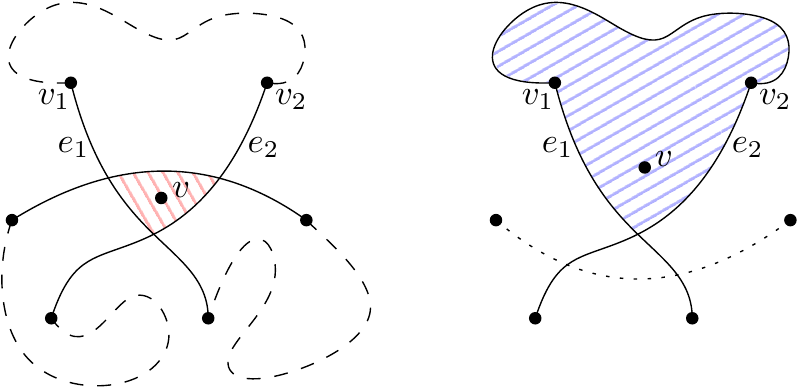}
	\caption {If the vertex $v$ in drawing $D$ (left) lies inside the  triangle $\Delta$ (shaded red) then its position w.r.t.\ an induced $K_{2,3}$ is fixed: in the drawing $D'$ (right) $v$ must be in the same \cell of this $K_{2,3}$ and can thus not be in $\Delta'$.%
	}
	\label{fig:nopointinsidetwice}
\end{figure}

\begin{proof}
	First, observe that out of the six endpoints of the three edges forming $\Delta$ in $D$, exactly three belong to a bipartition class of the vertices of $K_{m,n}$, and the other three to the other class.
	Consider the three curves connecting these endpoints in the way as depicted as dashed lines on the left side of \cref{fig:nopointinsidetwice}. At least one of these curves connects two points from different bipartition classes of the vertices of $K_{m,n}$, and is thus a valid edge of the complete bipartite drawing. Otherwise, at least four of the endpoints would have to belong to the same bipartition class - a contradiction to the first observation.
	
	W.l.o.g.\ let $v_1$ and $v_2$ be the endpoints of the edges $e_1$ and $e_2$, respectively, which are connected this way. Consider the drawing of a $K_{2,2}$ induced by the vertices of $e_1$ and $e_2$, see \cref{fig:nopointinsidetwice}(right). Together with $v$  this gives a drawing of a $K_{2,3}$ and by \cref{prop:Ksmall} the vertex $v$ lies in a fixed \cell of this drawing. In the drawing $D$ this is the area inside~$\Delta$, which is read shaded in \cref{fig:nopointinsidetwice}. Thus in the drawing $D'$, the vertex $v$ must lie in the same area of the $K_{2,3}$ induced by $e_1$, $e_2$ (shaded blue in \cref{fig:nopointinsidetwice}) and cannot lie inside of $\Delta'$.
\end{proof} 

\begin{lemma}\label{lem:point_inside_one_triangle}
	For $m,n\geq 3$ and $m+n \geq 7$ 
	let $D$ and $D'$ be two drawings of $K_{m,n}$ on the sphere with the same \ERS and let $\Delta$ and $\Delta'$ be triangles in $D_1$ and $D_2$, respectively, such that $\Delta'$ is an inverted triangle of $\Delta$. Let $v$ be a vertex that in $D$ lies inside $\Delta$. Then in~$D'$, $v$~cannot lie outside $\Delta'$. 
\end{lemma}

\begin{proof}	
	\begin{figure}[htb]
		\centering
		\includegraphics[page=1]{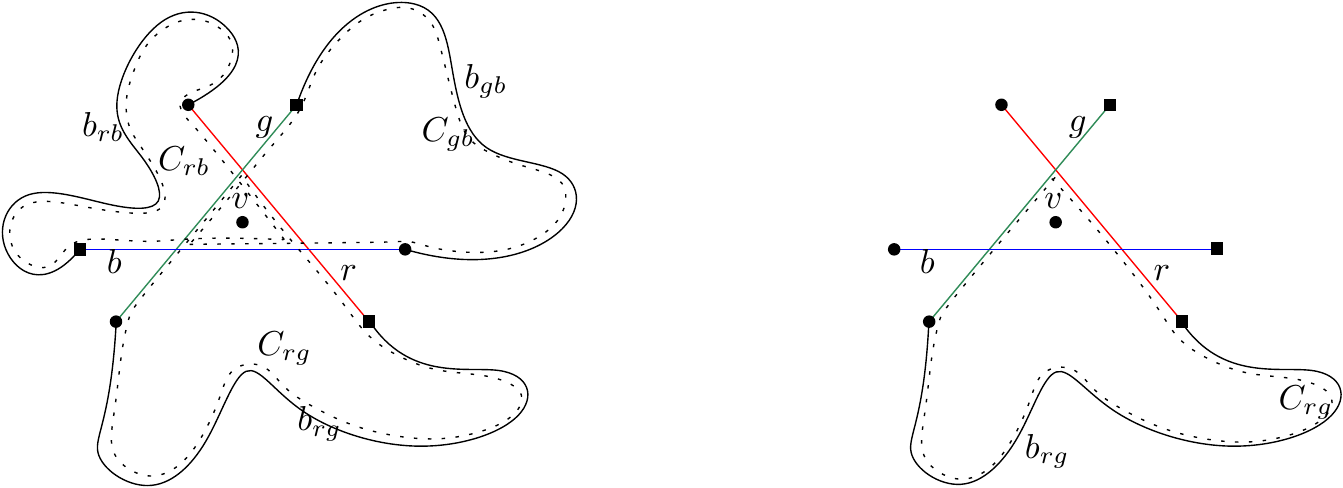}
		\caption {Drawing $D$ where the  triangle $\Delta$ has a vertex $v$ on the inside of $\Delta$. 
			The vertices of the two bipartition classes of vertices are drawn as circles and squares, respectively. 
			On the left, the endvertices of the edges of~$\Delta$ belong to the two bipartition classes of the vertices of $K_{m,n}$ alternatingly, 
			while on the right, they form two consecutive blocks.}
		\label{fig:pointinsidetriangle1}
	\end{figure}
	
	Consider the endvertices of $\Delta$. 
	There are two different possibilities how they belong to the two  bipartition classes of the vertices of $K_{m,n}$. Either they are alternating (Case~\ref{inside:case:1}, see \cref{fig:pointinsidetriangle1}(left)) or there are two blocks of three vertices each from the same class (Case~\ref{inside:case:2}, see \cref{fig:pointinsidetriangle1}(right)). To simplify the description, we color the three edges that form $\Delta$ in blue (edge~$b$), red (edge~$r$), and green (edge~$g$), and consider three of the edges connecting their endpoints, namely the black edges $b_{rg}$, $b_{gb}$, and $b_{br}$ as shown in the drawings in \cref{fig:pointinsidetriangle1}.
	Note that we do not assume anything on how the three black edges are drawn, except that they belong to a simple drawing. For example, $b_{rg}$ may intersect $b_{gb}$ and $b$. 
	
	In order to obtain a contradiction, in both cases, we assume that, in drawing $D$ the vertex~$v$ lies on the inside of~$\Delta$, and that in drawing $D'$, the vertex~$v$ lies on the outside of~$\Delta'$. A drawing of $D'$ is shown in \cref{fig:pointinsidetriangle2}. 
	As in \cref{fig:pointinsidetriangle1}, we do not assume anything on how the three black edges are drawn beyond simplicity (and there might be intersections not drawn in the figure).

		\begin{figure}[htb]
			\centering
			\includegraphics[page=2]{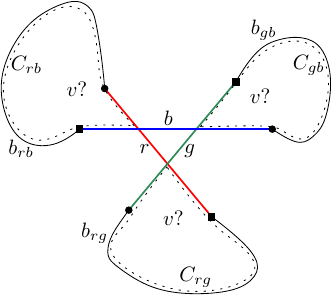}
			\caption {Drawing $D'$ where the vertex $v$ must lie in the three indicated cycles outside of $\Delta'$.}
			\label{fig:pointinsidetriangle2}
		\end{figure}

	\begin{case}[Alternating endpoints of the classes]
		\label{inside:case:1}
		Consider the drawing of $K_{2,3}$ induced by the endpoints of $r$ and $g$ and the vertex~$v$ and the drawing of $K_{2,2}$ induced by the endpoints of $r$ and $g$. 
		Applying \cref{prop:Ksmall} to the drawing of $K_{2,3}$ above, it follows that the vertex~$v$ in $\Delta$ also lies in the \cell of the drawing of $K_{2,2}$ that is bounded by a part of the edge $r$, a part of the edge $g$, and the edge $b_{rg}$, as indicated by the dotted cycle in the drawing in \cref{fig:pointinsidetriangle1}(left). We denote this bounding cycle by~$C_{rg}$. Similarly, $v$ also lies in the \cell of the drawing of $K_{2,2}$ induced by $r$ and $b$ bounded by the dotted cycles $C_{rb}$, and that corresponding to $g$ and $b$ bounded by the dotted cycle $C_{gb}$.

		Consider the subgraph induced by $v$ and the endpoints of $r$ and $b$. Since $D$ and $D'$ have the same \ERS, $v$ has to lie in the same \cell of any drawing of this subgraph. Thus, it has to lie on the same side of the cycle $C_{rb}$ in $D$ and $D'$. Analogously, it has to lie on the same side of $C_{rg}$ and $C_{gb}$ in $D$ and $D'$.
		But since the three cycles do not have a common area inside~$\Delta'$ anymore, there must be some other place where the intersection of the relevant area of the three cycles (that is, the area where $v$ lies in) is non-empty; cf~\cref{fig:pointinsidetriangle3}.

		\begin{figure}[htb]
			\centering
			\includegraphics[page=3]{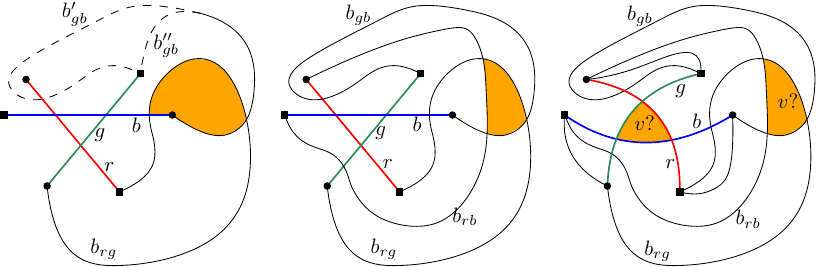}
			\caption {Drawing $D'$ where the intersection of the three regions is non empty so that $v$ can be placed there.}
			\label{fig:pointinsidetriangle3}
		\end{figure}

		Let us start with the intersection of $C_{rg}$ and $C_{gb}$. The possible crossings for these two cycles are $b_{rg} \cap b$, $b_{gb} \cap r$, and $b_{rg} \cap b_{gb}$. 
		As the two cycles must have a common area, it follows by a simple parity argument, that we have precisely two crossings between these two cycles. By symmetry, we can w.l.o.g.\ assume that one of these crossings is $b_{rg} \cap b$. 
		Note that there is only one possible crossing rotation for these two edges (\cref{fig:pointinsidetriangle3}(left)).
		
		Now $b_{gb}$ cannot cross the part of $r$ which is part of $C_{rg}$ without also crossing $b_{rg}$. So the second crossing is $b_{rg} \cap b_{gb}$. Again there is only one possible rotation for this crossing. $b_{gb}$ might cross $r$ outside of $C_{rg}$ (drawn as $b_{gb}'$ in \cref{fig:pointinsidetriangle3}(left)) or might not cross $r$ outside of~$C_{rg}$ (drawn as $b_{gb}''$ \cref{fig:pointinsidetriangle3}(left)).
		
	This gives an area where vertex $v$ could potentially lie; it is draw in orange in \cref{fig:pointinsidetriangle3}. Thus also $b_{rb}$ must intersect this region. As it cannot intersect $b$, it has to intersect both, $b_{rg}$ and $b_{gb}$. If we draw $b_{gb}''$ then this is not possible, as $g$ can only be crossed once by $b_{rb}$. So $b_{gb}$ is drawn as $b_{gb}'$, that is, crossing~$r$.
		
		Then there is only one unique way to draw $b_{rb}$ (\cref{fig:pointinsidetriangle3}(middle)).

		So far the drawing has already 9 crossings, the maximum for a drawing of $K_{3,3}$. So the remaining three edges have to be drawn without crossings, which completes the drawing (\cref{fig:pointinsidetriangle3}(middle)).
		
		As the edges on the boundary of this \cell are disjoint from the edges on the boundary of $\Delta'$, we can have a drawing $\tilde{D}$ that is identical to the drawing of this $K_{3,3}$ in $D'$, except that the triangle $\tilde{\Delta}$ corresponding to $\Delta'$ is flipped (\cref{fig:pointinsidetriangle3}(right)). Observe that the \ERS restricted to this $K_{3,3}$ is the same in both $\tilde{D}$ and $D'$. However, now there are two \cells where $v$ can lie, inside the triangle $\tilde{\Delta}$ and the \cell where $v$ lies in $D'$ (see shaded \cells in \cref{fig:pointinsidetriangle3}(right)).
		But this is a contradiction to \cref{prop:rot_cell}. 
		So it follows that in this setting, $v$ cannot lie on the inside of $\Delta'$.
	\end{case}
	
	\begin{case}[Blocks of three endpoints of each class]\label{inside:case:2}
		For convenience, we label the vertices and the crossings among $b, r, g$ as in \cref{fig:pointinsidetriangle4}; w.l.o.g.{} vertex $v$ and the vertices with subscript~2 are in the same bipartition class.
		In the drawing of~$K_{2,2}$ induced by the endvertices of $r$ and~$g$, consider the tricell enclosed by $b_{rg}$, $r$, and $g$, indicated by dotted cycle in the drawing in \cref{fig:pointinsidetriangle4}(left).
		Let $C_{rg}$ be the bounding cycle of this cell, and we will refer to this cell as the fixed side of~$C_{rg}$.
		Observe that $v$ lies in this fixed side.
		Similarly, in the drawing of~$K_{2,2}$ induced by the endvertices of $r$ and $b$ (resp. $g$ and $b$), we denote by~$C_{rb}$ (resp.~$C_{gb}$) the bounding cycle of the tricell that has $b_{rb}$ (resp.~$b_{gb}$) on its boundary, also indicated by dotted cycle in the drawing in \cref{fig:pointinsidetriangle4}(left).
		We call the cell the fixed side of~$C_{rb}$ (resp.~$C_{gb}$), and $v$ does not lie in this fixed side.
		Following a similar argument as in Case~\ref{inside:case:1}, in the drawing of~$\Delta'$ in $D'$, $v$ lies in the corresponding fixed side of $C_{rg}$ and not in those of $C_{gb}$ and $C_{rb}$; see \cref{fig:pointinsidetriangle4}(right).
		
		In $D'$, since $v$ lies in the fixed side of $C_{rg}$ but not in the fixed side of $C_{gb}$ and since the edge $b_{gb}$ cannot cross $b_{rg}$ due to the simple drawing assumption, $b_{gb}$ has to cross the edge fragment ${\crossing}_{rg} r_1$, where ${\crossing}_{rg}$ is the crossing point of $r$ and $g$. 
		Consider the \cell of the drawing of $K_{2,2}$ induced by $r$ and $b_{gb}$ in $D'$ that $v$ lies.
		In particular, this \cell is formed by $b_{rg}$, $b_{gb}$, and $r$, and hence $b_1$ does not lie in the \cell.
		Hence, the edge $e$ that connects $v$ and $b_1$ has to cross the boundary of the \cell an odd number of times.
		As $e$ cannot cross $b_{gb}$, it has to cross that boundary exactly once.
		We consider two subcases:
		
		\begin{figure}
			\centering
			\includegraphics[page=4]{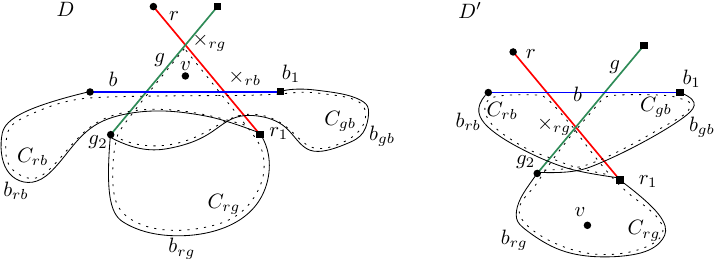}
			\caption {In $D$ (left), $v$ must lie inside the indicated cycle $C_{rg}$ but outside the indicated cycles $C_{gb}$ and $C_{rb}$. Hence, the same must hold for $D'$ (right).}
			\label{fig:pointinsidetriangle4}
		\end{figure}

		\enlargethispage{3ex}	
		\newsavebox{\meinebox}\savebox{\meinebox}[24pt][l]{$r_1 {\crossing}_{rg}$}
		\begin{subcase}[$e$ crosses  \usebox{\meinebox} in $D'$ and does not cross $b_{rg}$]
			In $D'$, because (i) $v$ and $b_1$ lie on the same side of $C_{rb}$, (ii) $e$ crosses $r_1 {\crossing}_{rg}$, and (iii) $e$ cannot cross $b$, we must have (iv) $e$ also crosses $b_{rb}$.
			Similarly, in $D$, because of (i), (iii), and (iv) in the previous statement, we can conclude that $e$ has to cross $r_1 {\crossing}_{rb}$.
			Since $v$ and $b_1$ lie in different sides of $\Delta$ in $D$, $e$ has to cross at least an edge of~$\Delta$.
			Since $e$ crosses $r_1 {\crossing}_{rb}$, it cannot cross ${\crossing}_{rg} {\crossing}_{rb}$.
			Coupled with the fact that $e$ cannot cross $b$, we conclude that $e$ has to cross $g$.
			However, consider now the cycle $C_{rg}$ in $D$.
			When we go along $e$ from $v$ to $b_1$, we first cross $g$ and since we cannot cross $b_{rg}$, we cannot cross ${\crossing}_{rb} r_1$ with the same crossing rotation as in $D'$.
			Hence, we have a contradiction.
		\end{subcase}
		
		\begin{subcase}[$e$ crosses $b_{rg}$ in $D'$]
			We have the situation in $D'$ as in \cref{fig:pointinsidetriangle4}(left).
			As observed at the beginning, $b_{gb}$ crosses $r$ in $D'$, and hence the same holds in $D$.
			As $b_{gb}$ cannot cross $g$, there are two ways to draw $b_{gb}$ in $D$.
			If $b_{gb}$ crosses the edge fragment $r_2 {\crossing}_{rg}$ as in \cref{fig:pointinsidetriangle4}(middle), observe that we cannot draw the edge $e$ that does not cross $b$ or $b_{gb}$ and at the same time crosses $b_{rg}$ with the same crossing rotation as in $D'$.
			Hence, $b_{gb}$ has to cross the edge fragment ${\crossing}_{rb} r_1$.
			However, as $e$ cannot cross $b_{gb}$, the only way we can draw $e$ from $v$ while respecting the rotation of the crossing $e \cap b_{rg}$ is as in \cref{fig:pointinsidetriangle4}(right).
			However, we then cannot complete the drawing of edge $e$ to $b_1$.
		\end{subcase}\noqed
	\end{case}
	This completes the proof of \cref{lem:point_inside_one_triangle}.
\end{proof}

\propflippable*
\begin{proof} 
	As $\Delta$ is an invertible triangle in $D$, there exists an inverted triangle $\Delta'$ in another drawing $D'$ of $K_{m,n}$ with the same \ERS.
	Suppose that there is a vertex $v$ that lies on the inside of $\Delta$. Then, due to \cref{lem:point_inside_both_triangles}, in $D'$ the vertex $v$ cannot lie on the inside of $\Delta'$, and, due to \cref{lem:point_inside_one_triangle}, $v$ also cannot lie on the outside of $\Delta'$.
	Hence, $v$ cannot be drawn anywhere in $D'$, a contradiction.   
	Thus, all vertices of $K_{m,n}$ lie on the outside of $\Delta$ in~$D$. 
\end{proof}

\section{Missing proof of Section~\ref{sec:numberofflips}: Lower bound on the flip distance}\label{sec:appnumberofflips_lower}
\begin{figure}[htb]
	\centering
	\includegraphics[page=2]{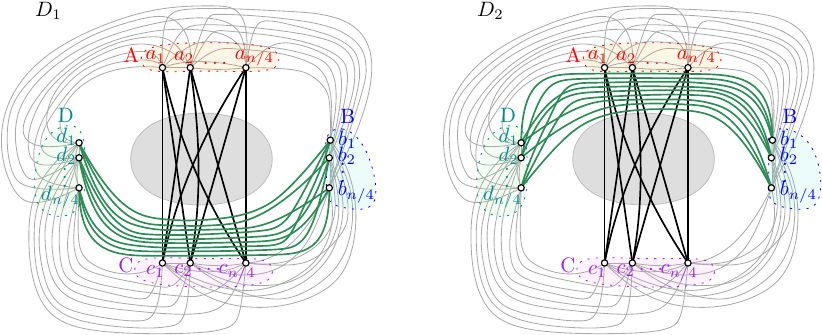}
	\caption {Two simple drawings of $K_n$ with the same \ERS. There are $\Theta(n^2)$ (green) edges between $B$ and $D$ which need to be moved over $\Theta(n^4)$ crossings of (black) edges between $A$ and $C$, resulting in a total of $\Theta(n^6)$ triangle flips.}
	\label{fig:applowerbound}
\end{figure}

To prove Theorem~\ref{thm:distance_lower}, the lower bound on the flip distance, we show that the two drawings of~$K_n$ in \cref{fig:applowerbound} are at flip distance~$\Theta(n^6)$. The figure is the same as \cref{fig:lowerbound} in Section~\ref{sec:numberofflips} and repeated for readability. We further repeat the theorem before its proof for convenience.

\distanceLower*

\begin{proof}
	We first show the bound for $G=K_n$ by giving a construction of $D_1$ and $D_2$; see \cref{fig:applowerbound} for a depiction.
	For convenience assume that~$n$ is divisible by four. Both drawings in \cref{fig:applowerbound} have the same extended rotation system. There are $\Theta(n^2)$ black edges connecting vertices of $A=\{a_1,a_2,\ldots,a_{\frac{n}{4}}\}$ with vertices of $C=\{c_1,c_2,\ldots,c_{\frac{n}{4}}\}$, which generate $\Theta(n^4)$ crossings. All these crossings are in the gray shaded area. Thus, there are $\Theta(n^4)$ crossings between black edges inside this area. All the green edges connecting vertices of $B=\{b_1,b_2,\ldots,b_{\frac{n}{4}}\}$ with vertices of $D=\{d_1,d_2,\ldots,d_{\frac{n}{4}}\}$ cross all the black edges. In the drawing $D_1$ (\cref{fig:applowerbound}(left)), all crossings between a green and a black edge are between the vertices of $C$ and the gray shaded area. In the drawing $D_2$ (\cref{fig:applowerbound}(right)), all crossings between a green and a black edge are between the vertices of $A$ and the gray shaded area. Thus, any two crossing black edges together with every green edge induce a triangle that has to be flipped to transform $D_1$ into $D_2$. Since there are $\Theta(n^2)$ green edges, at least $\Theta(n^6)$ triangles need to be flipped. 
	
	The same arguments hold for any subdrawing which contains the black edges between $A$ and $C$ and the green edges between $B$ and $D$, where an arbitrary subset of the gray edges can be included. Thus, the lower bound holds for a more general class of complete multipartite graphs, as long as it includes complete bipartite subgraphs both between~$A$ and~$C$ and between~$B$ and~$D$. 
\end{proof}
\end{document}